\documentclass{aa}
\usepackage{graphicx,lscape}

\usepackage{ulem}
\usepackage{color}

\usepackage{txfonts}
\usepackage{natbib}
\usepackage{graphicx}
\usepackage{longtable}
\bibpunct{(}{)}{;}{a}{}{,}

\definecolor{red}{rgb}{0.75,0.0,0.0}
\definecolor{yel}{rgb}{0.65,0.65,0.0}
\definecolor{grn}{rgb}{0.0,0.75,0.0}
\definecolor{blu}{rgb}{0.0,0.0,0.75}
\definecolor{gry}{rgb}{0.75,0.75,0.75}

\begin{document}
\def\eps@scaling{.95}
\def\epsscale#1{\gdef\eps@scaling{#1}}
\def\plotone#1{{\centering \leavevmode
    \epsfxsize=\eps@scaling\columnwidth \hbox{\epsfbox{#1}}}}
\def\plottwo#1#2{{\centering \leavevmode
    \epsfxsize=.45\columnwidth \hbox{\epsfbox{#1}} \hfil
    \epsfxsize=.45\columnwidth \hbox{\epsfbox{#2}}}}
\def\plotfiddle#1#2#3#4#5#6#7{{\centering \leavevmode
    \vbox to#2{\rule{0pt}{#2}}
    \includegraphics{#1}}}
\def\annrev{{ARA\&A}}
\def\araa{{ARA\&A}}
\def\aa{{A\&A}}
\def\aasup{{A\&AS}}
\def\aasupp{{A\&AS}}
\def\aj{{AJ}}
\def\apj{{ApJ}}
\def\apjlett{{ApJ}}
\def\apjl{{ApJ}}
\def\apjsupp{{ApJS}}
\def\apjs{{ApJS}}
\def\baas{{BAAS}}
\def\ban{{BAN}}
\def\mn{{MNRAS}}
\def\mnras{{MNRAS}}
\def\nature{{Nature}}
\def\pasj{{PASJ}}
\def\pasp{{PASP}}
\def\apjref#1;#2;#3;#4 {\par\pp#1, {#2}, #3, #4 \par}
\def\refindent{\par\noindent\parskip=2pt\hangindent=3pc\hangafter=1 }
\def\refp#1#2#3#4{\refindent{#1,} {#2}, #3, #4}
\def\refb#1#2#3{\refindent{#1}{ {#2}, }{#3}}
\def\refx#1{\refindent{#1}}

\def\HST{{\it HST}}
\def\ISO{{\it ISO}}
\def\deg{$^{\rm o}$}
\def\degC{$^{\rm o}$C}
\def\arcsec{\ifmmode '' \else $''$\fi}
\def\arcmin{$'$}
\def\arcsecpoint{\ifmmode ''\!. \else $''\!.$\fi}
\def\arcminpoint{$'\!.$}
\def\kms{\ifmmode {\rm km\ s}^{-1} \else km s$^{-1}$\fi}
\def\Msun{\ifmmode {\rm M}_{\odot} \else M$_{\odot}$\fi}
\def\Lsun{\ifmmode {\rm L}_{\odot} \else L$_{\odot}$\fi}
\def\Zsun{\ifmmode {\rm Z}_{\odot} \else Z$_{\odot}$\fi}
\def\ergsAcm{ergs\,s$^{-1}$\,cm$^{-2}$\,\AA$^{-1}$}
\def\ergscm2{ergs\,s$^{-1}$\,cm$^{-2}$}
\def\icm3{{\rm cm}^{-3}}
\def\icm2{{\rm cm}^{-2}}
\def\qo{\ifmmode q_{\rm o} \else $q_{\rm o}$\fi}
\def\Ho{\ifmmode H_{\rm o} \else $H_{\rm o}$\fi}
\def\ho{\ifmmode h_{\rm o} \else $h_{\rm o}$\fi}
\def\ltsim{\raisebox{-.5ex}{$\;\stackrel{<}{\sim}\;$}}
\def\gtsim{\raisebox{-.5ex}{$\;\stackrel{>}{\sim}\;$}}
\def\vFWHM{\ifmmode v_{\mbox{\tiny FWHM}} \else
            $v_{\mbox{\tiny FWHM}}$\fi}
\def\CCF{\ifmmode F_{\it CCF} \else $F_{\it CCF}$\fi}
\def\ACF{\ifmmode F_{\it ACF} \else $F_{\it ACF}$\fi}
\def\Halpha{\ifmmode {\rm H}\alpha \else H$\alpha$\fi}
\def\Hbeta{\ifmmode {\rm H}\beta \else H$\beta$\fi}
\def\Hgamma{\ifmmode {\rm H}\gamma \else H$\gamma$\fi}
\def\Hdelta{\ifmmode {\rm H}\delta \else H$\delta$\fi}
\def\Lya{\ifmmode {\rm Ly}\alpha \else Ly$\alpha$\fi}
\def\Lyb{\ifmmode {\rm Ly}\beta \else Ly$\beta$\fi}
\def\Lyg{\ifmmode {\rm Ly}\beta \else Ly$\gamma$\fi}
\def\fei{Fe\,{\sc i}}
\def\feii{Fe\,{\sc ii}}
\def\feiii{Fe\,{\sc iii}}
\def\feiv{Fe\,{\sc iv}}
\def\fexx{Fe\,{\sc xx}}
\def\hi{H\,{\sc i}}
\def\hii{H\,{\sc ii}}
\def\hei{He\,{\sc i}}
\def\heii{He\,{\sc ii}}
\def\ci{C\,{\sc i}}
\def\cii{C\,{\sc ii}}
\def\ciii{\ifmmode {\rm C}\,{\sc iii} \else C\,{\sc iii}\fi}
\def\civ{\ifmmode {\rm C}\,{\sc iv} \else C\,{\sc iv}\fi}
\def\cv{\ifmmode {\rm C}\,{\sc v} \else C\,{\sc v}\fi}
\def\cvi{\ifmmode {\rm C}\,{\sc vi} \else C\,{\sc vi}\fi}
\def\ni{N\,{\sc i}}
\def\nii{N\,{\sc ii}}
\def\niii{N\,{\sc iii}}
\def\niv{N\,{\sc iv}}
\def\nv{N\,{\sc v}}
\def\nvi{N\,{\sc vi}}
\def\nvii{N\,{\sc vii}}
\def\oi{O\,{\sc i}}
\def\oii{O\,{\sc ii}}
\def\oiii{O\,{\sc iii}}
\def\o5007{[O\,{\sc iii}]\,$\lambda5007$}
\def\oiv{O\,{\sc iv}}
\def\ov{O\,{\sc v}}
\def\ovi{O\,{\sc vi}}
\def\ovii{O\,{\sc vii}}
\def\oviii{O\,{\sc viii}}
\def\neiii{Ne\,{\sc iii}}
\def\nev{Ne\,{\sc v}}
\def\nevi{Ne\,{\sc vi}}
\def\neviii{Ne\,{\sc viii}}
\def\neix{Ne\,{\sc ix}}
\def\nex{Ne\,{\sc x}}
\def\naix{Na\,{\sc ix}}
\def\mgi{Mg\,{\sc i}}
\def\mnii{Mn\,{\sc ii}}
\def\Niii{Ni\,{\sc ii}}
\def\mgii{Mg\,{\sc ii}}
\def\mgx{Mg\,{\sc x}}
\def\siiv{Si\,{\sc iv}}
\def\siIII{Si\,{\sc iii}}
\def\siII{Si\,{\sc ii}}
\def\sixii{Si\,{\sc xii}}
\def\si{S\,{\sc i}}
\def\sii{S\,{\sc ii}}        
\def\siii{S\,{\sc iii}}
\def\siv{S\,{\sc iv}}
\def\sv{S\,{\sc v}}
\def\svi{S\,{\sc vi}}
\def\ariii{Ar\,{\sc iii}}
\def\ariv{Ar\,{\sc iv}}
\def\arv{Ar\,{\sc v}}
\def\arvi{Ar\,{\sc vi}}
\def\arvii{Ar\,{\sc vii}}
\def\arviii{Ar\,{\sc viii}}
\def\caii{Ca\,{\sc ii}}
\def\fei{Fe\,{\sc i}}
\def\feii{Fe\,{\sc ii}}
\def\feiii{Fe\,{\sc iii}}
\def\alii{Al\,{\sc ii}}
\def\aliii{Al\,{\sc iii}}
\def\piii{P\,{\sc iii}}
\def\piv{P\,{\sc iv}}
\def\pv{P\,{\sc v}}
\def\cliv{Cl\,{\sc iv}}
\def\clv{Cl\,{\sc v}}
\def\nai{Na\,{\sc i}}
\def\o{\o}
\newcommand{\pdf}[2]{\frac{\partial #1}{\partial #2}}

\newcommand{\vi}{v_{\scriptscriptstyle 0}}    
\newcommand{\vz}[1]{#1_{\scriptscriptstyle 0}}  
\newcommand{\vy}[2]{#1_{\scriptscriptstyle #2}}
\newcommand{\vu}[1]{#1^{\scriptscriptstyle 0}}  
\def\gtorder{\mathrel{\raise.3ex\hbox{$>$}\mkern-14mu
             \lower0.6ex\hbox{$\sim$}}}
\def\ltorder{\mathrel{\raise.3ex\hbox{$<$}\mkern-14mu
             \lower0.6ex\hbox{$\sim$}}}
\def\proptwid{\mathrel{\raise.3ex\hbox{$\propto$}\mkern-14mu
             \lower0.6ex\hbox{$\sim$}}}

\newcommand{\km}{km s$^{-1}$}
\newcommand{\Ly}{Ly$\alpha$}

\title{Anatomy of the AGN in NGC 5548:  II. The Spatial, Temporal \\
and Physical Nature of the Outflow from HST/COS Observations}

\author{N. Arav\inst{1}
	\and
	C. Chamberlain\inst{1}
	\and
	G.A. Kriss\inst{2,3}
	\and
	J.S. Kaastra\inst{4,5}
	\and
	M. Cappi\inst{6}
	\and
	M. Mehdipour\inst{4,7}
	\and
	P.-O. Petrucci\inst{8,9}
	\and
	K.C. Steenbrugge\inst{10,11}
	\and
	E. Behar\inst{12}
	\and
	S. Bianchi\inst{13}
	\and
	R. Boissay\inst{14}
	\and
	G. Branduardi-Raymont\inst{7}
	\and
	E. Costantini\inst{4}
	\and
	J.C. Ely\inst{2}
	\and
	J. Ebrero\inst{4}
	\and
	L. di Gesu\inst{4}
	\and
	F.A. Harrison\inst{15}
	\and
	S. Kaspi\inst{12}
	\and
	J. Malzac\inst{16,17}
	\and
	B. De Marco\inst{18}
	\and
	G. Matt\inst{13}
	\and
	K.P. Nandra\inst{18}
	\and
	S. Paltani\inst{14}
	\and
	B.M. Peterson\inst{19,20}
	\and
	C. Pinto\inst{21}
	\and
	G. Ponti\inst{18}
	\and
	F. Pozo Nu\~{n}ez\inst{22}
	\and
	A. De Rosa\inst{23}
	\and
	H. Seta\inst{24}
	\and
	F. Ursini\inst{8,9}
	\and
	C.P. de Vries\inst{4}
	\and
	D.J. Walton\inst{15}
	\and
	M. Whewell\inst{7}
}

\offprints{arav@vt.edu}

\institute{
	Department of Physics, Virginia Tech, Blacksburg, VA 24061, USA.
	\and
	Space Telescope Science Institute, 3700 San Martin Drive, Baltimore, MD 21218, USA.
	\and
	Department of Physics and Astronomy, The Johns Hopkins University, Baltimore, MD 21218, USA.
	\and
	SRON Netherlands Institute for Space Research, Sorbonnelaan 2, 3584 CA Utrecht, the Netherlands.
	\and
	Leiden Observatory, Leiden University, Post Office Box 9513, 2300 RA Leiden, Netherlands.
	\and
	INAF-IASF Bologna, Via Gobetti 101, I-40129 Bologna, Italy.
	\and
	Mullard Space Science Laboratory, University College London, Holmbury St. Mary, Dorking, Surrey, RH5 6NT, UK.
	\and
	Univ. Grenoble Alpes, IPAG, F-38000 Grenoble, France.
	\and
	CNRS, IPAG, F-38000 Grenoble, France.
	\and
	Instituto de Astronom\'{i}a, Universidad Cat\'{o}lica del Norte, Avenida Angamos 0610, Casilla 1280, Antofagasta, Chile.
	\and
	Department of Physics, University of Oxford, Keble Road, Oxford, OX1 3RH, UK.
	\and
	Department of Physics, Technion-Israel Institute of Technology, 32000 Haifa, Israel.
	\and
	Dipartimento di Matematica e Fisica, Universit\`{a} degli Studi Roma Tre, via della Vasca Navale 84, 00146 Roma, Italy.
	\and
	Department of Astronomy, University of Geneva, 16 Ch. d'Ecogia, 1290 Versoix, Switzerland.
	\and
	Cahill Center for Astronomy and Astrophysics, California Institute of Technology, Pasadena, CA 91125, USA.
	\and
	Universit\'{e} de Toulouse, UPS-OMP, IRAP, Toulouse, France.
	\and
	CNRS, IRAP, 9 Av. colonel Roche, BP 44346, 31028 Toulouse Cedex 4, France.
	\and
	Max-Planck-Institut f\"{u}r extraterrestrische Physik, Giessenbachstrasse, D-85748 Garching, Germany.
	\and
	Department of Astronomy, The Ohio State University, 140 W 18th Avenue, Columbus, OH 43210, USA.
	\and
	Center for Cosmology \& AstroParticle Physics, The Ohio State University, 191 West Woodruff Avenue, Columbus, OH 43210, USA.
	\and
	Institute of Astronomy, University of Cambridge, Madingley Rd, Cambridge, CB3 0HA, UK.
	\and
	Astronomisches Institut, Ruhr-Universit\"{a}t Bochum, Universit\"{a}tsstra{\ss}e 150, 44801, Bochum, Germany.
	\and
	INAF/IAPS - Via Fosso del Cavaliere 100, I-00133 Roma, Italy.
	\and
	Research Center for Measurement in Advanced Science, Faculty of Science, Rikkyo University 3-34-1 Nishi-Ikebukuro,Toshima-ku, Tokyo, Japan.
}

\date{\today}

\abstract
{AGN outflows are thought to influence the evolution of their host galaxies and super massive black holes. Our deep multiwavelength campaign on NGC 5548 revealed an unusually strong X-ray obscuration. The resulting dramatic decrease in incident ionizing flux on the outflow, allowed us to construct a comprehensive physical, spatial and temporal picture for the long-studied AGN wind in this object.}
{To determine the distance of the outflowing components from the central source; their total column density and the mechanism responsible for the observed absorption trough variability.} 
{We study the UV spectra acquired during the campaign as well as from  four previous epochs, where the outflows  are detected as blue-shifted absorption troughs in the spectra of the object. Our principal analysis tools are ionic column density extraction techniques,  photoinization models based on the code CLOUDY, and collisional excitation simulations.} 
{A simple model based on a fixed total column-density absorber, reacting to changes in ionizing illumination, matches the very different ionization states seen in five spectroscopic epochs spanning 16 years. The main outflow component is situated at 3.5$\pm$1 pc from the central source. Three other components  are situated between 5-70 pc  and two are further than 100 pc. The wealth of observational constraints and the disparate relationship of the observed X-ray and UV flux between different epochs make our physical model a leading contender for interpreting trough variability data of quasar outflows.}
{This campaign, in combination with prior data, yields the first simple model that can explain the physical characteristics and the substantial variability observed  in an AGN outflow.}


\titlerunning{NGC 5548}
\authorrunning{N. Arav et al.}

\maketitle


\section{Introduction}
AGN outflows are detected as blueshifted absorption troughs, with respect to the object systemic redshift. Such outflows in powerful quasars can expel sufficient gas from their host galaxies to halt star formation, limit their growth and lead to the co-evolution of the size of the host and the mass of its central super massive black holes \citep[e.g.,][]{Ostriker10,Hopkins10,Soker11,Ciotti10,Faucher-Giguere12,Borguet13,Arav13}. Therefore, deciphering the properties of AGN outflows is crucial for testing their role in galaxy evolution. 

Nearby bright AGN are excellent laboratories for studying these outflows as they yield: a) high-resolution UV data, which allow us to study the outflow kinematics and can yield diagnostics for their distance from the central source; and b) high quality X-ray spectra that give the physical conditions for the bulk of the outflowing material 
\citep[e.g.,][]{Steenbrugge05,Gabel05b,Arav07,Costantini07,Kaastra12}. Thus, such observations are a vital stepping stone for quantifying outflows from the luminous (but distant) quasars, for which high quality X-ray data are not available.

For these reasons, we embarked on a deep multiwavelength campaign on the prototypical AGN outflow seen in the intensively studied Seyfert 1 galaxy NGC 5548. For the past 16 years, this outflow has shown 6 kinematic components in the UV band \citep[labeled in descending order of velocity, following][]{Crenshaw03b}, and their associated X-ray warm absorber (WA). Our 2013 campaign revealed a new X-ray obscurer accompanied by broad UV absorption \citep[analyzed in][]{Kaastra14}.  The appearance of the obscurer allows us to derive a comprehensive physical picture of the long-term observed outflow, which we report here. 

The plan of the paper is as follows: In \S~\ref{sect:obs} we describe the observations and data reduction; in \S~ \ref{sect:comp1} we analyze the key component of the outflow;  in \S~\ref{sect:comp2-6} we discuss the remaining 5 components; in \S~\ref{sect:WA} we connect the results of the UV analysis with those of the X-ray warm absorber of the same outflow; and in \S~\ref{sect:discussion} we compare our results with previous studies, discuss the implication of our results to the variability of AGN outflow troughs in general, and elaborate on the connection between the X-ray obscurer and the persisting outflow; In \S~\ref{sect:summary} we summarize our results.

\section{Observations and Data Reduction}\label{sect:obs}

Our 2013 multiwavelength campaign on NGC 5548 included coordinated
observations using XMM-Newton, HST,  Swift, INTEGRAL, and NuSTAR. 
 \citet{Kaastra14} describe the overall structure of the campaign. A full log of all
the observations is given by \citet{Mehdipour14}. Here we present a more
detailed analysis of the UV observations we obtained using the Cosmic Origins
Spectrograph (COS) \citep{Green2012} onboard {\it HST}. We obtained five COS observations
simultaneously with five of the XMM-Newton observations between 2013-06-22 and
2013-08-01. Each two-orbit observation used gratings G130M and G160M at multiple
central wavelength settings and multiple focal-plane positions (FP-POS) to cover
the wavelength range from 1132 \AA\ to 1801 \AA\ at a resolving power of
$\sim$15,000. Table~\ref{table:observations} lists the observation dates of the individual visits, the
exposure times, and the continuum flux measured at 1350 \AA\ in the rest frame,
as well as corresponding information for archival HST observations of NGC 5548
that are also used in this analysis. The five observations from the summer
of 2013 were optimally weighted to produce an average spectrum that we use for
our analysis. \citet{Kaastra14} describe the data reduction process from the
calibration of the data to the production of this average spectrum. 

The 2013 average HST/COS spectrum with all identified absorption features is
shown in Figure A1. As described in Kaastra et al. (2014), we
modeled the emission from NGC 5548 using a reddened power law \citep[with extinction
fixed at $\rm E(B - V) = 0.02$,][]{Schlegel98}, weak \feii\ emission
longward of 1550 \AA\ in the rest frame, broad and narrow emission lines modeled
with several Gaussian components, blue-shifted broad absorption on all permitted
transitions in NGC 5548, and a Galactic damped Ly$\alpha$ absorption line. Using
this emission model, we normalized the average spectrum to facilitate our
analysis of the narrow intrinsic absorption lines in NGC 5548. Figure 1 shows
normalized spectra for absorption lines produced by \siIII~$\lambda1206$, 
\siiv~$\lambda\lambda1394,1403$, \civ~$\lambda\lambda1548,1550$, \nv~$\lambda\lambda1238,1242$, 
and Ly~$\alpha$ as a function of rest-frame velocity relative to a systemic redshift of
$z = 0.017175$ \citep{deVaucouleurs91} via the NASA/IPAC Extragalactic Database
(NED).

As shown in Table~\ref{table:observations}, high-resolution UV spectra of NGC 5548 using HST cover an
additional four epochs stretching back to 1998. We use the calibrated data sets
for each of these observations as obtained from the Mikulski Archive for Space
Telescopes (MAST) at the Space Telescope Science Institute (STScI). We compare
the strengths of the narrow UV absorption troughs for each of these epochs with
our new data set from 2013 in Figs. A2 and A3.

\begin{figure}
\begin{center}
\includegraphics[width=8.5truecm,angle=0]{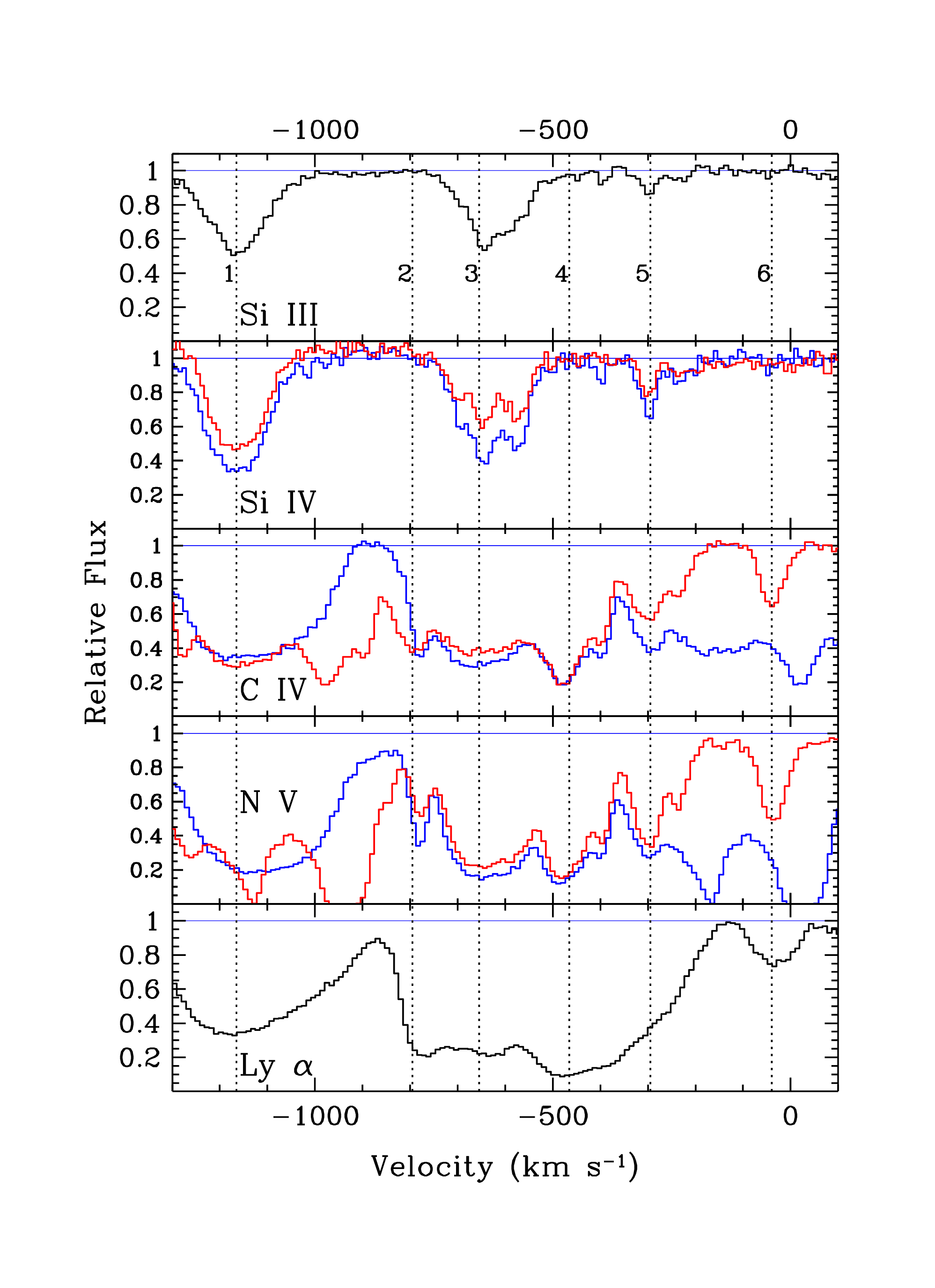}
\caption{Intrinsic absorption features in the 2013 COS spectrum of NGC 5548.
Normalized relative fluxes are plotted as a function of velocity relative to
the systemic redshift of $z=0.017175$, top to bottom: \siIII~$\lambda1206$, 
\siiv~$\lambda\lambda1394,1403$, \civ~$\lambda\lambda1548,1550$, \nv~$\lambda\lambda1238,1242$, 
and Ly~$\alpha$, as a function of rest-frame velocity. For the doublets the red and blue components are shown in red and blue, respectively. 
Dotted vertical lines indicate the velocities of the absorption
components numbered as in Crenshaw et al. (2003).
}

  \label{fig_cosvel}
\end{center}
\end{figure}


\section{Physical and temporal characteristics of Component 1} \label{sect:comp1}

The key for building a coherent picture of the long-seen outflow is component 1: the strongest and highest velocity outflow component (centered at $-$1160 km~s$^{-1}$). Due to the strong suppression of incident ionizing flux by the obscurer, the 2013 HST/COS data of component 1 show a wealth of absorption troughs from ions never before observed in the NGC 5548 outflow. These data allow us to decipher the
 physical characteristics of this component.  In \S~\ref{NU} we use the column density measurements of \pv, \piii, \feiii, and \siII\ as input in photoionization models, and derive the total hydrogen column density for component 1 of $\log(N_H)=21.5_{-0.2}^{+0.4}$ cm$^{-2}$, and an ionization parameter of $\log(U_H)=-1.5_{-0.2}^{+0.4}$. In \S~\ref{Density} we use the column density measurements of \ciii* and \siIII* to infer the electron number density $\log(n_e)=4.8\pm0.1$ cm$^{-3}$, which combined with  the value of the incident $U_H$ yields a distance $R=3.5\pm1$ parsec between component 1 and the central source. In \S~\ref{temporal} we construct a simple model based on a fixed total column-density absorber, reacting to changes in ionizing illumination, that matches the very different ionization states seen in five HST high-resolution spectroscopic epochs spanning 16 years.

\begin{figure}
\begin{center}
\includegraphics[width=8.5truecm,angle=0]{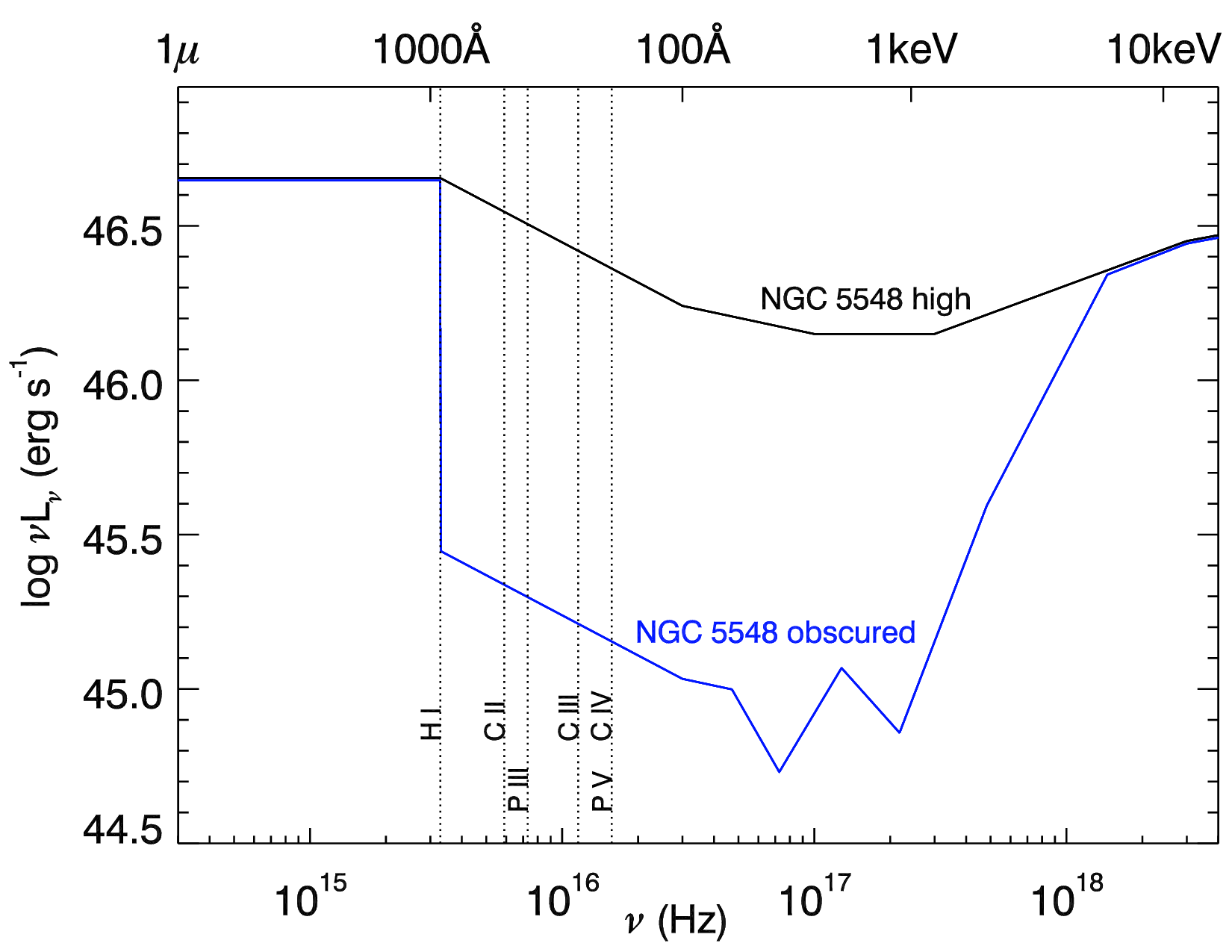}
\caption{The adopted NGC 5548 SEDs. In black we show the SED for NGC 5548 in 2002 at an unobscured X-ray flux \citep[hereafter ``high SED,'' from][]{Steenbrugge05}. In blue we show the SED appropriate to the 2013 epoch, where for the same flux at 1000 \AA, the new X-ray obscurer reduced the ionizing flux by a factor of 17 between 1 Ry and 1 keV (obscured SED). The ionization potentials for destruction of some of the prominent observed species are shown by the vertical lines.
}
\label{fig:SED}
\end{center}
\end{figure}

\subsection{Total Column Density $\left(N_H\right)$ and Ionization Parameter ($U_H$)}\label{NU}

In the Appendix we describe the methods we use to derive the ionic column densities $\left(N_{ion}\right)$ from the outflow absorption troughs.
Table~2 gives the $N_{ion}$ measurements for all observed troughs of all 6 outflow components in the 5 HST epochs (spanning 16 years) of high-resolution UV spectroscopy. The upper and lower limits were derived using the Apparent Optical Depth (AOD) method. All the reported measurements were done using the Partial Covering (PC) method. We used the Power-Law (PL) method only on the \ciii* in order to quantify the systematic error due to different measuring methods (see \S~\ref{Density} and Fig.~\ref{fig:density}).

The $N_{ion}$ we measure are a
result of the ionization structure of the outflowing material, and  can be compared to photoionization models to determine the
physical characteristics of the absorbing gas.  As boundary conditions for the photoionization models, we need to specify the choice of incident Spectral Energy Distribution (SED), and the chemical abundances of the outflowing gas. 

We make the simple (and probably over-restrictive) assumption that the shape of the SED emitted from the accretion disk did not change over the 16 years of high-resolution UV spectroscopy.  Specifically, we assume that the emitted SED has the same shape it had when we obtained simultaneous X-ray/UV observations in 2002 
  \citep[the ``high'' SED in Figure~\ref{fig:SED}, determined by][]{Steenbrugge05}. In 2013 the obscurer absorbed much of the soft ionizing photon flux from this SED before it reached component 1. We model the incident SED on component 1 as the ``obscured'' SED in Figure~\ref{fig:SED}, and further justify its specific shape in \S~\ref{temporal}
For abundances, we use pure proto-Solar abundances given by \citet{Lodders09}.

With the choice of SED and chemical abundances, two main parameters 
govern the photoionization structure of the absorber: the total
hydrogen column density $\left(N_H\right)$ and the ionization
parameter
\begin{equation}
U_H\equiv\frac{{\displaystyle Q_H}}{{\displaystyle 4\pi R^2 \vy{n}{H} c}} \label{UEqn}
\end{equation}
where $Q_H$ is the rate of hydrogen-ionizing photons emitted by the
object, $c$ is the speed of light, $R$ is the distance from the
central source to the absorber and $\vy{n}{H}$ is the total hydrogen
number density. We model the photoionization structure and predict
the resulting ionic column densities by self-consistently solving the
ionization and thermal balance equations with version  13.01 of the
spectral synthesis code {\sc Cloudy}, last described in
\cite{Ferland13}.  We assume a plane-parallel geometry for a gas of
constant $\vy{n}{H}$.

 To find the pair of
$\left(U_H,N_H\right)$ that best predicts the set
of observed column densities, we vary $U_H$ and $N_H$ in 0.1 dex steps
to generate a grid of models \citep[following the same approach
  described in][]{Borguet12a} and perform a minimization of the
function
\begin{equation}
\chi^2=\sum_i\left(\frac{\log (N_{i,mod}) - \log (N_{i,obs})}{\log (N_{i,obs}) - 
\log \left(N_{i,obs} \pm \sigma_i \right)} \right)^2 
\label{chi2_linear}
\end{equation}
where, for ion $i$, $N_{i,obs}$ and $N_{i,mod}$ are the observed and
modeled column densities, respectively, and $\sigma_i$ is the error in
the observed column density.  The measurement errors are not symmetric. We use the positive error ($+\sigma_i$) when 
$\log (N_{i,mod}) > \log (N_{i,obs})$ and the negative error ($-\sigma_i$) when $\log (N_{i,mod}) < \log (N_{i,obs})$.

\begin{figure}
\begin{center}
\includegraphics[width=8.5truecm,angle=0]{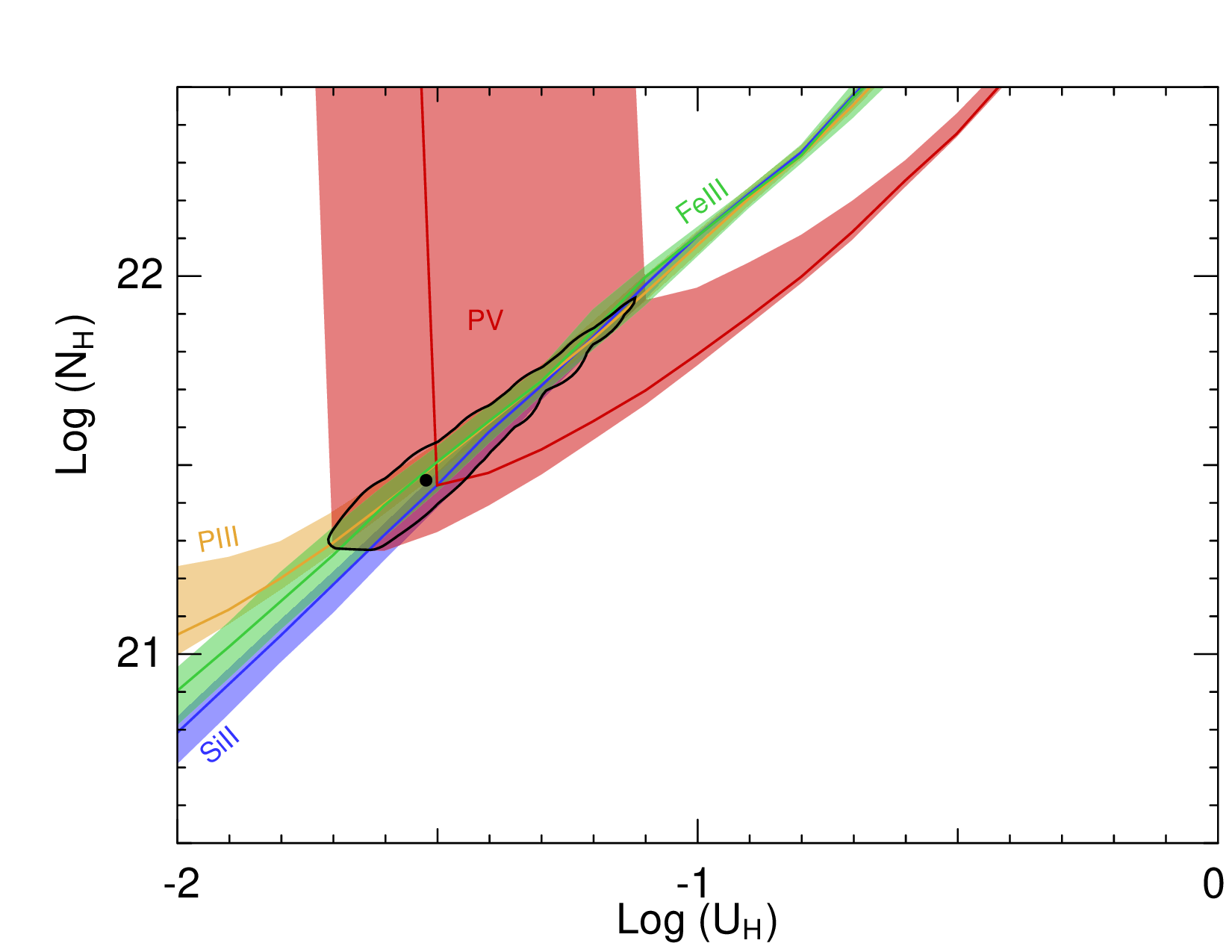}
\caption{A photoionization phase plot showing the ionization solution  for component 1  epoch 2013. We use the obscured SED and assumed proto-solar metalicity \citep{Lodders09}.  Solid lines and associated colored bands represent the locus of  $U_H,N_H$ models, which predict the 
measured $N_{ion}$, and their 1$\sigma$ uncertainties. The black dot is the best solution and is surrounded by a 1$\sigma \ \ \chi^2$ contour. 
}
\label{fig:2013_only_phase_plot}
\end{center}
\end{figure}

The ionization solution  for component 1 at the 2013 epoch is shown in figure \ref{fig:2013_only_phase_plot}. We only show constraints from $N_{ion}$ measurements, and note that all the lower limits reported in Table~2 are satisfied by this solution.
We find $\log(N_H)=21.5_{-0.2}^{+0.4}$ cm$^{-2}$, and an ionization parameter of $\log(U_H)=-1.5_{-0.2}^{+0.4}$, where the errors are strongly correlated as illustrated by the 1$\sigma$ $\chi^2$ contour.

\subsection{Number Density and Distance}\label{Density}

\begin{figure}
\begin{center}
\includegraphics[width=8.5truecm,angle=0]{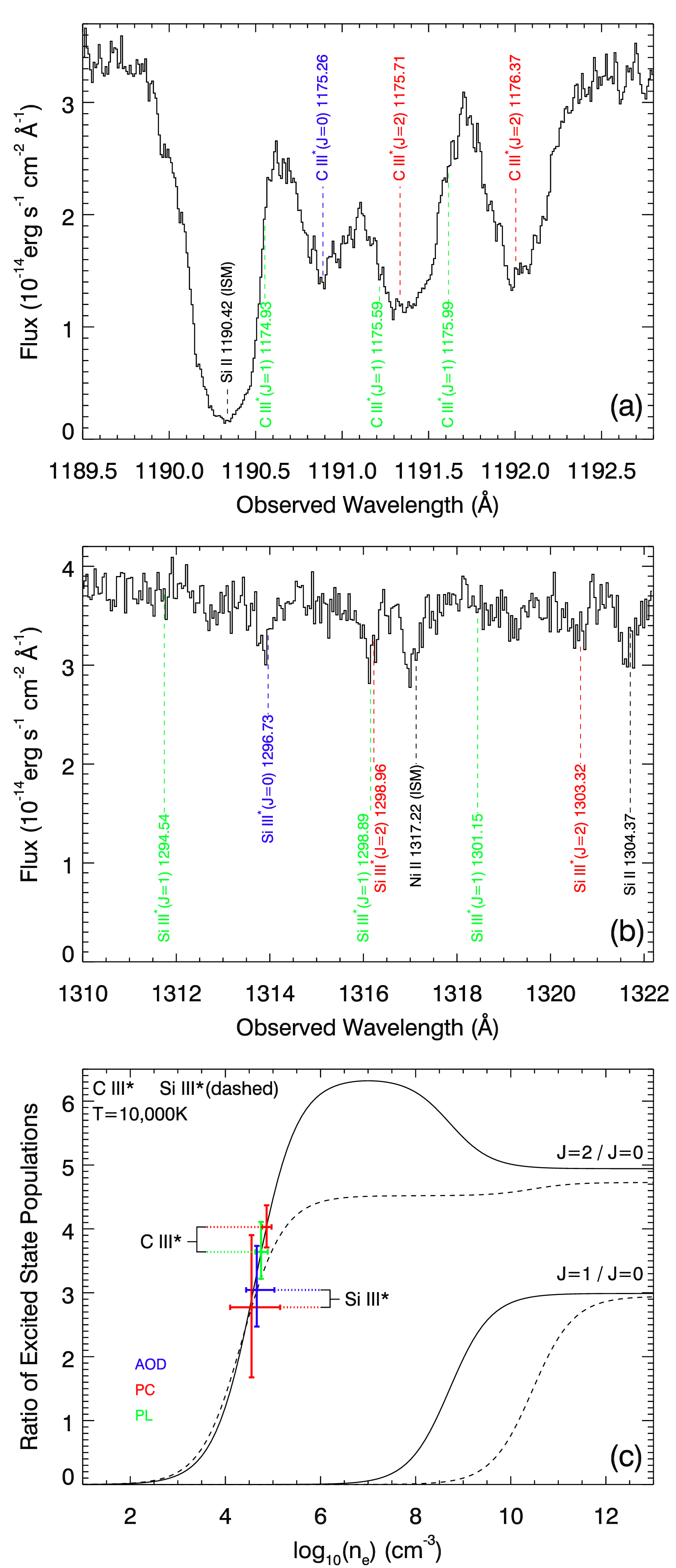}
\caption{
{\bf a}) Absorption spectrum for the \ciii* 1175 \AA\ multiplet. The 2013 COS spectrum shows clear and relatively unblended individual troughs from the J=2 and J=0 levels, but no contribution from the J=1 level that is populated at higher densities (see panel c).  {\bf b)} Absorption spectrum for the \siIII* 1298 \AA\ multiplet (similar in level structure to the \ciii* 1175 \AA\ multiplet). The 2013 COS spectrum shows shallow but highly significant individual troughs from the J=2 and J=0 levels, but again no contribution from the J=1 level that is populated at higher densities. {\bf c)} \ciii* and \siIII* level population ratios, theory and measurements. The computed populations for the J=2/J=0 and J=1/J=0 are plotted as a function of electron number density for both ions (see text for elaboration).  The crosses show the measured ratios  for the J=2/J=0 ratio of both ions. From the \ciii* ratios we infer $\log(n_e)=4.8\pm0.1$ cm$^{-3}$, where the error includes both statistical and systematic effects. This value is fully consistent with the one inferred from the \siIII* ratio, where in the case of \siIII* the statistical error is larger since the troughs are much shallower.
}
\label{fig:density}
\end{center}
\end{figure}

As shown in Figure \ref{fig:density}, we detect absorption troughs from the \ciii* 1175 \AA\ multiplet, arising from the metastable 3P$_j$ levels of the 2s2p term. As detailed in \citet{Gabel05b}, the excited  \ciii* 1175 \AA\ multiplet comprises six lines arising from three J levels. The J=0 and J=2 levels have significantly lower radiative transition probabilities to the ground state than the J=1 level and are thus populated at much lower densities than the latter. In particular, figure 5 in \citet{Borguet12b} shows that the relative populations of the three levels are a sensitive probe to a wide range of $n_e$ while being insensitive to temperature. 
The two high-S/N troughs from the J=2 level allow us to accurately account for the mild saturation in these troughs and therefore to derive reliable $N_{ion}$ for both levels. From these measurements we infer $\log(n_e)=4.8\pm0.1$ cm$^{-3}$ (see Fig.~\ref{fig:density}).
We also detect two shallow troughs from the same metastable level of \siIII*  (see panel b of Fig.~\ref{fig:density}), which we use to measure an independent and consistent value for $n_e$ (see panel c of Fig.~\ref{fig:density}). The collisional excitation simulations shown in Figure \ref{fig:density}c were performed using version 7.1.3 of CHIANTI \citep{Dere97,Landi13}, with a temperature of 10,000K (similar to the predicted temperature of our CLOUDY model for component 1 during the 2013 epoch).

In addition to the \ciii* and \siIII* troughs, the 2013 COS spectra of component 1 show several additional troughs from excited states:
\cii*, \siII*, \piii*, \siii* and \feiii* (as well as their associated resonance transitions). Careful measurement of those troughs show that in these cases, the deduced $n_e$ is a lower limit that is larger than the critical density of the involved excited and zero energy levels.  In all cases the critical densities are below $\log(n_e)=4.8$ cm$^{-3}$, thus they are consistent with the $n_e$ measurement we derive from the \ciii* and \siIII* troughs.

As can be seen from the definition of the ionization parameter $U_H$ (Equation \ref{UEqn}), knowledge of the hydrogen number density 
$\vy{n}{H}$ for a given  $U_H$ and $N_H$ allows us to derive the distance $R$. Our photoionization models show that for component 1, $\log(U_H)=-1.5$ and since $n_e\simeq1.2\vy{n}{H}$  (as is the case for highly ionized plasma), $\vy{n}{H}=5.3\times10^{4}$  cm$^{-3}$. To determine the $Q_H$ that affects component 1, we first calculate the bolometric luminosity using the average flux at 1350\AA\ for visits 1-5 in 2013, the redshift of the object and the obscured SED (see Fig.~\ref{fig:SED}). We find $L_{bol}=2.6\times10^{44}$  erg s$^{-1}$ and from it, $Q_H=6.9 \times10^{52}$ s$^{-1}$. Therefore, equation (\ref{UEqn}) yields $R=3.5_{-1.2}^{+1.0} $ pc, where the error is determined from propagating the errors of the contributing quantities. 

Assuming the canonical 50\% global covering factor for Seyfert outflows  \citep{Crenshaw99}, and using equation (1) in \citet{Arav13} we find that the mass flux associated with the UV manifestation of component 1  is $1.0_{-0.5}^{+2.0}$ solar masses per year, and that the kinetic luminosity is $4_{-2}^{+8}\times10^{41}$ erg s$^{-1}$. We note that most of the $N_H$ in the various outflow components is associated with the higher ionization X-ray phase of the outflow. Therefore, we defer a full discussion of the total mass flux and kinetic luminosity of the outflow to a future paper that will present a combined  analysis of the UV and X-ray data sets.

\subsection{Modeling the Temporal Behavior of the Outflow}\label{temporal}

The absorption troughs of component 1 change drastically between the five HST high-resolution spectroscopic epochs spanning 16 years (see figures A2 and A3).
After finding the location and physical characteristics of component 1 using the 2013 data, the next step is to derive a self-consistent temporal picture for this component. There are two general models that explain trough variability in AGN outflows
\citep [e.g.,][and references therein]{Barlow92,Gabel03,Capellupo12,Arav12,Filiz13}. 
One model attributes the trough variability to changes of the ionizing flux experienced by the outflowing gas. In its simplest form, this model assumes that the total $N_H$ along the line of sight does not change as  a function of time.
A second model invokes material moving across the line of sight, which in general causes changes of $N_H$ along the line of sight as a function of time to explain the observed trough changes.

In the case of component 1, we have enough constraints to exclude the model of material moving across our line of sight. The outflow is situated at 3.5 pc from the central source, which combined with the estimated mass of the black hole in NGC 5548 
\citep[4$\times10^7$ solar masses;][]{Pancoast13}, yields a Keplerian speed of $1.9\times10^7$ cm~s$^{-1}$ at that distance.
As can be seen from figure A1, 2/3  of the emission at the wavelength of component 1 arises from the \civ\ Broad Emission Line (BEL). Therefore, the transverse motion model crucially depends on the ability of gas clouds to cross most of the projected size of the Broad Line Region (BLR) in the time spanning the observations epochs. Reverberation studies \citep{korista95} give the diameter of the \civ\ BLR as 15 light days  or $3.9\times10^{16}$ cm, which for $v_{\bot}=1.9\times10^7$ cm~s$^{-1}$, yields a crossing time of $2.0\times10^9$ seconds, or 65 years. 

Thus, in the 16 years between our epochs, material that moves at the Keplerian velocity, 3.5 pc away from the NGC 5548 black hole, will cross only about 25\% of the projected size of the \civ\ BLR. Therefore, the much larger change in the residual intensity of the component 1 \civ\ trough cannot be attributed to new material appearing due to transverse motion at this distance.
We note that 25\% motion across the projected size of the \civ\ BLR is a highly conservative limit for two reasons: 1) at certain velocities there are changes of 50\% in the residual intensity in the component 1 \civ\ trough  between the 2011 and 2013 epochs; and in the elapsing 2 years,  transverse motion will only cover 3\% of the projected size of the \civ\ BLR; 2) material that moves away from the central source under the influence of radial forces should conserve its angular momentum. Therefore, if it moved to distances that are much larger compared with its initial distance, its $v_{\bot}\ll v_{kep}$ at its current distance.
We conclude that even under favorable conditions, the transverse motion model of gas into or out of the line of sight cannot explain the observed behavior of component 1 over the 5 observed epochs.

Can changes of the ionizing flux experienced by the outflowing gas explain the observed trough changes?
We construct such a model under the simplest and restrictive assumption that the $N_H$ of component 1 did not change over the 16 years spanning the 5 high resolution UV spectral epochs. Furthermore, for $\log(n_e)=4.8$ (cm$^{-3}$), the absorber should react to changes in incident ionizing flux on time-scales of ~5 days \citep[see eq.\ \ref{eq:trec} here, and discussion in][]{Arav12}. Therefore, for component 1 we use the restrictive assumption of a simple photoionization equilibrium, determined by the flux level of the specific observation.

\begin{figure}
\begin{center}
\includegraphics[width=8.5truecm,angle=0]{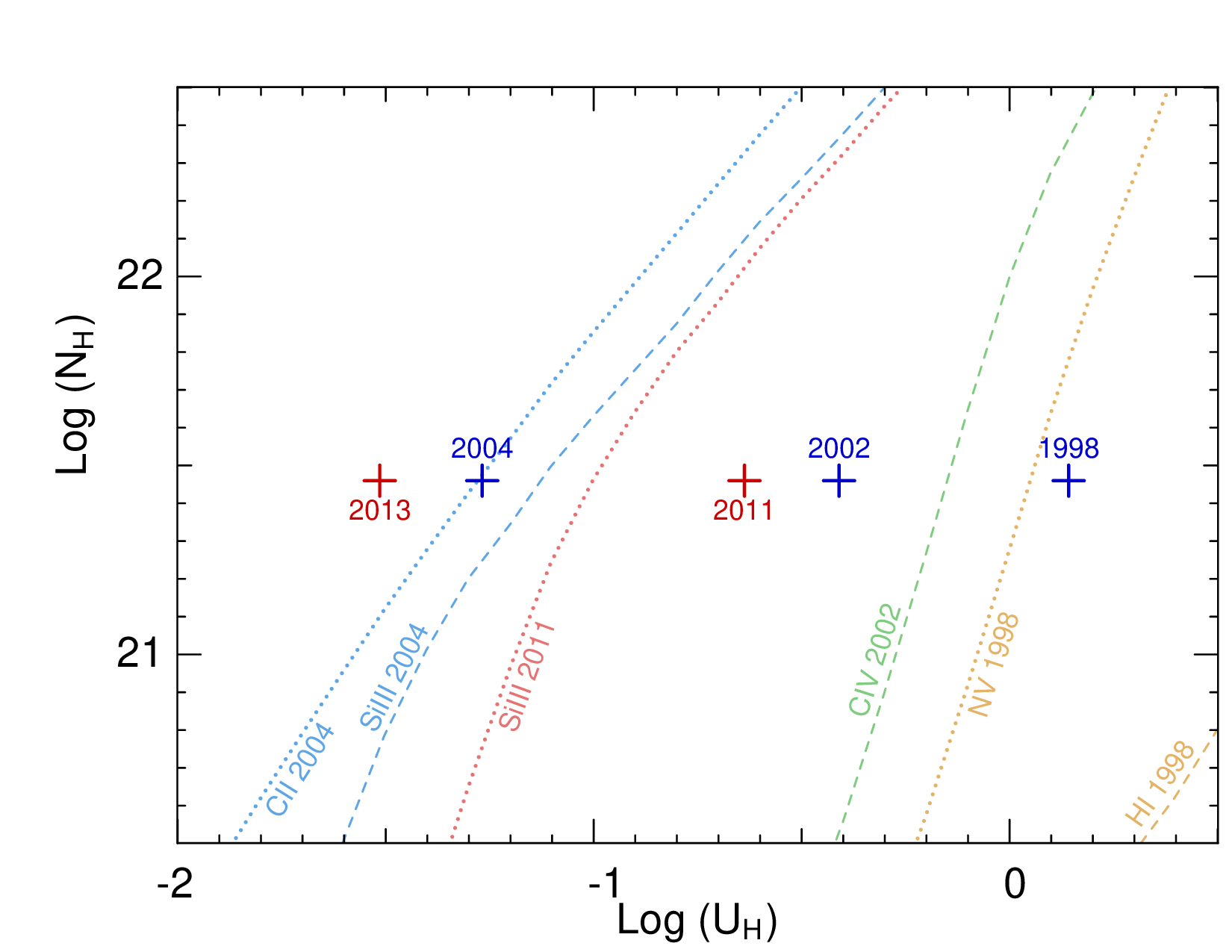}
\caption{A photoionization phase plot showing the ionization solutions  for component 1  for all 5 epochs. The 2013 epoch solution is identical to the one shown in Fig.~\ref{fig:2013_only_phase_plot}.  For the 1998, 2002, and 2004 epochs we used the high SED (see Fig.~\ref{fig:SED})  with the same abundances, and their ionization solutions are shown in blue crosses. Dashed lines represent $N_{ion}$ lower-limit that allow the phase-space above the line, while the dotted lines are upper-limits that allow the phase-space below the line (only the most restrictive constraints are shown). $N_H$ is fixed for all epochs at the value determined from the 2013 solution. For 1998, 2002 and 2004 the difference in $U_H$ values is determined by the ratio of fluxes at 1350\AA\ and the actual value is anchored by the observed $N_{ion}$ constraints. As explained in \S~\ref{temporal}, the $U_H$ position of the 2011 epoch is less tightly constrained. The solution for each epoch satisfies all the $N_{ion}$ constraints for that epoch.
}
\label{fig:5epochs_phase_plot}
\end{center}
\end{figure}

In 1998 the AGN was in a high flux level of $F$=6 (measured at 1350 \AA\ rest-frame and given in units of 
$10^{-14}$ ergs~s$^{-1}$~cm$^{-2}$~\AA$^{-1}$). At that epoch the absorber only showed a \Ly\ trough necessitating $\log(U_H)\gtorder0.1$ (otherwise a \nv\  trough would be detected, see Fig.~\ref{fig:5epochs_phase_plot}). In 2002 the AGN was in a medium flux level with $F$=2, at which time the absorber showed \civ\ and \nv\ troughs in addition to \Ly. In 2004 the AGN was at a historically low flux of $F$=0.25. In that epoch a \siIII\ trough appeared in addition to the \civ, \nv\ and \Ly\ troughs; however, a \cii\ trough did not appear. The combination of the \siIII\ and \cii\ constraints necessitates $-1.3<\log(U_H)<-1.15$ (see Fig.~\ref{fig:5epochs_phase_plot}). The change in $\log(U_H)$ required by the photoionization models agrees remarkably well with the change in flux between the 1998 and 2004 epochs as 
$\log(F_{2004}/F_{1998})=-1.4$. Thus, a constant $N_H$ absorber yields an excellent fit for the absorption features from two epochs with the same spectral energy distribution (high SED in Figure Fig.~\ref{fig:5epochs_phase_plot}), but with very different $U_H$ values. Comparison of several key troughs between the 5 epochs is shown in Figs. A2 and A3.  The 1350 \AA\ flux measurements, plus observation details are given in Table~\ref{table:observations} and the derived column densities for all the outflow features are given in Table~\ref{table:Nion}.

In 2013 the AGN flux was $F$=3. With this flux level and assuming the same SED, the $U_H$ value should have been 50\% higher than in 2002, and we would expect to see only \civ, \nv\ and \Ly\ troughs. Instead we also detect \siIII, \cii, \siII\ and  \alii. Therefore, the incident SED for component 1 must have changed, and indeed the 2013 soft X-ray flux is 25 times lower compared to that of the 2002 epoch 
\citep[see Fig.~1 in][]{Kaastra14}. This drop is caused by the newly observed obscurer close to the AGN, which does not fully cover the source \citep{Kaastra14}. We found a good match to the UV absorption and soft X-ray flux with an SED that is similar to that of the high flux one longward of 1 Ry, but abruptly drops to 6\% of that flux between 1~Ry and 1~keV (see Fig.~\ref{fig:SED}). This picture is consistent with the transmitted flux resulting from the low-ionization, partial covering model of the obscurer derived in \citet{Kaastra14}. To complete the UV picture for component 1, in 2011 NGC 5548 showed $F$=6, equal to that of the 1998 epoch. However, \civ\ and \nv\ are clearly seen in the 2011 epoch but not any of the other ionic species seen in 2013. To explain this situation we assume that the obscurer was present at the 2011 epoch, but it only blocked somewhere between 50--90\% of the emitted ionizing radiation between 1 Ry and 1 keV. The possible presence of a weaker X-ray obscurer is also suggested by broad absorption on the blue wing of the C IV emission line in 2011 that is weaker than that seen in 2013.

 To summarize, a simple model based on a fixed total column-density absorber, reacting to changes in ionizing illumination, matches the very different phenomenology seen in all high-resolution UV spectra of component 1 spanning 16 years. Figure \ref{fig:schematics} gives a schematic illustration of the temporal model.

\section{Components 2-6} \label{sect:comp2-6}
What are the distances and physical conditions ($N_H$ and $U_H$) for the other 5 outflow components (2-6)? As we show below, we can derive distances (or interesting limits of) for all these UV components and some loose constraints on the $N_H$ and $U_H$ of components 3 and 5. 
 Figure \ref{fig:schematics} gives  the velocities and distances of all 6 components, as well as a 
schematic illustration for their temporal  behavior.

{\bf Constraining the Distances:} components 2-6 do not show absorption from excited levels (Except for component 3, whose \cii/\cii* troughs only yield a lower limit for $n_e$.). However, components 3 and 5 show clear variations in their \siIII\ and \siiv\ troughs between our 2013/6/22 and 2013/8/1 observations, but not between the 2013/7/24 and 2013/8/1 ones, when we do see changes in component 1. Using the formalism given in \S~4 of \citet{Arav12}, we can deduce the $n_e$ of these components from the observed time lags.
Suppose an absorber in photoionization equilibrium experiences a sudden change
in the incident ionizing flux such that $I_i(t>0) = (1+f)I_i(t=0)$, where $-1
\le f \le \infty$. Then the timescale for change in the ionic fraction is given by:
\begin{equation}
 t^* = \left[ -f \alpha_i n_e \left( \frac{n_{i+1}}{n_i}-
\frac{\alpha_{i-1}}{\alpha_i} \right) \right]^{-1},
\label{eq:trec}
\end{equation}
where $\alpha_i$ is the recombination rate of ion $i$ and $n_i$ is the fraction of a given element in ionization stage $i$ .
From the 40 days that separate epochs showing trough changes and the 7 days separating epochs with no change, we can deduce $3.5<\log(n_e)<4.5$ (cm$^{-3}$) for both components, otherwise their troughs would not react to changes in incident ionizing flux in the observed way, despite the large changes in incident ionizing flux over that period. Assuming a similar $U_H$ to that of component 1 (see discussion below), this $n_e$ range yields distances of $5<R<15$ parsec. Similarly, component 6 shows \civ\ and \nv\ troughs in 2011 but not in 2002. This nine years timescale yields $R<100$ parsec. Component 2 and component 4 do not show  changes in the UV absorption between any of the epochs. Therefore, we can derive a lower limit for their distance of $R>130$ parsec.  

{\bf Constraining $N_H$ and $U_H$:} It is not feasible to put physically interesting constraints on components 2, 4 and 6. First, they only show troughs from \civ, \nv, and \Lya, which (based on our analysis of component 1) are probably highly saturated. Second,  the ionization time-scales of components 2 and 4 are larger than 16 years.  Therefore, even if a $U_H$ can be deduced from the measurements, it will only be a representative average value for a period of time larger than 16 years.

\begin{figure}
\begin{center}
\includegraphics[width=8.5truecm,angle=0]{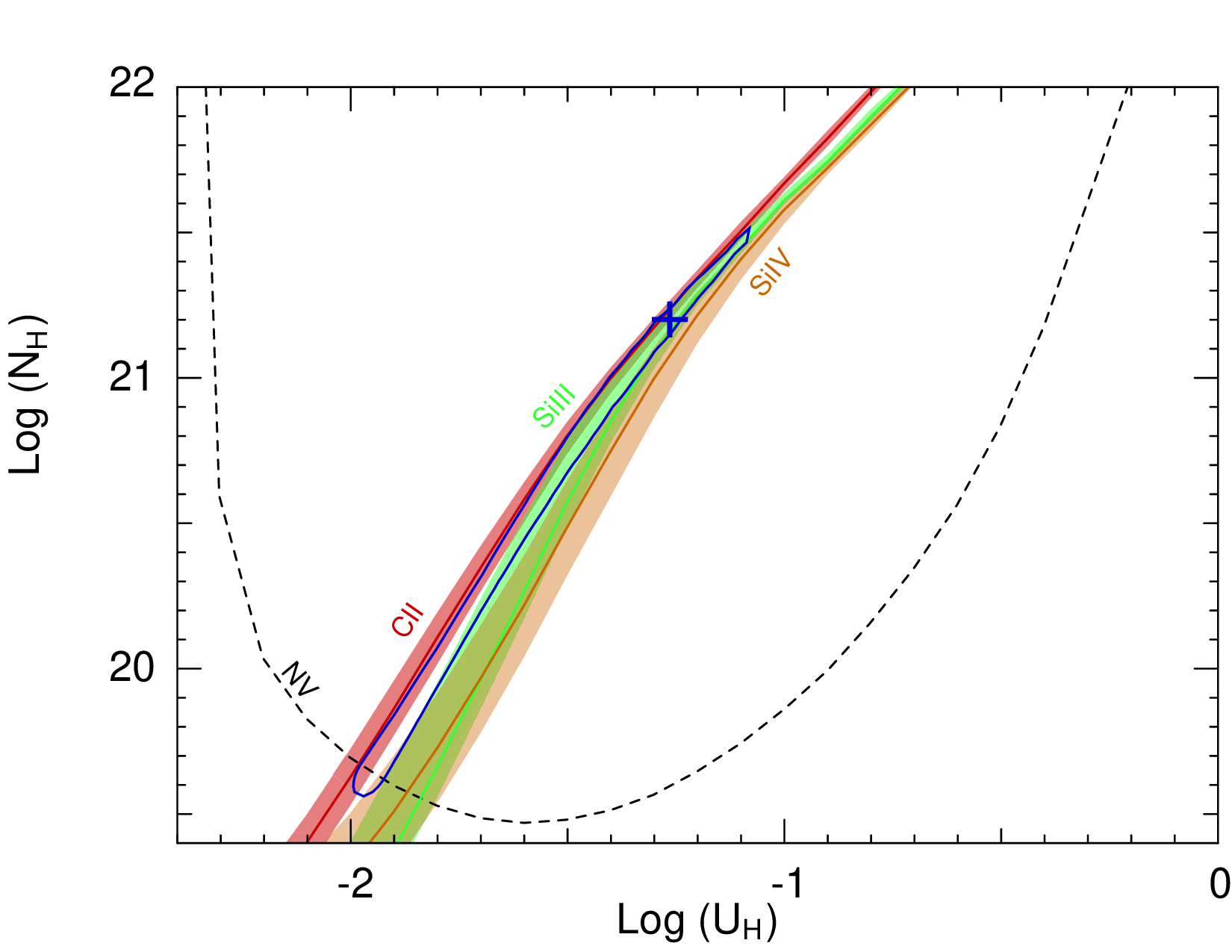}
\caption{A photoionization phase plot showing the ionization solution  for component 3  epoch 2013. As for component 1, we use the obscured SED and assumed proto-solar metalicity \citep{Lodders09}.  Solid lines and associated colored bands represent the locus of  $U_H,N_H$ models, which predict the 
measured $N_{ion}$, and their 1$\sigma$ uncertainties, while the dashed line is the lower limit on the \nv\ column density that permits the phase-space above it. The blue cross is the best $\chi^2$ solution and is surrounded by a 1$\sigma \ \ \chi^2$ blue contour. 
}
\label{fig:comp3_2013_phase_plot}
\end{center}
\end{figure}

\begin{figure*}
\begin{center}
\includegraphics[width=\textwidth]{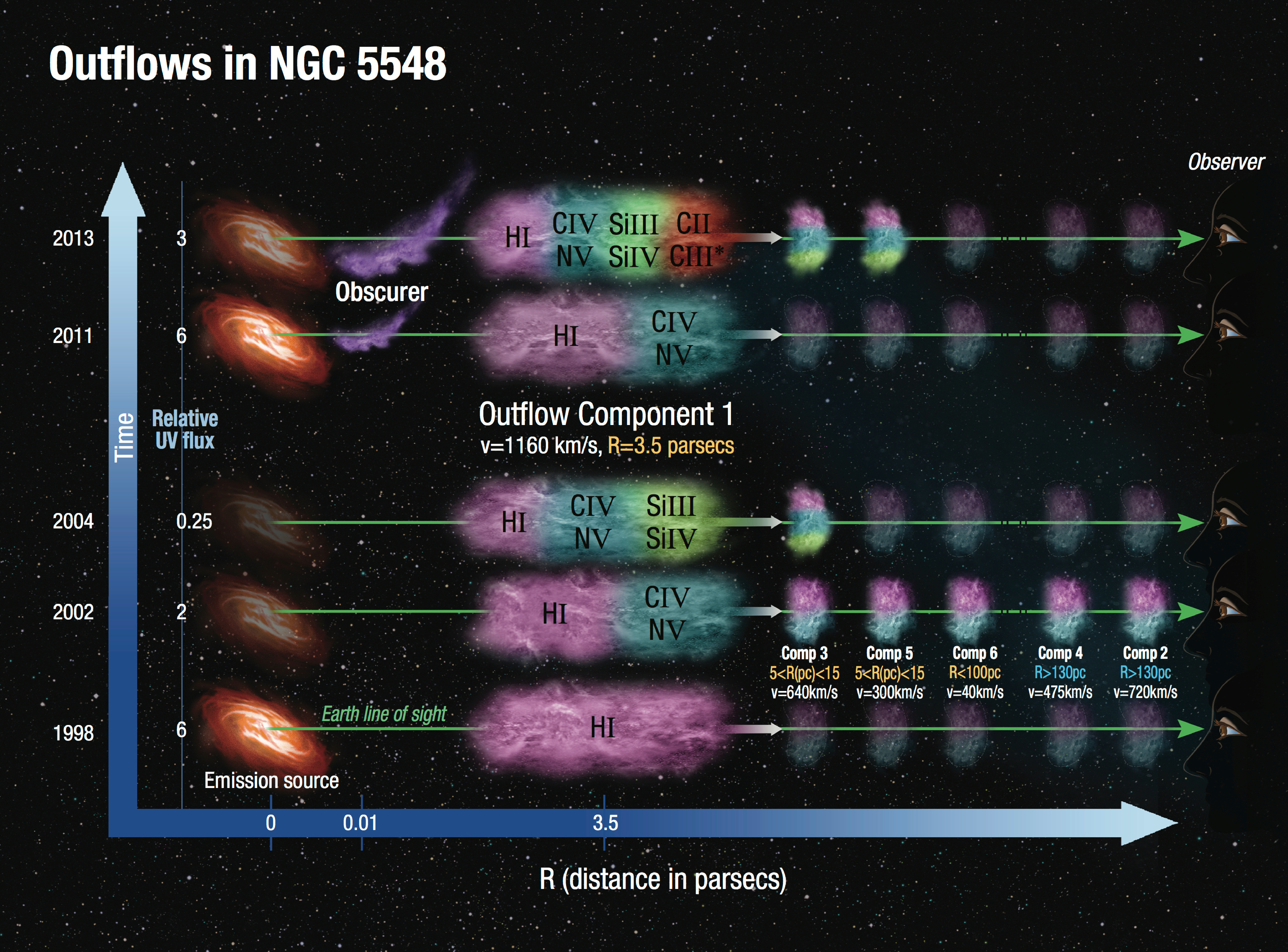} 
\caption{An illustration of the physical, spatial and temporal conditions of the outflows seen in NGC 5548. Along the time axis we show the behavior of the emission source at the 5 UV epochs and  give its  UV flux values (measured at 1350 \AA\ rest-frame and given in units of 
$10^{-14}$ ergs~s$^{-1}$~cm$^{-2}$~\AA$^{-1}$). The obscurer is situated at roughly 0.01 pc from the emission source and is only seen in 2011 and 2013 (it is much stronger in 2013). Outflow component 1 shows the most dramatic changes in its absorption troughs. Different observed ionic species are represented as colored zones within the absorbers. The trough changes are fully explained by our physical model shown in Figure \ref{fig:5epochs_phase_plot}. Using component 1 \ciii* troughs, which are only seen in the 2013 epoch, we determine its number density (see Figure \ref{fig:density}) to be $\log(n_e)=4.8\pm0.1$~cm$^{-3}$, and therefore its distance, R=3.5 parsec. The distances for the other components are discussed in \S~\ref{sect:comp2-6}. Dimmer clouds represent epochs where components 2-6 did not show new absorption species compared with the 2002 epoch.
}
\label{fig:schematics}
\end{center}
\end{figure*}

Figure \ref{fig:comp3_2013_phase_plot} shows the $N_H$--$U_H$ phase plot for component 3 based on the $N_{ion}$ reported in 
Table~\ref{table:Nion} (the \Lya\ and \civ\ $N_{ion}$ lower limits are trivially satisfied by the lower limit shown for the \nv\ $N_{ion}$). The phase plot constraints given by the $N_{ion}$ measurements are mostly parallel to each other. Therefore, the 
$N_H$--$U_H$ constraints are rather loose, allowing a narrow strip from $\log(N_H)=19.6$ and $\log(U_H)=-2$, to 
$\log(N_H)=21.5$ and $\log(U_H)=-1.1$.  If we take the most probable value of $\log(U_H)=-1.3$, the distance estimate for component 3 will drop by 30\% compared with the estimate  of $5<R<15$ parsec, which used the $\log(U_H)=-1.5$ of component 1.
 For component 5, the situation is rather similar, as the detected \siIII\ and \siiv\ allow a narrow strip from $\log(N_H)=19.2$ and $\log(U_H)=-1.8$, to 
$\log(N_H)=20.7$ and $\log(U_H)=-1.2$, while the lowest $\chi^2$ is achieved at $\log(N_H)=20.7$ and $\log(U_H)=-1.3$.

\section{Comparison with the Warm Absorber Analysis} \label{sect:WA}

How do the physical characteristics inferred from the outflows' UV diagnostics compare to the properties of the X-ray manifestation of the outflow known as the Warm Absorber? Since the soft X-ray flux in our 2013 data is very low due to the appearance of the obscurer, we cannot characterize the WA that is connected with the 6 UV outflow components at that epoch. 
Our main inferences about the WA are from the 2002 epoch when we obtained simultaneous X-ray and UV spectra of the outflow (when no obscurer was present) that gave a much higher soft X-ray flux (compared with the 2013 observations)  and allowed a detail modeling of the WA in that epoch \citep{Steenbrugge05,Kaastra14}. Due to the inherent complications of comparing analyses on different spectral regions (X-ray and UV) separated by 11 years (2002 and 2013), of a clearly time-dependent phenomenon, we defer a full comparison to a later paper (Ebrero et al 2015).  Here we outline some of the main points in such a comparison, based on the analysis presented here and the published analysis of the WA \citep{Steenbrugge05,Kaastra14} and discuss some of the similarities and challenges of such a combined analysis.

{\bf 1. Kinematic Similarity.}
There is kinematic correspondence between the UV absorption troughs in components 1--5 and the six ionization components (A--F) of the X-ray WA \citep[][see Table \ref{UVWA} here]{Kaastra14}. X-ray components F and C  span the width of UV component 1. X-ray component E   matches UV component 2. The lowest ionization X-ray components, A and B comprise the full width of the blended UV troughs of components 3 and 4 . Finally, X-ray component D kinematically matches UV component 5.  However, as we show in point 2 below, this kinematic matching is physically problematic as the ionization parameters of WA components C, D, E and F are too high 
to produce observed troughs from \civ\ and \nv\ that are observed in all the UV components.  

\begin{table}[h!]
\caption{Comparison between the UV and WA components}\label{UVWA}
\begin{tabular}{lccc}\hline
Component    &Velocity$^a$    &$\log(N_\mathrm{H})$    &$\log(U_\mathrm{H})$\\
        &$(\mathrm{km~s}^{-1})$    &$(\mathrm{cm}^{-2})$    &(2002)\\\hline
UV 1$^b$    &-1160            &\multicolumn{1}{c}{$21.5^{+0.4}_{-0.2}$}    &$-0.4^{+0.4}_{-0.2}$\\
UV 2        &-720            &\multicolumn{1}{c}{---}    &\multicolumn{1}{c}{---}\\
UV 3        &-640            &\multicolumn{1}{c}{$19.6-21.5$}    &$(-0.9)-0.0\;\;\;\;\;\;$\\
UV 4        &-475            &\multicolumn{1}{c}{---}    &\multicolumn{1}{c}{---}\\
UV 5        &-300            &\multicolumn{1}{c}{$19.2-20.7$}    &$(-0.7)-(-0.1)$\\
UV 6        &-40            &\multicolumn{1}{c}{---}    &\multicolumn{1}{c}{---}\\\hline
WA A$^c$    &$-588\pm34$    &$20.30\pm0.12$                    &$-0.82\pm0.08\;\;\;$\\
WA B        &$-547\pm31$    &$20.85\pm0.06$                    &$-0.09\pm0.05\;\;\;$\\
WA C        &$-1148\pm20\;\:$   &$21.18\pm0.08$                    &$0.55\pm0.03$\\
WA D        &$-255\pm25$    &$21.03\pm0.07$                    &$0.76\pm0.03$\\
WA E        &$-792\pm25$    &$21.45\pm0.12$                    &$1.34\pm0.08$\\
WA F        &$-1221\pm25\;\:$   &$21.76\pm0.13$    &$1.53\pm0.05$\\\hline
\multicolumn{4}{l}{\parbox{0.95\columnwidth}{$\,^a$velocity centroid of the component}}\\
\multicolumn{4}{l}{\parbox{0.95\columnwidth}{$\,^b$UV components 1-6 are arranged by decreasing absolute velocity.}}\\
\multicolumn{4}{l}{\parbox{0.95\columnwidth}{$\,^c$Parameters for Warm Absorber components A-F are from Table~S2 of \cite{Kaastra14}. They are arranged by increasing ionization parameter.}}
\end{tabular}
\end{table}

\noindent {\bf 2. Comparing similar ionization phases.} \\
We note that the X-ray analysis of the WA in the Chandra 2002 observations uses a different ionization parameter ($\xi$) than the $U_H$ we use here; where $\xi\equiv L/(\vy{n}{H}r^2)$ (erg cm) with $\vy{n}{H}$ being the hydrogen number density, $L$ the ionizing luminosity between 13.6 eV and 13.6 keV and $r$ the distance from the central source. For the high SED, $log(U_H) = \log(\xi)+1.6$.
In Table \ref{UVWA}, we give the $log(U_H)$ for the WA components.
From the WA analysis and figure \ref{fig:5epochs_phase_plot} here, we deduce that 90\% of the WA material 
\citep[components C, D , E and F in Table S2 of][]{Kaastra14} is in too high an ionization stage to produce measurable lines from  the UV observed ions (e.g., \civ, \nv).  Only component A and B of the WA are at low enough ionization states to give rise to the UV observed material.

\noindent {\bf 3. Assuming constant $N_H$ for the UV components and components A and B of the WA.} \\
Our temporal model for component 1 has a constant $N_H$ value in all the observed epochs, including 2002. The model also predicts the $U_H$ of the 2002 epoch (see Fig.~\ref{fig:5epochs_phase_plot}). We can therefore compare the predictions of this model to the results of the reanalysis of the 2002 WA \citep{Kaastra14}, provided that the  $N_H$ for  components A and B of the WA also did not change over the 11 years between the epochs. We note that since UV component 1 is the closest to the central source, the assumption 
of constant $N_H$ for the other UV components, over this 11 years time period, is reasonable (see discussion in \S~\ref{temporal}).
Therefore, we use the same ionization assumptions for UV components 3 and 5 as for component 1. That is, their $N_H$ is fixed to the 2013 value and their $\log(U_H)_{2002}=\log(U_H)_{2013}+1.1$, which are the values we list in   Table \ref{UVWA}.
We do not have empirical constraints on the distances of WA components A-F from the central source.

\noindent {\bf 4. Comparing UV components 1 and 3 to components A and B of the WA.} \\
Using proto-Solar abundances \citep{Lodders09}, our 2002 model prediction for UV component 1 has $\log(N_H)=21.5_{-0.2}^{+0.4}$ cm$^{-2}$, and an ionization parameter of $\log(U_H)=-0.4_{-0.2}^{+0.4}$ (see \S~\ref{NU} and Fig.~\ref{fig:5epochs_phase_plot}).
This model gives a good match for the UV data of that epoch (2002) and its $\log(U_H)$  is in between those of WA components A [$\log(U_H)=-0.8$] and B [$\log(U_H)=-0.1$]. However, there are two inconsistencies between the models. First, components A and B have a total $\log(N_H)=20.95\pm0.1$ or about 2$\sigma$ disagreement with that of UV component 1. This discrepancy is mainly due to the limit on the \ovii\ $N_{ion}$ that arises from the bound-free edge of this ion in the X-ray data. In the WA model, about 95\% of the  \ovii\ $N_{ion}$ arises from components A and B.
Second, the reported velocity centroids for WA components A and B ($-588\pm34$ km $s^{-1}$ and $-547\pm31$ km $s^{-1}$, respectively) are in disagreement with the velocity centroids of UV component 1 ($-1160$ km $s^{-1}$) and its 300 km $s^{-1}$ width.

UV component 3 has a velocity centroid at --640 km $s^{-1}$  and a width of $\sim$200 km $s^{-1}$, and therefore  is  a better velocity match with WA components A and B. The large uncertainties in the inferred  $N_H$ and $U_H$ for UV component 3 (see Fig.~\ref{fig:comp3_2013_phase_plot} and Table \ref{UVWA}), make these values consistent with the $N_H$ and $U_H$ deduced for WA components A and B. However, the uncertainties also allow UV component 3 to have a negligible $N_H$ compared to WA components A and B.

We note that with the current analyses, the better the agreement between UV component 3 and WA components A and B, the worse is the disagreement between UV component 1 and WA components A and B. This is because UV component 1 already predicts higher values of 
 $N_H$ and  \ovii\ $N_{ion}$ than are measured in WA components A and B, and the kinematics of the deduced \ovii\ $N_{ion}$ disagree considerably. In points 5 and 6 below we identify two possible ways to alleviate and even eliminate these apparent discrepancies.

\noindent {\bf 5. Existence of considerable \ovii\ $N_{ion}$ at the velocity of UV component 1.}
The 2002 X-ray spectra presented by \citet{Steenbrugge05} consist of two different data sets that were acquired in the same week: 170 ks observations taken with the High Energy Transmission Grating Spectrometer (HETGS) and 340 ks observations with the
Low Energy Transmission Grating Spectrometer (LETGS). Figure 2 in \citet{Steenbrugge05} shows some of the low ionization WA troughs in velocity presentation, where the dotted lines show the position of the UV components (with somewhat different velocity values than we use here due to the use of a slightly different systemic redshift for the object). The LETGS data of the \ovii\ and \ov\ troughs are consistent with one main kinematic component matching the velocity of UV component 3. However, the more noisy but higher resolution HETGS data for the same transitions, show two sub troughs one corresponding to UV component 1 and one to UV component 3. Therefore, it is possible that much of the \ovii\ $N_{ion}$ is associated with UV component 1. 

\noindent {\bf 6. Abundances considerations:}
As mentioned in \S~\ref{sect:comp1}, for the UV analysis, we use pure proto-Solar abundances \citep{Lodders09}, which well-match the measured $N_{ion}$ from the UV data. But these models 
produce considerably more \ovii\ $N_{ion}$ in the 2002 epoch, than the measured \ovii\ $N_{ion}$ in the warm absorber data. However, the $N_H$ (and therefore also the \ovii\ $N_{ion}$) of UV component 1 is critically dependent on the assumed phosphorus abundance. Ionization models with all elements having proto-Solar abundances except phosphorus, for which we assume twice proto-Solar abundance, preserve the fit to the UV data (at 1/3 the $N_H$ value) and at the same time match the \ovii\ $N_{ion}$ measured in the X-ray warm absorber at the 2002 epoch. Larger over-abundances of phosphorus  further reduce the $N_H$ value and therefore the predicted \ovii\ $N_{ion}$ for the 2002 epoch. 

But are such assumed abundances physically reasonable? AGN outflows are known to have abundances higher by factors of two \citep{Arav07} and even ten \citep{Gabel06} compared with the proto-Solar values. In particular, phosphorus abundance in AGN outflows, relative to other metals, can be a factor of several higher than in proto-Solar abundances  \citep[see \S4.1 in][]{Arav01}.  Furthermore, the theoretical expectations for the value of chemical abundances in an AGN environment as a function of metalicity are highly varied. The leading models can differ about relative abundances values by factors of three or more 
\citep[e.g., comparing the values of][]{Hamann93,Ballero08}.

We conclude that if roughly half of the observed \ovii\ $N_{ion}$ is associated with UV component 1 (as discussed in point 5),
and if the relative abundance of phosphorus to oxygen is twice solar or larger, than the $N_H$, $U_H$ and velocity distribution  of the WA and UV outflowing gas can be consistent.

\section{Discussion} \label{sect:discussion}

\subsection{Comparison With Previous Work} \label{sect:previous}

How do these results compare with the previous extensive work on the “enduring” outflow in NGC 5548? For the first time a simple model of a constant $N_H$ absorber yields a physical picture that is consistent with the substantial trough variation seen in all epochs of high resolution UV spectral observations. The trough changes are explained solely by the observed differences in the incident ionizing flux. In addition, we determine robust distances for (or limits on) all 6 kinematic components. Our results differ considerably from those previously found for this outflow  \citep[e.g.,][]{Crenshaw09}. This is due to the powerful diagnostics that were revealed during the 2013 campaign, recognizing that many of the observed troughs are highly saturated \citep[e.g., the existence of similar depth \civ\ and \pv\ troughs in component 1; see discussion in][]{Borguet12b} and the fact that the previous work did not account for the emergence of \siIII\ troughs associated with components 1 and 3 in the 2004 data. 

\subsection{Implications for BAL Variability Studies} \label{sect:BAL_variability}

Our multiwavelength campaign has significant consequences for studies of absorption trough variability in quasar outflows, and in particular for the intensive studies of trough variability in Broad Absorption Line (BAL) quasars \citep[e.g.,][and references therein]{Barlow92,Capellupo12,Filiz13}. As discussed in \S~\ref{temporal}, the two main proposed mechanisms for trough variability are (1) reaction to changes in ionizing flux of a constant absorber (which is the model we successfully use to explain the NGC 5548 outflow trough changes); and (2) absorber motion across the line of sight \citep[e.g.,][]{Gabel03}, 
which as we demonstrated, cannot  explain the variability of the NGC 5548 outflow (see \S~\ref{temporal}). 

In some BAL cases the rest-frame UV flux around 1350 \AA\ does not change appreciably between the studied epochs while significant trough variability is observed. This behavior is taken as an argument against mechanism (1) 
\citep[e.g.,][]{Barlow92,Filiz13} as the ionizing flux (below 912\AA) is assumed to correlate with the longer wavelength UV flux. However, our simultaneous UV/X-ray observations show a clear case where the ionizing photon flux drops by a factor of 25 between the 2002 and 2013 epochs, while the 1350 \AA\ UV flux actually increases by 50\%. 

The spectroscopic data set of NGC 5548 gives high S/N and high-resolution spectra of many UV absorption troughs, and simultaneously yields the crucial soft X-ray flux that is responsible for the ionization of the outflow. Such data can constrain trough variability models far better than the standard variability data sets where one or two rest-frame UV BAL are observed at two (and less frequently in more) epochs \citep[e.g.,][]{Capellupo12,Filiz13}. 

\subsection{Implications for the X-ray Obscurer} \label{sect:obscurer}

What do our results about the “enduring” outflow tell us about the X-ray obscurer and the broad UV absorption discovered by \citet{Kaastra14}?  The derived transmission for the X-ray obscurer is consistent with the SED required for the 2013 spectrum, thus showing that the obscurer is closer to the super massive black holes than 3.5 pc., and that its shadow influences the conditions in the more distant narrow UV absorbers.

\section{Summary} \label{sect:summary}

In 2013 we executed the most comprehensive multiwavelength spectroscopic campaign on any AGN to date, directed at NGC 5548.
This paper  presents the  analysis' results from our HST/COS data of the often observed UV outflow, which is detected in 6 distinct kinematic components.
Our campaign revealed an unusually strong X-ray obscuration. The resulting dramatic decrease in incident ionizing flux on the outflow allowed us to construct a comprehensive physical, spatial and temporal picture for the well-studied AGN wind in this object.  Our main findings are listed below (see Fig.~\ref{fig:schematics} for a graphic illustration of our results).

\begin{enumerate}
 \item  Our best constraints are obtained for component 1 (the highest velocity component): It is situated at 3.5$\pm$1 pc from the central source, has a total hydrogen column density of $\log(N_H)=21.5_{-0.2}^{+0.4}$ cm$^{-2}$,  an ionization parameter of $\log(U_H)=-1.5_{-0.2}^{+0.4}$, and an electron number density of $\log(n_e)=4.8\pm0.1$ cm$^{-3}$. This component probably carries the largest $N_H$ associated with the UV outflow. 
See \S~\ref{NU} and \S~\ref{Density} for elaboration.

\item For component 1 a simple model based on a fixed total column-density absorber, reacting to changes in ionizing illumination, matches the very different ionization states seen at five spectroscopic epochs spanning 16 years. See \S~\ref{temporal} for elaboration.

\item The wealth of observational constraints  makes our changes-of-ionization model a leading contender for interpreting trough variability data of quasar outflows, in particular Broad Absorption Line (BAL) variability. 
See \S~6.2 for elaboration.

 \item Components 3 and 5 are situated between 5-15 pc from the central source, component 6 is closer than 100 pc and
  components 2 and 4 are further out than 130 pc. See \S~\ref{sect:comp2-6} for elaboration.

 \item A detailed comparison of the physical characteristics inferred from the outflows' UV diagnostics with those of  the X-ray  Warm Absorber is deferred to a future paper. Here we outline some of the main points in such a comparison, and discuss some of the similarities and challenges of such a combined analysis. 
 See \S~\ref{sect:WA} for elaboration.

\end{enumerate}

\begin{acknowledgements}

This work was supported by NASA through grants for HST program number 13184 from
the Space Telescope Science Institute, which is operated by the Association of
Universities for Research in Astronomy, Incorporated, under NASA contract
NAS5-26555. SRON is supported financially by NWO, the Netherlands Organization
for Scientific Research. M.M. acknowledges the support of a Studentship
Enhancement Programme awarded by the UK Science \& Technology Facilities Council
(STFC). P.-O.P. and F.U. thanks financial support from the CNES and from the CNRS/PICS. 
F.U. acknowledg PhD funding from the VINCI program of the French-Italian University.
K.C.S. wants to acknowledge financial support from the Fondo Fortalecimiento de
la Productividad Científica VRIDT 2013. E.B. is supported by grants from
Israel's MoST, ISF (1163/10), and I-CORE program (1937/12). J.M. acknowledges
funding from CNRS/PNHE and CNRS/PICS in France. G.M. and F.U. acknowledge
financial support from the Italian Space Agency under grant ASI/INAF
I/037/12/0-011/13. B.M.P. acknowledges support from the US NSF through grant
AST-1008882. M.C, S.B, G.M and A. D. R., acknowledge INAF/PICS
support. G.P. acknowledges support via an EU Marie Curie Intra-European
fellowship under contract no. FP-PEOPLE-2012-IEF-331095. M. W. acknowledges the
support of a PhD studentship awarded by the UK Science \& Technology Facilities
Council (STFC). 

The data used in this research are stored in the public archives of the
satellites that are involved. We thank the International Space Science Institute
(ISSI) in Bern for support. This work is based on observations obtained with
XMM-Newton, an ESA science mission with instruments and contributions directly
funded by ESA Member States and the USA (NASA). It is also based on observations
with INTEGRAL, an ESA project with instrument and science data center funded by
ESA member states (especially the PI countries: Denmark, France, Germany, Italy,
Switzerland, Spain), Czech Republic, and Poland and with the participation of
Russia and the USA. This work made use of data supplied by the UK Swift Science
Data Centre at the University if Leicester. This research made use of the
Chandra Transmission Grating Catalog and archive (http://tgcat.mit.edu). This
research has made use of data obtained with the NuSTAR mission, a project led by
the California Institute of Technology (Caltech), managed by the Jet Propulsion
Laboratory (JPL) and funded by NASA, and has utilized the NuSTAR Data Analysis
Software (NUSTARDAS) jointly developed by the ASI Science Data Center (ASDC,
Italy) and Caltech (USA). Figure \ref{fig:schematics} was created by Ann Feild from STScI.

\end{acknowledgements}  

\bibliography{astro}

\begin{thebibliography}{44}
\expandafter\ifx\csname natexlab\endcsname\relax\def\natexlab#1{#1}\fi

\bibitem[{{Arav} {et~al.}(1999){Arav}, {Becker}, {Laurent-Muehleisen}, {Gregg},
  {White}, {Brotherton}, \& {de Kool}}]{Arav99b}
{Arav}, N., {Becker}, R.~H., {Laurent-Muehleisen}, S.~A., {et~al.} 1999, \apj,
  524, 566

\bibitem[{{Arav} {et~al.}(2013){Arav}, {Borguet}, {Chamberlain}, {Edmonds}, \&
  {Danforth}}]{Arav13}
{Arav}, N., {Borguet}, B., {Chamberlain}, C., {Edmonds}, D., \& {Danforth}, C.
  2013, \mnras, 436, 3286

\bibitem[{{Arav} {et~al.}(2001){Arav}, {de Kool}, {Korista}, {Crenshaw}, {van
  Breugel}, {Brotherton}, {Green}, {Pettini}, {Wills}, {de Vries}, {Becker},
  {Brandt}, {Green}, {Junkkarinen}, {Koratkar}, {Laor}, {Laurent-Muehleisen},
  {Mathur}, \& {Murray}}]{Arav01}
{Arav}, N., {de Kool}, M., {Korista}, K.~T., {et~al.} 2001, \apj, 561, 118

\bibitem[{{Arav} {et~al.}(2012){Arav}, {Edmonds}, {Borguet}, {Kriss},
  {Kaastra}, {Behar}, {Bianchi}, {Cappi}, {Costantini}, {Detmers}, {Ebrero},
  {Mehdipour}, {Paltani}, {Petrucci}, {Pinto}, {Ponti}, {Steenbrugge}, \& {de
  Vries}}]{Arav12}
{Arav}, N., {Edmonds}, D., {Borguet}, B., {et~al.} 2012, \aap, 544, A33

\bibitem[{{Arav} {et~al.}(2007){Arav}, {Gabel}, {Korista}, {Kaastra}, {Kriss},
  {Behar}, {Costantini}, {Gaskell}, {Laor}, {Kodituwakku}, {Proga}, {Sako},
  {Scott}, \& {Steenbrugge}}]{Arav07}
{Arav}, N., {Gabel}, J.~R., {Korista}, K.~T., {et~al.} 2007, \apj, 658, 829

\bibitem[{{Arav} {et~al.}(2005){Arav}, {Kaastra}, {Kriss}, {Korista}, {Gabel},
  \& {Proga}}]{Arav05}
{Arav}, N., {Kaastra}, J., {Kriss}, G.~A., {et~al.} 2005, \apj, 620, 665

\bibitem[{{Arav} {et~al.}(2002){Arav}, {Korista}, \& {de Kool}}]{Arav02}
{Arav}, N., {Korista}, K.~T., \& {de Kool}, M. 2002, \apj, 566, 699

\bibitem[{{Arav} {et~al.}(2008){Arav}, {Moe}, {Costantini}, {Korista}, {Benn},
  \& {Ellison}}]{Arav08}
{Arav}, N., {Moe}, M., {Costantini}, E., {et~al.} 2008, \apj, 681, 954

\bibitem[{{Ballero} {et~al.}(2008){Ballero}, {Matteucci}, {Ciotti}, {Calura},
  \& {Padovani}}]{Ballero08}
{Ballero}, S.~K., {Matteucci}, F., {Ciotti}, L., {Calura}, F., \& {Padovani},
  P. 2008, \aap, 478, 335

\bibitem[{{Barlow} {et~al.}(1992){Barlow}, {Junkkarinen}, {Burbidge},
  {Weymann}, {Morris}, \& {Korista}}]{Barlow92}
{Barlow}, T.~A., {Junkkarinen}, V.~T., {Burbidge}, E.~M., {et~al.} 1992, \apj,
  397, 81

\bibitem[{{Borguet} {et~al.}(2013){Borguet}, {Arav}, {Edmonds}, {Chamberlain},
  \& {Benn}}]{Borguet13}
{Borguet}, B.~C.~J., {Arav}, N., {Edmonds}, D., {Chamberlain}, C., \& {Benn},
  C. 2013, \apj, 762, 49

\bibitem[{{Borguet} {et~al.}(2012{\natexlab{a}}){Borguet}, {Edmonds}, {Arav},
  {Benn}, \& {Chamberlain}}]{Borguet12b}
{Borguet}, B.~C.~J., {Edmonds}, D., {Arav}, N., {Benn}, C., \& {Chamberlain},
  C. 2012{\natexlab{a}}, \apj, 758, 69

\bibitem[{{Borguet} {et~al.}(2012{\natexlab{b}}){Borguet}, {Edmonds}, {Arav},
  {Dunn}, \& {Kriss}}]{Borguet12a}
{Borguet}, B.~C.~J., {Edmonds}, D., {Arav}, N., {Dunn}, J., \& {Kriss}, G.~A.
  2012{\natexlab{b}}, \apj, 751, 107

\bibitem[{{Capellupo} {et~al.}(2012){Capellupo}, {Hamann}, {Shields},
  {Rodr{\'{\i}}guez Hidalgo}, \& {Barlow}}]{Capellupo12}
{Capellupo}, D.~M., {Hamann}, F., {Shields}, J.~C., {Rodr{\'{\i}}guez Hidalgo},
  P., \& {Barlow}, T.~A. 2012, \mnras, 422, 3249

\bibitem[{{Ciotti} {et~al.}(2010){Ciotti}, {Ostriker}, \& {Proga}}]{Ciotti10}
{Ciotti}, L., {Ostriker}, J.~P., \& {Proga}, D. 2010, \apj, 717, 708

\bibitem[{{Costantini} {et~al.}(2007){Costantini}, {Kaastra}, {Arav}, {Kriss},
  {Steenbrugge}, {Gabel}, {Verbunt}, {Behar}, {Gaskell}, {Korista}, {Proga},
  {Quijano}, {Scott}, {Klimek}, \& {Hedrick}}]{Costantini07}
{Costantini}, E., {Kaastra}, J.~S., {Arav}, N., {et~al.} 2007, \aap, 461, 121

\bibitem[{{Crenshaw} {et~al.}(1999){Crenshaw}, {Kraemer}, {Boggess}, {Maran},
  {Mushotzky}, \& {Wu}}]{Crenshaw99}
{Crenshaw}, D.~M., {Kraemer}, S.~B., {Boggess}, A., {et~al.} 1999, \apj, 516,
  750

\bibitem[{{Crenshaw} {et~al.}(2003){Crenshaw}, {Kraemer}, {Gabel}, {Kaastra},
  {Steenbrugge}, {Brinkman}, {Dunn}, {George}, {Liedahl}, {Paerels}, {Turner},
  \& {Yaqoob}}]{Crenshaw03b}
{Crenshaw}, D.~M., {Kraemer}, S.~B., {Gabel}, J.~R., {et~al.} 2003, \apj, 594,
  116

\bibitem[{{Crenshaw} {et~al.}(2009){Crenshaw}, {Kraemer}, {Schmitt}, {Kaastra},
  {Arav}, {Gabel}, \& {Korista}}]{Crenshaw09}
{Crenshaw}, D.~M., {Kraemer}, S.~B., {Schmitt}, H.~R., {et~al.} 2009, \apj,
  698, 281

\bibitem[{{de Kool} {et~al.}(2002){de Kool}, {Korista}, \& {Arav}}]{deKool02}
{de Kool}, M., {Korista}, K.~T., \& {Arav}, N. 2002, \apj, 580, 54

\bibitem[{{de Vaucouleurs} {et~al.}(1991){de Vaucouleurs}, {de Vaucouleurs},
  {Corwin}, {Buta}, {Paturel}, \& {Fouqu{\'e}}}]{deVaucouleurs91}
{de Vaucouleurs}, G., {de Vaucouleurs}, A., {Corwin}, Jr., H.~G., {et~al.}
  1991, {Third Reference Catalogue of Bright Galaxies. Volume I: Explanations
  and references. Volume II: Data for galaxies between 0$^{h}$ and 12$^{h}$.
  Volume III: Data for galaxies between 12$^{h}$ and 24$^{h}$.}

\bibitem[{{Dere} {et~al.}(1997){Dere}, {Landi}, {Mason}, {Monsignori Fossi}, \&
  {Young}}]{Dere97}
{Dere}, K.~P., {Landi}, E., {Mason}, H.~E., {Monsignori Fossi}, B.~C., \&
  {Young}, P.~R. 1997, \aaps, 125, 149

\bibitem[{{Edmonds} {et~al.}(2011){Edmonds}, {Borguet}, {Arav}, {Dunn},
  {Penton}, {Kriss}, {Korista}, {Costantini}, {Steenbrugge},
  {Gonzalez-Serrano}, {Aoki}, {Bautista}, {Behar}, {Benn}, {Crenshaw},
  {Everett}, {Gabel}, {Kaastra}, {Moe}, \& {Scott}}]{Edmonds11}
{Edmonds}, D., {Borguet}, B., {Arav}, N., {et~al.} 2011, \apj, 739, 7

\bibitem[{{Faucher-Gigu{\`e}re} {et~al.}(2012){Faucher-Gigu{\`e}re},
  {Quataert}, \& {Murray}}]{Faucher-Giguere12}
{Faucher-Gigu{\`e}re}, C.-A., {Quataert}, E., \& {Murray}, N. 2012, \mnras,
  420, 1347

\bibitem[{{Ferland} {et~al.}(2013){Ferland}, {Porter}, {van Hoof}, {Williams},
  {Abel}, {Lykins}, {Shaw}, {Henney}, \& {Stancil}}]{Ferland13}
{Ferland}, G.~J., {Porter}, R.~L., {van Hoof}, P.~A.~M., {et~al.} 2013, \rmxaa,
  49, 137

\bibitem[{{Filiz Ak} {et~al.}(2013){Filiz Ak}, {Brandt}, {Hall}, {Schneider},
  {Anderson}, {Hamann}, {Lundgren}, {Myers}, {P{\^a}ris}, {Petitjean}, {Ross},
  {Shen}, \& {York}}]{Filiz13}
{Filiz Ak}, N., {Brandt}, W.~N., {Hall}, P.~B., {et~al.} 2013, \apj, 777, 168

\bibitem[{{Gabel} {et~al.}(2006){Gabel}, {Arav}, \& {Kim}}]{Gabel06}
{Gabel}, J.~R., {Arav}, N., \& {Kim}, T. 2006, \apj, 646, 742

\bibitem[{{Gabel} {et~al.}(2003){Gabel}, {Crenshaw}, {Kraemer}, {Brandt},
  {George}, {Hamann}, {Kaiser}, {Kaspi}, {Kriss}, {Mathur}, {Mushotzky},
  {Nandra}, {Netzer}, {Peterson}, {Shields}, {Turner}, \& {Zheng}}]{Gabel03}
{Gabel}, J.~R., {Crenshaw}, D.~M., {Kraemer}, S.~B., {et~al.} 2003, \apj, 583,
  178

\bibitem[{{Gabel} {et~al.}(2005){Gabel}, {Kraemer}, {Crenshaw}, {George},
  {Brandt}, {Hamann}, {Kaiser}, {Kaspi}, {Kriss}, {Mathur}, {Nandra}, {Netzer},
  {Peterson}, {Shields}, {Turner}, \& {Zheng}}]{Gabel05b}
{Gabel}, J.~R., {Kraemer}, S.~B., {Crenshaw}, D.~M., {et~al.} 2005, \apj, 631,
  741

\bibitem[{{Green} {et~al.}(2012){Green}, {Froning}, {Osterman}, {Ebbets},
  {Heap}, {Leitherer}, {Linsky}, {Savage}, {Sembach}, {Shull}, {Siegmund},
  {Snow}, {Spencer}, {Stern}, {Stocke}, {Welsh}, {B{\'e}land}, {Burgh},
  {Danforth}, {France}, {Keeney}, {McPhate}, {Penton}, {Andrews},
  {Brownsberger}, {Morse}, \& {Wilkinson}}]{Green2012}
{Green}, J.~C., {Froning}, C.~S., {Osterman}, S., {et~al.} 2012, \apj, 744, 60

\bibitem[{{Hamann} {et~al.}(1997){Hamann}, {Barlow}, {Junkkarinen}, \&
  {Burbidge}}]{Hamann97b}
{Hamann}, F., {Barlow}, T.~A., {Junkkarinen}, V., \& {Burbidge}, E.~M. 1997,
  \apj, 478, 80

\bibitem[{{Hamann} \& {Ferland}(1993)}]{Hamann93}
{Hamann}, F. \& {Ferland}, G. 1993, \apj, 418, 11

\bibitem[{{Hopkins} \& {Elvis}(2010)}]{Hopkins10}
{Hopkins}, P.~F. \& {Elvis}, M. 2010, \mnras, 401, 7

\bibitem[{{Kaastra} {et~al.}(2012){Kaastra}, {Detmers}, {Mehdipour}, {Arav},
  {Behar}, {Bianchi}, {Branduardi-Raymont}, {Cappi}, {Costantini}, {Ebrero},
  {Kriss}, {Paltani}, {Petrucci}, {Pinto}, {Ponti}, {Steenbrugge}, \& {de
  Vries}}]{Kaastra12}
{Kaastra}, J.~S., {Detmers}, R.~G., {Mehdipour}, M., {et~al.} 2012, \aap, 539,
  A117

\bibitem[{{Kaastra et al.}(2014)}]{Kaastra14}
{Kaastra et al.}, J.~S. 2014, Science

\bibitem[{{Korista} {et~al.}(1995){Korista}, {Alloin}, {Barr}, {Clavel},
  {Cohen}, {Crenshaw}, {Evans}, {Horne}, {Koratkar}, {Kriss}, {Krolik},
  {Malkan}, {Morris}, {Netzer}, {O'Brien}, {Peterson}, {Reichert},
  {Rodriguez-Pascual}, {Wamsteker}, {Anderson}, {Axon}, {Benitez}, {Berlind},
  {Bertram}, {Blackwell}, {Bochkarev}, {Boisson}, {Carini}, {Carrillo},
  {Carone}, {Cheng}, {Christensen}, {Chuvaev}, {Dietrich}, {Dokter},
  {Doroshenko}, {Dultzin-Hacyan}, {England}, {Espey}, {Filippenko}, {Gaskell},
  {Goad}, {Ho}, {Huchra}, {Jiang}, {Kaspi}, {Kollatschny}, {Laor}, {Luminet},
  {MacAlpine}, {MacKenty}, {Malkov}, {Maoz}, {Martin}, {Matheson}, {McCollum},
  {Merkulova}, {Metik}, {Mignoli}, {Miller}, {Pastoriza}, {Pelat}, {Penfold},
  {Perez}, {Perola}, {Persaud}, {Peters}, {Pitts}, {Pogge}, {Pronik}, {Pronik},
  {Ptak}, {Rawley}, {Recondo-Gonzalez}, {Rodriguez-Espinosa}, {Romanishin},
  {Sadun}, {Salamanca}, {Santos-Lleo}, {Sekiguchi}, {Sergeev}, {Shapovalova},
  {Shields}, {Shrader}, {Shull}, {Silbermann}, {Sitko}, {Skillman}, {Smith},
  {Smith}, {Snijders}, {Sparke}, {Stirpe}, {Stoner}, {Sun}, {Thiele}, {Tokarz},
  {Tsvetanov}, {Turnshek}, {Veilleux}, {Wagner}, {Wagner}, {Wanders}, {Wang},
  {Welsh}, {Weymann}, {White}, {Wilkes}, {Wills}, {Winge}, {Wu}, \&
  {Zou}}]{korista95}
{Korista}, K.~T., {Alloin}, D., {Barr}, P., {et~al.} 1995, \apjs, 97, 285

\bibitem[{{Landi} {et~al.}(2013){Landi}, {Young}, {Dere}, {Del Zanna}, \&
  {Mason}}]{Landi13}
{Landi}, E., {Young}, P.~R., {Dere}, K.~P., {Del Zanna}, G., \& {Mason}, H.~E.
  2013, \apj, 763, 86

\bibitem[{{Lodders} {et~al.}(2009){Lodders}, {Palme}, \& {Gail}}]{Lodders09}
{Lodders}, K., {Palme}, H., \& {Gail}, H.-P. 2009, in ''Landolt-B{\"o}rnstein -
  Group VI Astronomy and Astrophysics Numerical Data and Functional
  Relationships in Science and Technology Volume, ed. J.~E. {Tr{\"u}mper}, 44

\bibitem[{{Mehdipour et al}(2014)}]{Mehdipour14}
{Mehdipour et al}, M. 2014

\bibitem[{{Ostriker} {et~al.}(2010){Ostriker}, {Choi}, {Ciotti}, {Novak}, \&
  {Proga}}]{Ostriker10}
{Ostriker}, J.~P., {Choi}, E., {Ciotti}, L., {Novak}, G.~S., \& {Proga}, D.
  2010, \apj, 722, 642

\bibitem[{{Pancoast} {et~al.}(2013){Pancoast}, {Brewer}, {Treu}, {Park},
  {Barth}, {Bentz}, \& {Woo}}]{Pancoast13}
{Pancoast}, A., {Brewer}, B.~J., {Treu}, T., {et~al.} 2013, ArXiv e-prints

\bibitem[{{Schlegel} {et~al.}(1998){Schlegel}, {Finkbeiner}, \&
  {Davis}}]{Schlegel98}
{Schlegel}, D.~J., {Finkbeiner}, D.~P., \& {Davis}, M. 1998, \apj, 500, 525

\bibitem[{{Soker} \& {Meiron}(2011)}]{Soker11}
{Soker}, N. \& {Meiron}, Y. 2011, \mnras, 411, 1803

\bibitem[{{Steenbrugge} {et~al.}(2005){Steenbrugge}, {Kaastra}, {Crenshaw},
  {Kraemer}, {Arav}, {George}, {Liedahl}, {van der Meer}, {Paerels}, {Turner},
  \& {Yaqoob}}]{Steenbrugge05}
{Steenbrugge}, K.~C., {Kaastra}, J.~S., {Crenshaw}, D.~M., {et~al.} 2005, \aap,
  434, 569

\end{thebibliography}

\pagebreak

\appendix
\section{Ionic Column Density Measurements}

\begin{table*}
\caption{Observations and flux values for all epochs}\label{table:observations}
\begin{tabular}{ll@{$\hspace{0.2cm}\mathrm{to}\hspace{0.2cm}$}llllr@{$\,\pm\,$}lr@{$\,\pm\,$}lr@{$\,\pm\,$}l}
\hline
\multicolumn{1}{c}{Epoch}	&\multicolumn{2}{c}{Observing Dates}	&\multicolumn{1}{c}{Instrument}	&\multicolumn{1}{c}{Grating}	&\multicolumn{1}{c}{Exposure}	&\multicolumn{2}{c}{avg $F_{1350}^a$}	&\multicolumn{2}{c}{$\mathrm{fit}$ $F_{1350}^b$}	&\multicolumn{2}{c}{$\mathrm{fit}$ $\alpha^c$}\\\hline
2013\_v1	&\multicolumn{2}{c}{2013/06/22}	&HST:COS	&G130M	&1.8ks	&1.95&0.36	&\multicolumn{4}{c}{}\\
		&\multicolumn{2}{c}{}		&		&G160M	&2.1ks	&\multicolumn{6}{c}{}\\
2013\_v2	&\multicolumn{2}{c}{2013/07/12}	&HST:COS	&G130M	&5.0ks	&2.22&0.34	&\multicolumn{4}{c}{}\\
		&\multicolumn{2}{c}{}		&		&G160M	&2.2ks	&\multicolumn{6}{c}{}\\
2013\_v3	&\multicolumn{2}{c}{2013/07/24}	&HST:COS	&G130M	&2.0ks	&3.70&0.44	&\multicolumn{4}{c}{}\\
		&\multicolumn{2}{c}{}		&		&G160M	&2.2ks	&\multicolumn{6}{c}{}\\
2013\_v4	&\multicolumn{2}{c}{2013/07/30}	&HST:COS	&G130M	&2.0ks	&3.51&0.44	&\multicolumn{4}{c}{}\\
		&\multicolumn{2}{c}{}		&		&G160M	&2.2ks	&\multicolumn{6}{c}{}\\
2013\_v5	&\multicolumn{2}{c}{2013/08/01}	&HST:COS	&G130M	&2.0ks	&3.28&0.40	&\multicolumn{4}{c}{}\\
		&\multicolumn{2}{c}{}		&		&G160M	&2.2ks	&\multicolumn{6}{c}{}\\
2013\_v6	&\multicolumn{2}{c}{2013/12/20}	&HST:COS	&G130M	&2.0ks	&3.26&0.40	&3.14&0.02	&-0.79&0.03\\
		&\multicolumn{2}{c}{}		&		&G160M	&2.2ks	&\multicolumn{6}{c}{}\\
2013\_v345	&2013/07/24&2013/08/01		&HST:COS	&G130M	&6.0ks	&3.51&0.24	&3.44&0.03	&-0.776&0.014\\
		&\multicolumn{2}{c}{}		&		&G160M	&6.6ks	&\multicolumn{6}{c}{}\\
2013\_v12345	&2013/06/22&2013/08/01		&HST:COS	&G130M	&12.8ks	&3.11&0.19	&3.051&0.015	&-0.736&0.026\\
		&\multicolumn{2}{c}{}		&		&G160M	&11.0ks	&\multicolumn{6}{c}{}\\
2011		&\multicolumn{2}{c}{2011/06/16}	&HST:COS	&G130M	&1.9ks	&6.17&0.52	&6.13&0.05	&-0.86&0.07\\
		&\multicolumn{2}{c}{}		&		&G160M	&2.4ks	&\multicolumn{6}{c}{}\\
2004		&2004/02/10&2004/02/13		&HST:STIS	&E140M	&52.2ks	&0.25&0.07	&0.203&0.017	&-1.86&0.29\\
2002		&2002/01/22&2002/01/23		&HST:STIS	&E140M	&15.3ks	&1.80&0.18	&1.80&0.04	&-1.46&0.06\\
1998		&\multicolumn{2}{c}{1998/03/11}	&HST:STIS	&E140M	&4.75ks	&6.41&0.75	&6.43&0.12	&-1.51&0.13\\\hline
\multicolumn{12}{l}{$\,^a$Flux at rest-frame 1350\AA\ in units of $10^{-14}~\mathrm{erg~s}^{-1}~\mathrm{cm}^{-2}~\mathrm{\AA}^{-1}$}\\
\multicolumn{12}{l}{$\,^b$Flux from a power-law fit of the form $F(\lambda)=F_{1350}(\lambda/1350)^\alpha$}\\
\multicolumn{12}{l}{$\,^c$Spectral index from the above power-law fit}
\end{tabular}
\end{table*}

The column density associated with a given ion
 as a function of the radial velocity $v$ is defined as:
    \begin{equation}
    \label{eqcoldens}
     {N_{ion}(v) = \frac{3.8 \times 10^{14}}{f_j \lambda_j}  <\tau_j(v)> ~~(\textrm{cm}^{-2}\textrm{ km}^{-1}\textrm{ s})}
   \end{equation} 
  where $f_j$, $ \lambda_j$ and $<\tau_j(v)>$ are respectively the oscillator strength, the rest wavelength
 and the average optical depth across the emission source of the line $j$ for which the optical depth solution
 is derived \citep[see][]{Edmonds11}. The optical depth solution across a trough is found for a given ion by assuming an
 absorber model. As shown in \citet{Edmonds11}, the major uncertainty on the derived column densities
 comes from the choice of absorption model. In this study we investigate the outflow properties using column densities
 derived from three common absorber models.

 Assuming a single, homogeneous emission source of intensity $F_0$, the simplest absorber model
is the one where a homogeneous absorber parameterized by a single optical depth fully covers the photon source.
 In that case, known as the apparent optical depth scenario (AOD), the optical depth of a line $j$
 as a function of the radial velocity $v$ in the trough is simply derived by the inversion of
 the Beer-Lambert law : $\tau_j(v)=-ln(F_j(v)/F_0(v))$, where $F_j(v)$ is the observed intensity
 of the line.

 Early studies of AGN outflows pointed out the inadequacy of such an absorber
 model, specifically its inability to account for the observed departure of measured optical depth ratio between
 the components of typical doublet lines from the expected laboratory
 line strength ratio $R = \lambda_i f_i /\lambda_j f_j$. Two parameter absorber
 models have been developed to explain such discrepancies.
 The partial covering model \citep[e.g.][]{Hamann97b,Arav99b,Arav02,Arav05} assumes
 that only a fraction $C$ of the emission source is covered by absorbing material with constant optical depth $\tau$.
 In that case, the intensity observed for a line $j$ of a chosen ion can be expressed as
 \begin{equation}
 \label{eqcov}
 { F_j(v) = F_0(v) (1+C(v)*(e^{-\tau{j}(v)}-1)).}
\end{equation}

 Our third choice are inhomogeneous absorber models.
 In that scenario, the emission source is totally covered by a smooth distribution of
 absorbing material across its spatial dimension $x$. Assuming the typical power law
 distribution of the optical depth $\tau(x) = \tau_{max} x^{a}$ \citep{deKool02,Arav05,Arav08},
 the observed intensity observed for a line $j$ of a chosen ion is given by
  \begin{equation}
    \label{eqpow}
    {F_j(v) = F_0(v)  \int^1_0  e^{-\tau_{max,j}(v) x^{ a(v)} } dx }
  \end{equation}

 Once the line profiles have been binned on a common velocity scale (we choose a resolution $dv =20~\kms$,
 slightly lower than the resolution of COS), a velocity dependent solution can be obtained for the couple of parameters
 $(C,\tau_{j})$ or $(a,\tau_{max})$ of both absorber models as long as one observes
 at least two lines from a given ion, sharing the same lower energy level.
 Once the velocity dependent solution is computed, the corresponding column
 density is derived using Equation (\ref{eqcoldens}) where $<\tau_j(v)> = C_{ion}(v) \tau_j(v)$ for
 the partial covering model and $<\tau_j(v)>= \tau_{max,j}(v)/(a_{ion}(v) + 1)$ for the power law distribution.
 Note that the AOD solution can be computed for any line (singlet or multiplet), without further
 assumption on the model, but will essentially give a lower limit on the column density when
the expected line strength ratio observed is different from the laboratory value.

\begin{table*}
\caption{UV column densities for the outflow components in NGC 5548}
\label{table:Nion}
\begin{tabular}{r@{$\,$}lr@{}c@{}lr@{}c@{}lr@{}c@{}lr@{}c@{}lr@{}c@{}lr@{}c@{}l}
\hline
\multicolumn{2}{c}{Ion}	&\multicolumn{3}{c}{$v_1$}		&\multicolumn{3}{c}{$v_2$}				&\multicolumn{3}{c}{$v_3$}				&\multicolumn{3}{c}{$v_4$}				&\multicolumn{3}{c}{$v_5$}				&\multicolumn{3}{c}{$v_6$}\\
&			&\multicolumn{3}{c}{$[-1450,-850]^a$}	&\multicolumn{3}{c}{$[-850,-750]$}			&\multicolumn{3}{c}{$[-750,-540]$}			&\multicolumn{3}{c}{$[-540,-360]$}			&\multicolumn{3}{c}{$[-360,-80]$}			&\multicolumn{3}{c}{$[-80,+50]$}\\\hline
\multicolumn{20}{c}{Epoch 2013}\\\hline
H&{\sc i}	&\color{blu}$>$&\color{blu}14.39&$\,\!^b$			&\color{blu}$>$&\color{blu}13.86&			&\color{blu}$>$&\color{blu}14.28&			&\color{blu}$>$&\color{blu}14.40&			&\color{blu}$>$&\color{blu}14.00&			&\color{blu}$>$&\color{blu}13.00&		\\
C&{\sc ii}	&\color{blu}$>$&\color{blu}14.52&			&\color{red}$<$&\color{red}12.81&			&&13.67&$^{+0.1}_{-0.1}$			&\color{red}$<$&\color{red}13.24&			&\color{red}$<$&\color{red}13.40&			&\color{red}$<$&\color{red}13.00&		\\
C&{\sc iii}$^*_{J=0}$    &&14.03&$^{+0.02}_{-0.02}$				&&&							&&&							&&&							&&&							&&&						\\
C&{\sc iii}$^*_{J=2}$    &&14.64&$^{+0.02}_{-0.02}$				&&&							&&&							&&&							&&&							&&&						\\
C&{\sc iv}	&\color{blu}$>$&\color{blu}14.60&			&&14.00&$^{+0.3}_{-0.1}$				&\color{blu}$>$&\color{blu}14.30&			&\color{blu}$>$&\color{blu}14.30&			&\color{blu}$>$&\color{blu}14.10&			&\color{blu}$>$&\color{blu}13.70&		\\
N&{\sc v}	&\color{blu}$>$&\color{blu}14.90&			&&13.90&$^{+0.2}_{-0.2}$				&\color{blu}$>$&\color{blu}14.90&			&\color{blu}$>$&\color{blu}14.80&			&\color{blu}$>$&\color{blu}14.60&			&\color{blu}$>$&\color{blu}14.10&		\\
Al&{\sc ii}	&\color{blu}$>$&\color{blu}13.05&			&&&							&&&							&&&							&&&							&&&						\\
Si&{\sc ii}	&&14.30&$^{+0.18}_{-0.3}$				&\color{red}$<$&\color{red}11.80&			&\color{red}$<$&\color{red}12.00&			&\color{red}$<$&\color{red}11.90&			&\color{red}$<$&\color{red}11.40&			&\color{red}$<$&\color{red}12.10&		\\
Si&{\sc iii}	&\color{blu}$>$&\color{blu}13.70&			&\color{red}$<$&\color{red}12.00&			&&13.10&$^{+0.3}_{-0.3}$				&\color{red}$<$&\color{red}12.20&			&&12.20&$^{+0.2}_{-0.2}$				&\color{red}$<$&\color{red}11.50&		\\
Si&{\sc iv}	&\color{blu}$>$&\color{blu}14.08&			&&12.30&$^{+0.2}_{-0.2}$				&&13.88&$^{+0.2}_{-0.2}$				&\color{red}$<$&\color{red}13.00&			&&13.13&$^{+0.2}_{-0.2}$				&\color{red}$<$&\color{red}12.30&		\\
P&{\sc iii}	&&14.04&$^{+0.16}_{-0.06}$				&&&							&&&							&&&							&&&							&&&						\\
P&{\sc v}	&&14.15&$^{+0.5}_{-0.15}$				&&&							&&&							&&&							&&&							&&&						\\
S&{\sc iii}	&&15.00&$^{+0.30}_{-0.22}$				&&&							&&&							&&&							&&&							&&&						\\
Fe&{\sc iii}	&&14.70&$^{+0.30}_{-0.30}$				&&&							&&&							&&&							&&&							&&&						\\\hline
\multicolumn{20}{c}{Epoch 2011}\\\hline
H&{\sc i}	&\color{blu}$>$&\color{blu}14.34&			&\color{blu}$>$&\color{blu}13.80&			&\color{blu}$>$&\color{blu}14.28&			&\color{blu}$>$&\color{blu}14.33&			&\color{blu}$>$&\color{blu}13.83&			&\color{blu}$>$&\color{blu}12.90&		\\
C&{\sc iv}	&\color{blu}$>$&\color{blu}14.20&			&\color{blu}$>$&\color{blu}13.70&			&\color{blu}$>$&\color{blu}14.30&			&&14.40&$^{+0.2}_{-0.2}$				&\color{blu}$>$&\color{blu}13.83&			&\color{blu}$>$&\color{blu}13.03&		\\
N&{\sc v}	&\color{blu}$>$&\color{blu}14.59&			&\color{blu}$>$&\color{blu}13.95&			&\color{blu}$>$&\color{blu}14.65&			&\color{blu}$>$&\color{blu}14.77&			&\color{blu}$>$&\color{blu}14.43&			&\color{blu}$>$&\color{blu}13.84&		\\
Si&{\sc iii}	&\color{red}$<$&\color{red}12.20&			&\color{red}$<$&\color{red}11.50&			&\color{red}$<$&\color{red}11.98&			&\color{red}$<$&\color{red}11.84&			&\color{red}$<$&\color{red}11.55&			&\color{red}$<$&\color{red}11.59&		\\
Si&{\sc iv}	&\color{red}$<$&\color{red}13.06&			&\color{red}$<$&\color{red}12.54&			&\color{red}$<$&\color{red}12.66&			&\color{red}$<$&\color{red}12.70&			&\color{red}$<$&\color{red}12.53&			&\color{red}$<$&\color{red}12.37&		\\\hline
\multicolumn{20}{c}{Epoch 2004}\\\hline
H&{\sc i}	&\color{blu}$>$&\color{blu}13.94&			&\color{blu}$>$&\color{blu}13.82&			&\color{blu}$>$&\color{blu}14.20&			&\color{blu}$>$&\color{blu}14.44&			&\color{blu}$>$&\color{blu}14.15&			&\color{blu}$>$&\color{blu}12.41&		\\
C&{\sc iv}	&\color{blu}$>$&\color{blu}14.59&			&\color{blu}$>$&\color{blu}13.83&			&\color{blu}$>$&\color{blu}13.95&			&\color{blu}$>$&\color{blu}14.25&			&\color{blu}$>$&\color{blu}14.20&			&\color{blu}$>$&\color{blu}13.16&		\\
N&{\sc v}	&\color{blu}$>$&\color{blu}14.65&			&\color{blu}$>$&\color{blu}13.83&			&\color{blu}$>$&\color{blu}14.20&			&\color{blu}$>$&\color{blu}14.35&			&\color{blu}$>$&\color{blu}13.75&			&\color{blu}$>$&\color{blu}13.20&		\\
Si&{\sc iii}	&\color{blu}$>$&\color{blu}13.61&			&\color{blu}$>$&\color{blu}12.47&			&\color{blu}$>$&\color{blu}13.29&			&\color{blu}$>$&\color{blu}12.41&			&\color{blu}$>$&\color{blu}12.66&			&\color{blu}$>$&\color{blu}12.33&		\\
Si&{\sc iv}	&\color{blu}$>$&\color{blu}14.18&			&\color{blu}$>$&\color{blu}12.91&			&\color{blu}$>$&\color{blu}13.65&			&\color{blu}$>$&\color{blu}12.71&			&\color{blu}$>$&\color{blu}13.32&			&\color{blu}$>$&\color{blu}12.76&		\\\hline
\multicolumn{20}{c}{Epoch 2002}\\\hline
H&{\sc i}	&\color{blu}$>$&\color{blu}14.34&			&\color{blu}$>$&\color{blu}13.91&			&\color{blu}$>$&\color{blu}14.29&			&\color{blu}$>$&\color{blu}14.46&			&\color{blu}$>$&\color{blu}13.91&			&\color{blu}$>$&\color{blu}13.20&		\\
C&{\sc iv}	&\color{blu}$>$&\color{blu}14.08&			&\color{blu}$>$&\color{blu}13.54&			&\color{blu}$>$&\color{blu}13.95&			&&14.30&$^{+0.2}_{-0.2}$				&\color{blu}$>$&\color{blu}13.60&			&\color{blu}$>$&\color{blu}13.23&		\\
N&{\sc v}	&\color{blu}$>$&\color{blu}14.54&			&&13.87&$^{+0.2}_{-0.2}$				&&14.50&$^{+0.2}_{-0.2}$				&\color{blu}$>$&\color{blu}14.76&			&\color{blu}$>$&\color{blu}14.12&			&\color{blu}$>$&\color{blu}13.21&		\\
Si&{\sc iii}	&\color{red}$<$&\color{red}12.66&			&\color{red}$<$&\color{red}11.80&			&\color{red}$<$&\color{red}11.97&			&\color{red}$<$&\color{red}11.91&			&\color{red}$<$&\color{red}12.14&			&\color{red}$<$&\color{red}11.97&		\\
Si&{\sc iv}	&\color{red}$<$&\color{red}13.31&			&\color{red}$<$&\color{red}12.39&			&\color{red}$<$&\color{red}12.81&			&\color{red}$<$&\color{red}12.74&			&\color{red}$<$&\color{red}13.13&			&\color{red}$<$&\color{red}12.58&		\\\hline
\multicolumn{20}{c}{Epoch 1998}\\\hline
H&{\sc i}	&\color{blu}$>$&\color{blu}14.06&			&\color{blu}$>$&\color{blu}13.66&			&\color{blu}$>$&\color{blu}14.28&			&\color{blu}$>$&\color{blu}14.36&			&\color{blu}$>$&\color{blu}13.75&			&\color{blu}$>$&\color{blu}12.96&		\\
C&{\sc iv}	&\color{red}$<$&\color{red}13.69&			&\color{blu}$>$&\color{blu}13.25&			&\color{blu}$>$&\color{blu}13.59&			&\color{blu}$>$&\color{blu}14.47&			&\color{red}$<$&\color{red}13.09&			&\color{red}$<$&\color{red}13.00&		\\
N&{\sc v}	&\color{red}$<$&\color{red}13.99&			&\color{blu}$>$&\color{blu}13.89&			&\color{blu}$>$&\color{blu}14.19&			&\color{blu}$>$&\color{blu}14.88&			&\color{blu}$>$&\color{blu}14.13&			&\color{red}$<$&\color{red}13.98&		\\
Si&{\sc iii}	&\color{red}$<$&\color{red}12.83&			&\color{red}$<$&\color{red}12.14&			&\color{red}$<$&\color{red}12.46&			&\color{red}$<$&\color{red}12.18&			&\color{red}$<$&\color{red}12.57&			&\color{red}$<$&\color{red}12.41&		\\
Si&{\sc iv}	&\color{red}$<$&\color{red}13.39&			&\color{red}$<$&\color{red}12.83&			&\color{red}$<$&\color{red}12.93&			&\color{red}$<$&\color{red}12.69&			&\color{red}$<$&\color{red}12.89&			&\color{red}$<$&\color{red}12.13&		\\\hline

\multicolumn{20}{l}{$\,^a$Integration limits in km~$\mathrm{s}^{-1}$.}\\
\multicolumn{20}{l}{$\,^b$Lower limit log column densities given in units of $\mathrm{cm}^{-2}$ are shown in {\color{blu} blue}.}\\
\multicolumn{20}{l}{\hspace{0.2cm}Upper limits are likewise shown in {\color{red} red}.}\\
\end{tabular}
\end{table*}

\renewcommand{\thefigure}{A.1.\alph{figure}}
\setcounter{figure}{0}
\begin{landscape}\begin{figure}
\includegraphics[width=0.95\textwidth,angle=90]{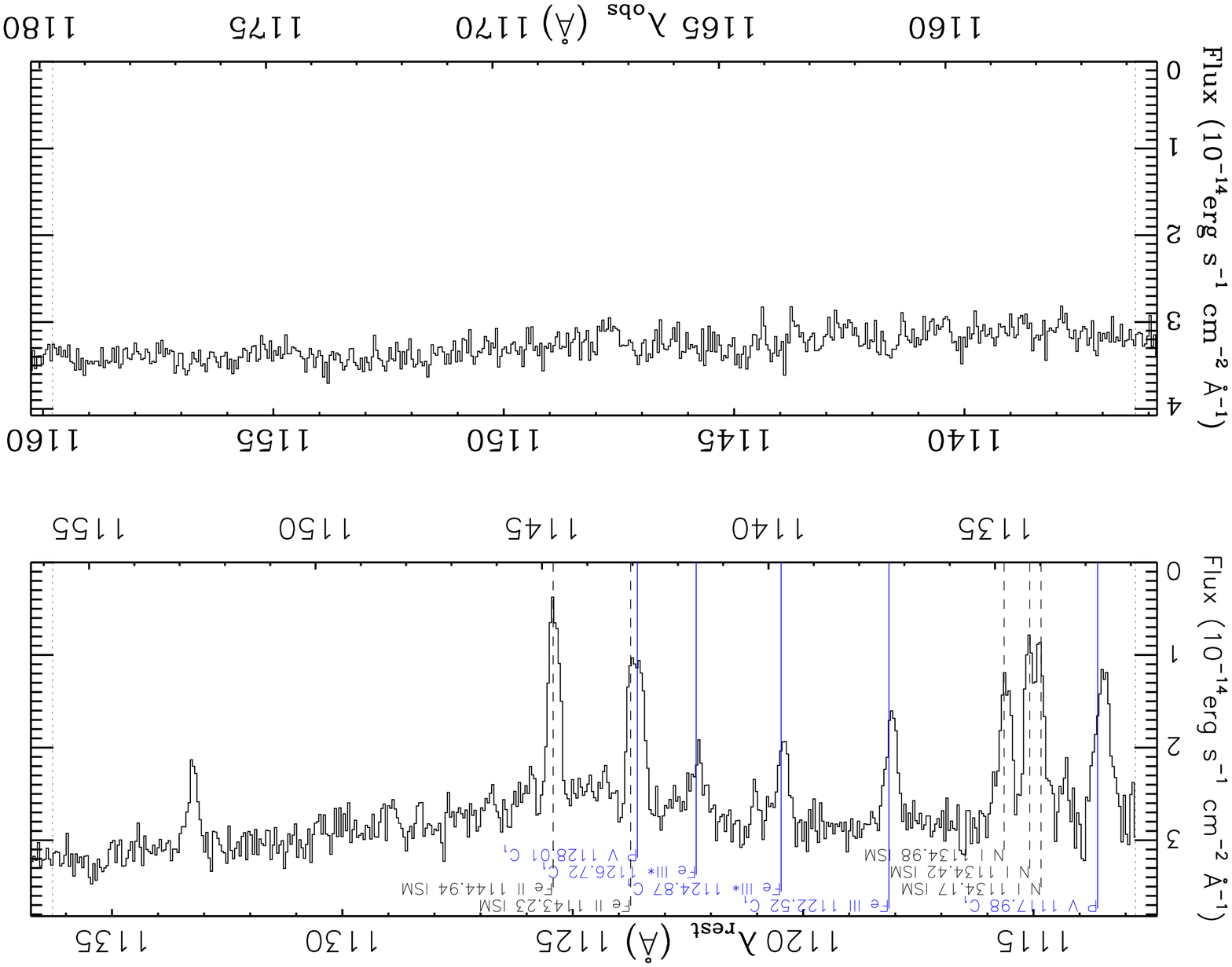}
\parbox{1.25\textheight}{\caption{Plot of the 2013 spectrum of NGC 5548.  The vertical axis is the flux in units of 10$^{-14}$ erg s$^{-1}$ cm$^{-2}$ \AA$^{-1}$, and the quasar-rest-frame and observer-frame wavelengths are given in Angstroms on the top and bottom of each sub-plot, respectively.  Each of the six kinematic components of the outflow shows absorption troughs from several ions.  We place a vertical mark at the expected center of each absorption trough (following the velocity template of \siiv\ and \nv) and state the ion, rest-wavelength and component number (C$_1$--C$_6$).  We also assign a color to each component number that ranges from blue (C$_1$) to red (C$_6$).  Absorption lines from the ISM are likewise marked in black with dashed lines.}}\end{figure}\end{landscape}

\pagebreak
\begin{landscape}\begin{figure}
\includegraphics[width=0.95\textwidth,angle=90]{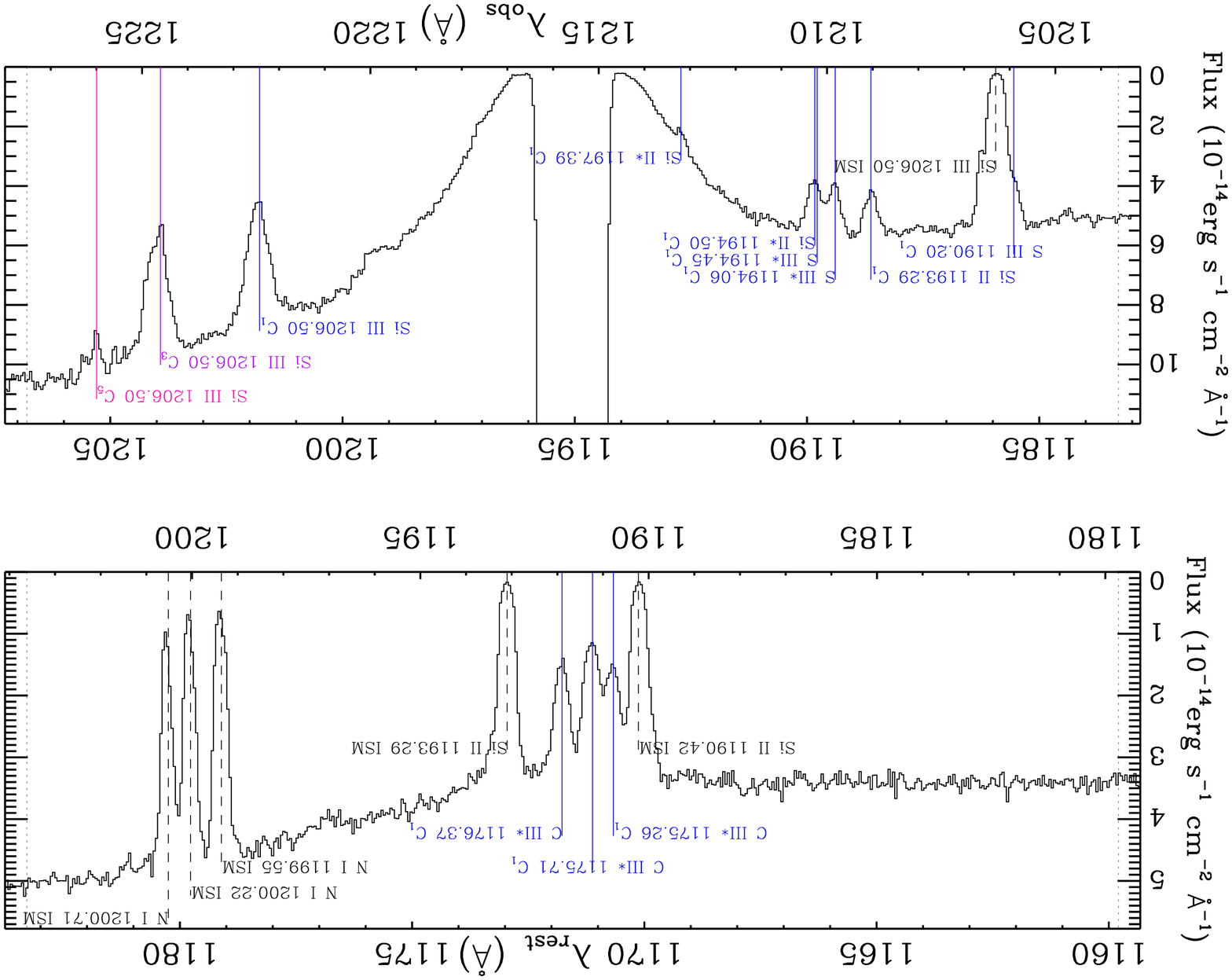}
\caption{continued.}\end{figure}\end{landscape}

\pagebreak
\begin{landscape}\begin{figure}
\includegraphics[width=0.95\textwidth,angle=90]{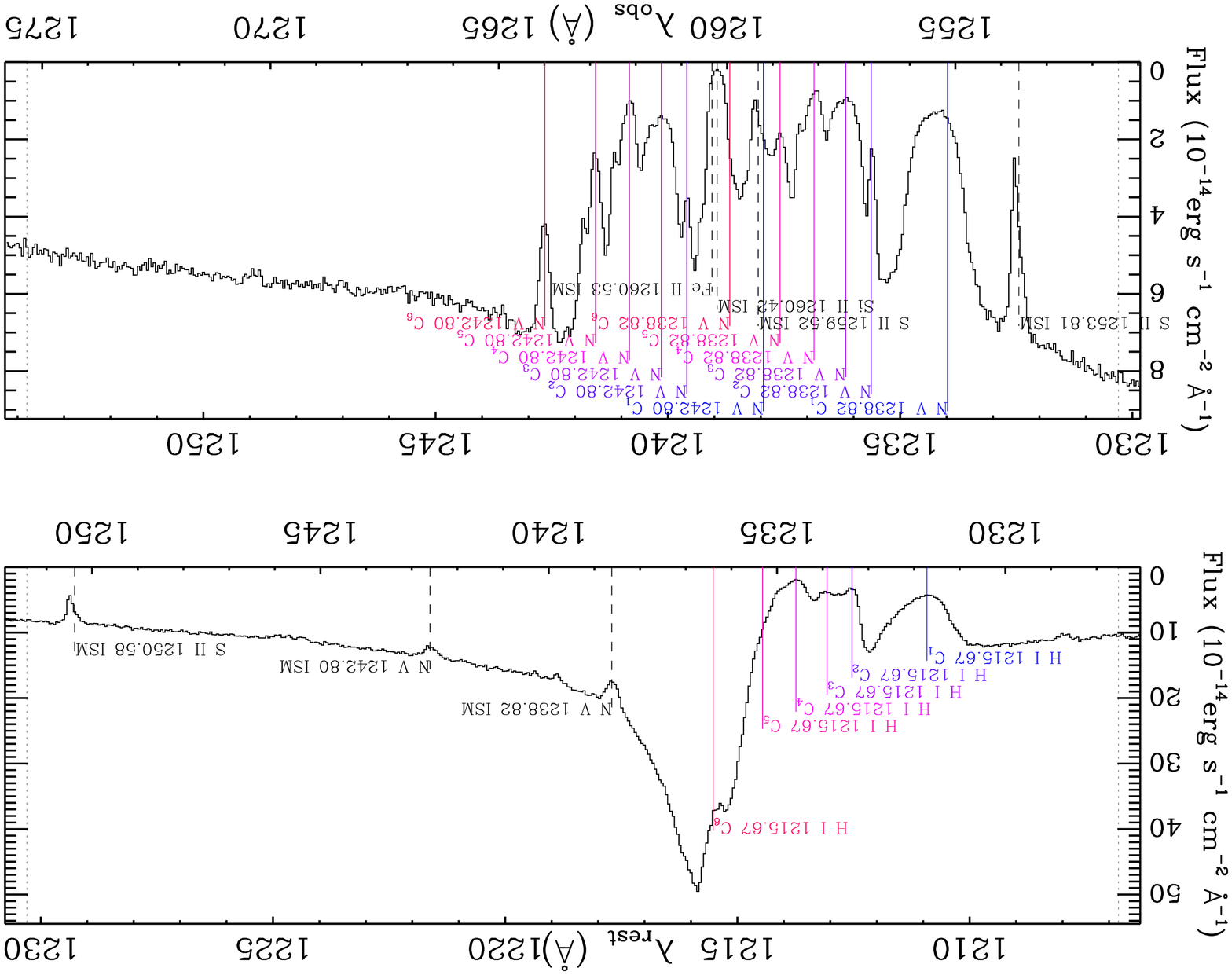}
\caption{continued.}\end{figure}\end{landscape}

\pagebreak
\begin{landscape}\begin{figure}
\includegraphics[width=0.95\textwidth,angle=90]{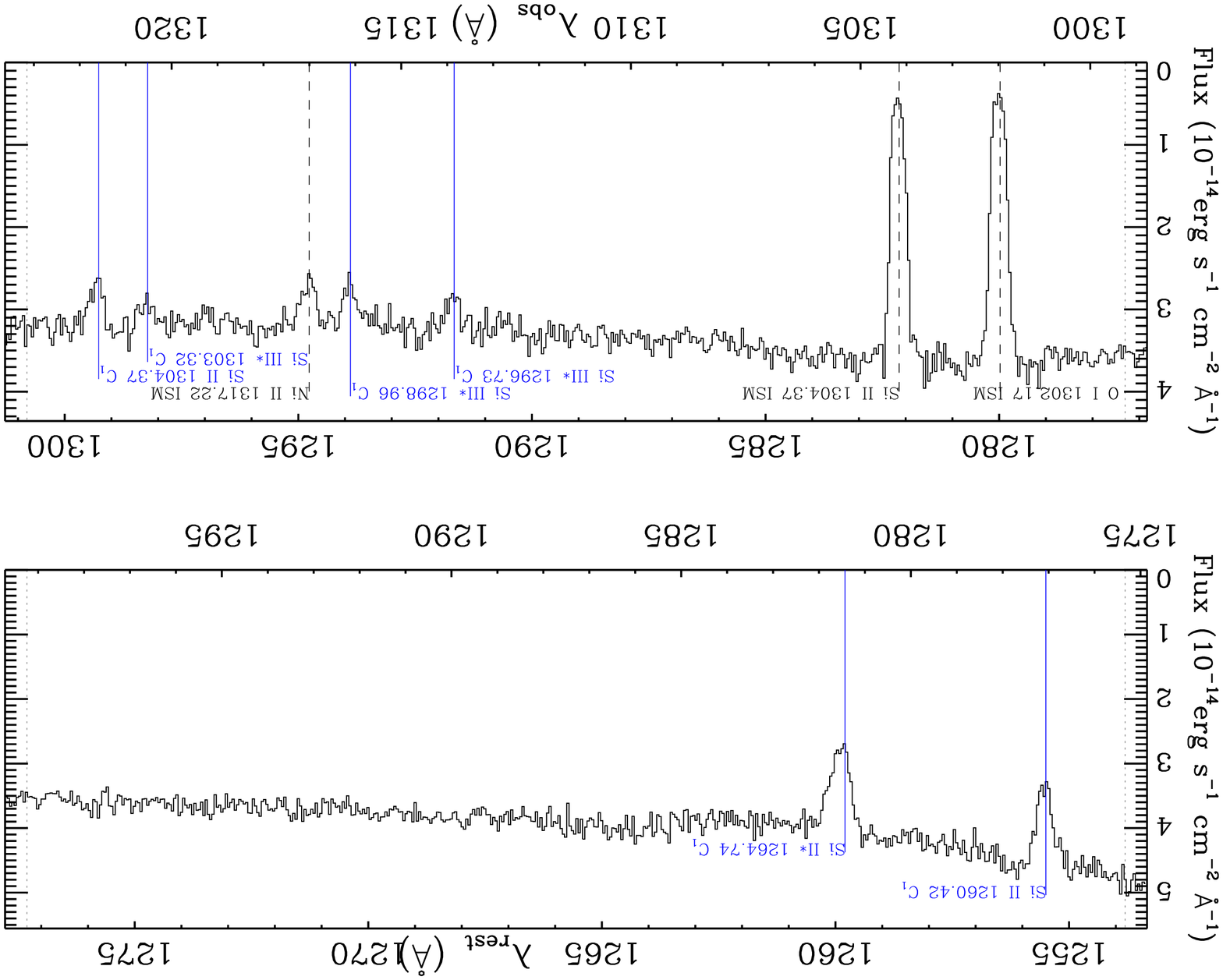}
\caption{continued.}\end{figure}\end{landscape}

\pagebreak
\begin{landscape}\begin{figure}
\includegraphics[width=0.95\textwidth,angle=90]{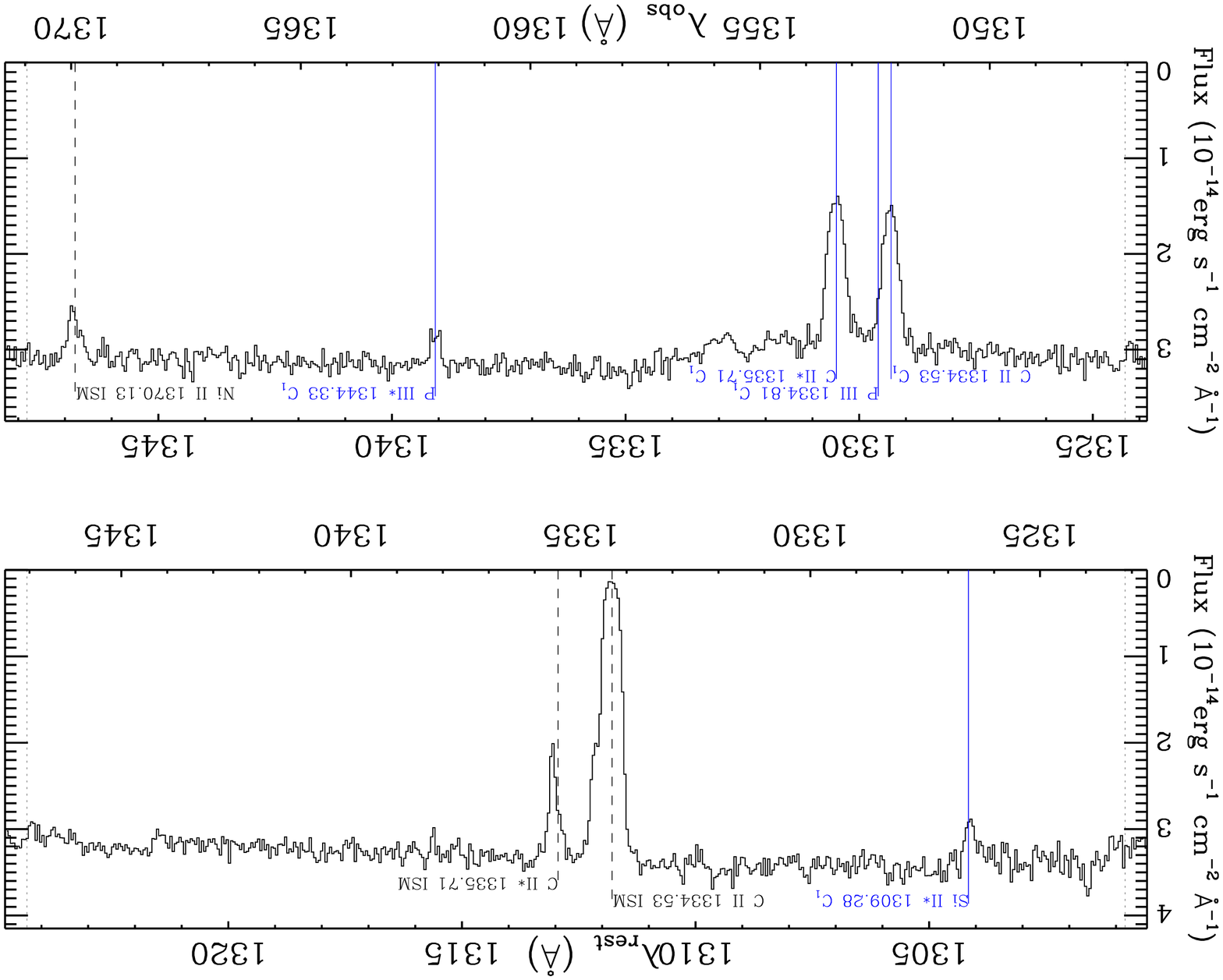}
\caption{continued.}\end{figure}\end{landscape}

\pagebreak
\begin{landscape}\begin{figure}
\includegraphics[width=0.95\textwidth,angle=90]{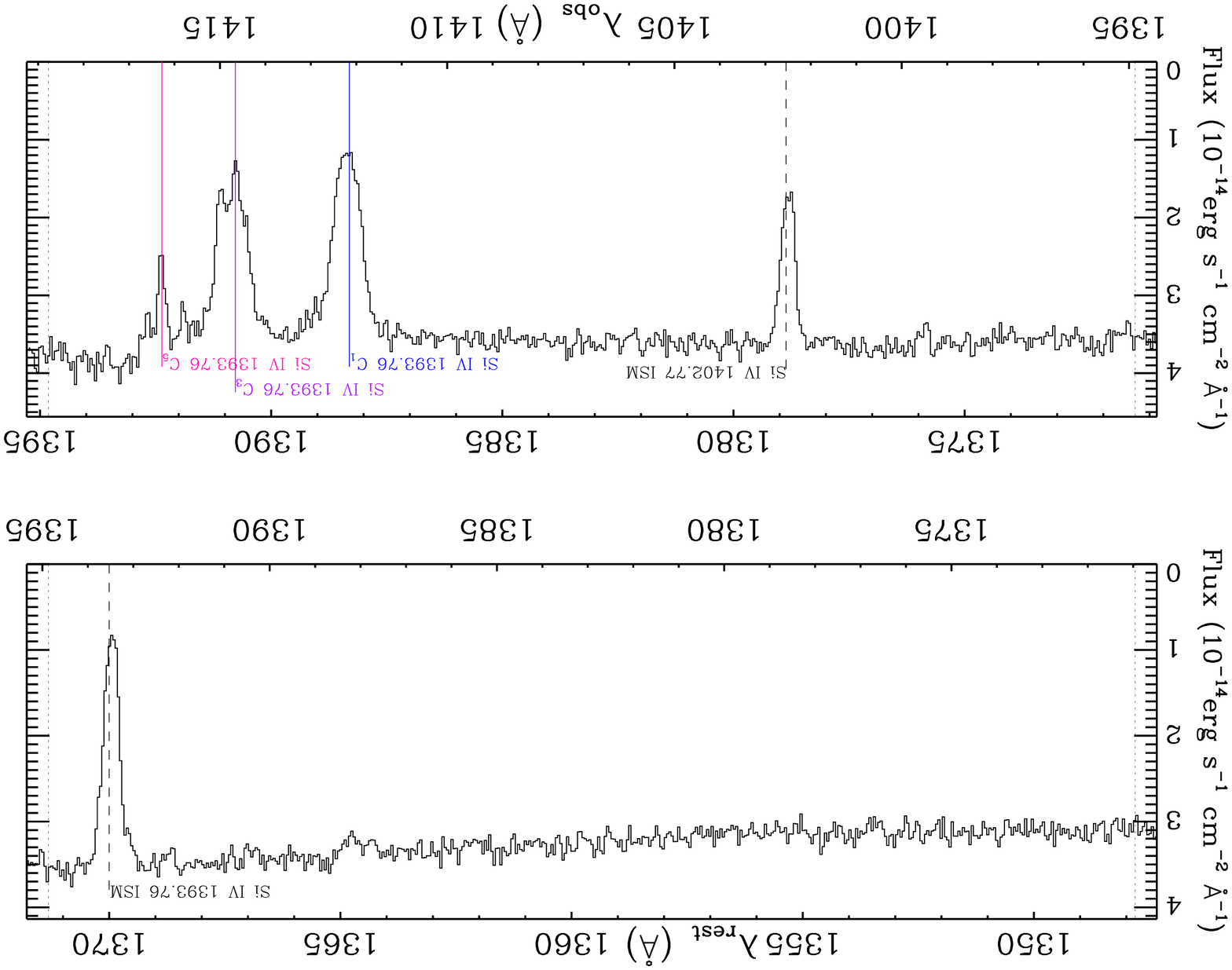}
\caption{continued.}\end{figure}\end{landscape}

\pagebreak
\begin{landscape}\begin{figure}
\includegraphics[width=0.95\textwidth,angle=90]{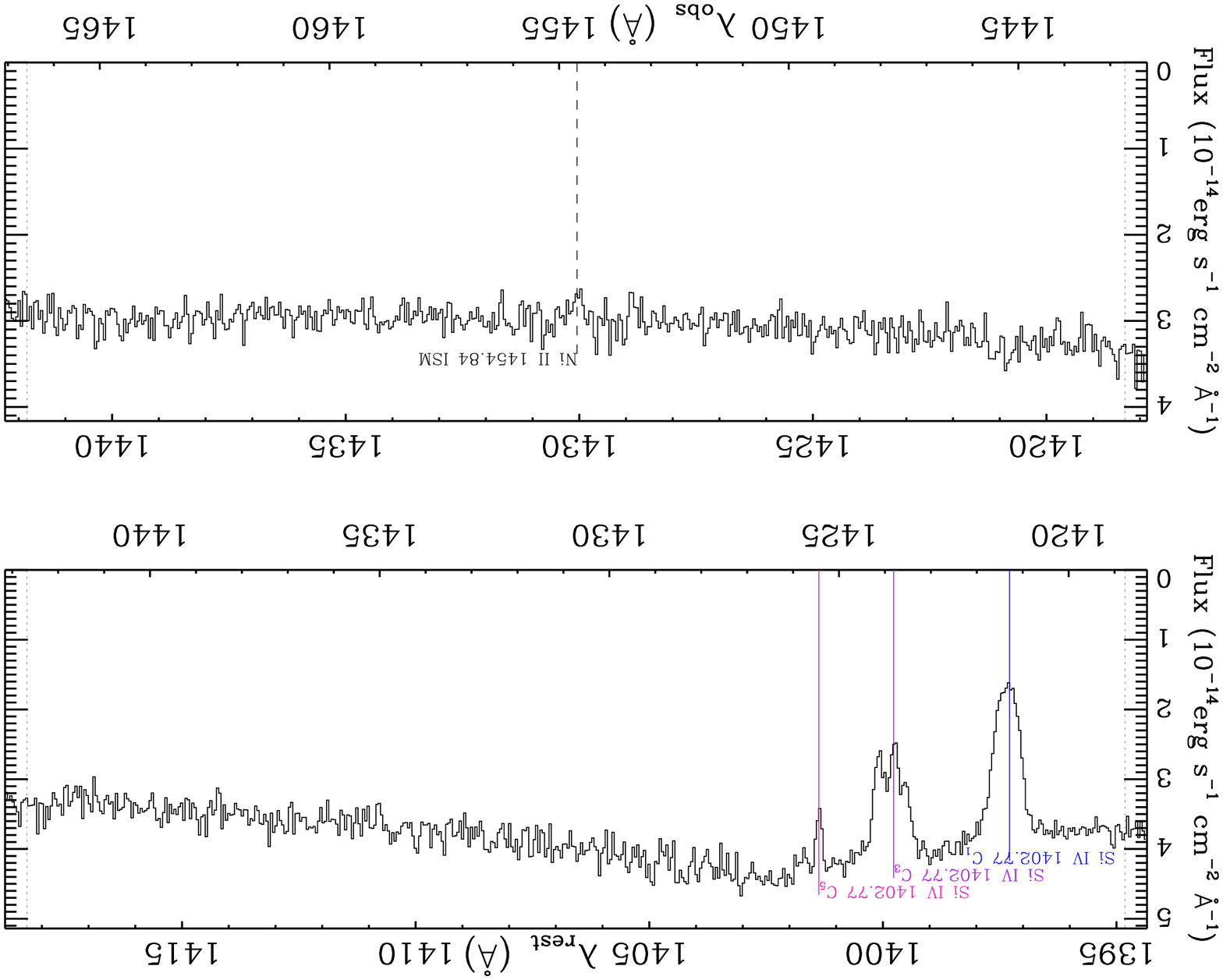}
\caption{continued.}\end{figure}\end{landscape}

\pagebreak
\begin{landscape}\begin{figure}
\includegraphics[width=0.95\textwidth,angle=90]{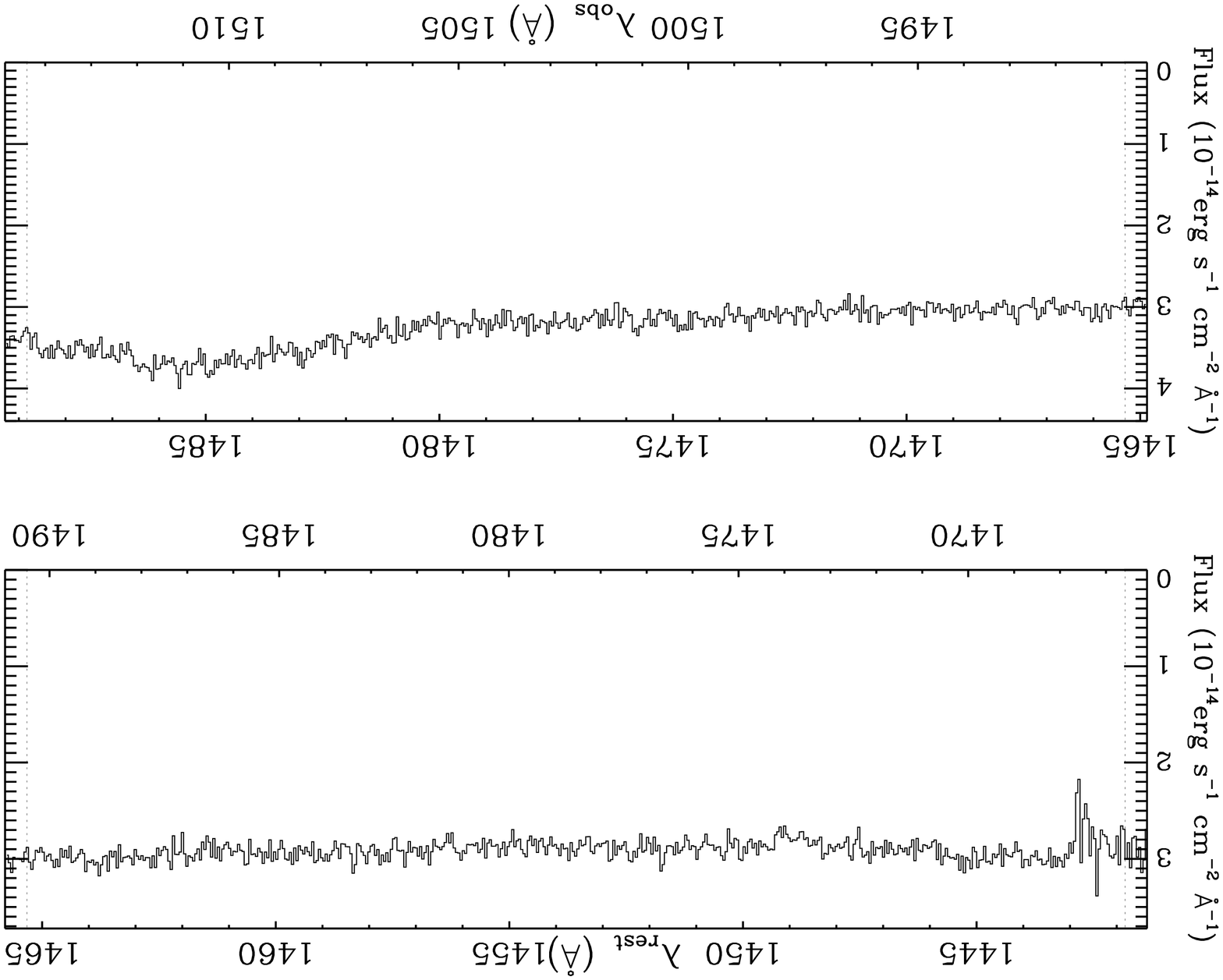}
\caption{continued.}\end{figure}\end{landscape}

\pagebreak
\begin{landscape}\begin{figure}
\includegraphics[width=0.95\textwidth,angle=90]{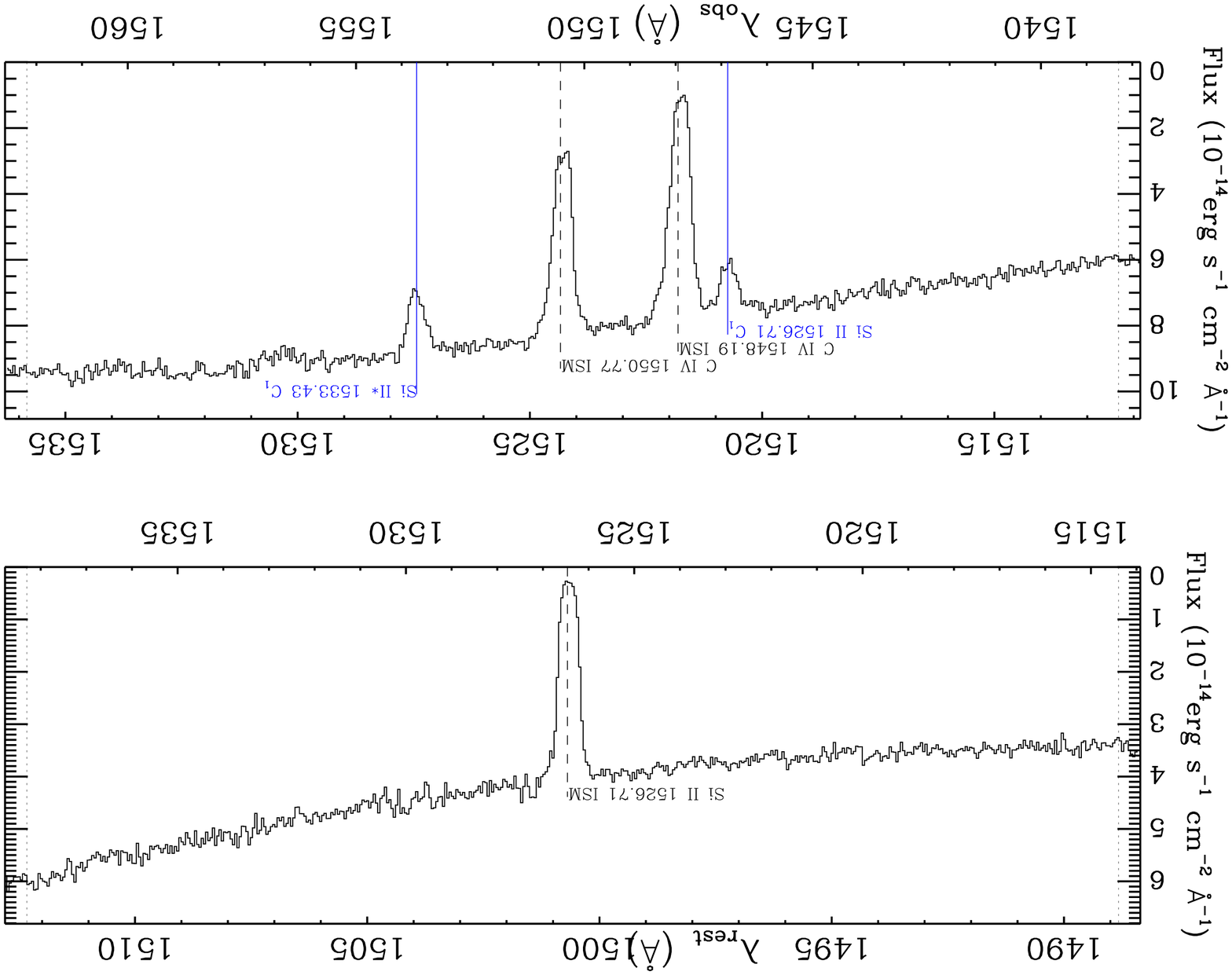}
\caption{continued.}\end{figure}\end{landscape}

\pagebreak
\begin{landscape}\begin{figure}
\includegraphics[width=0.95\textwidth,angle=90]{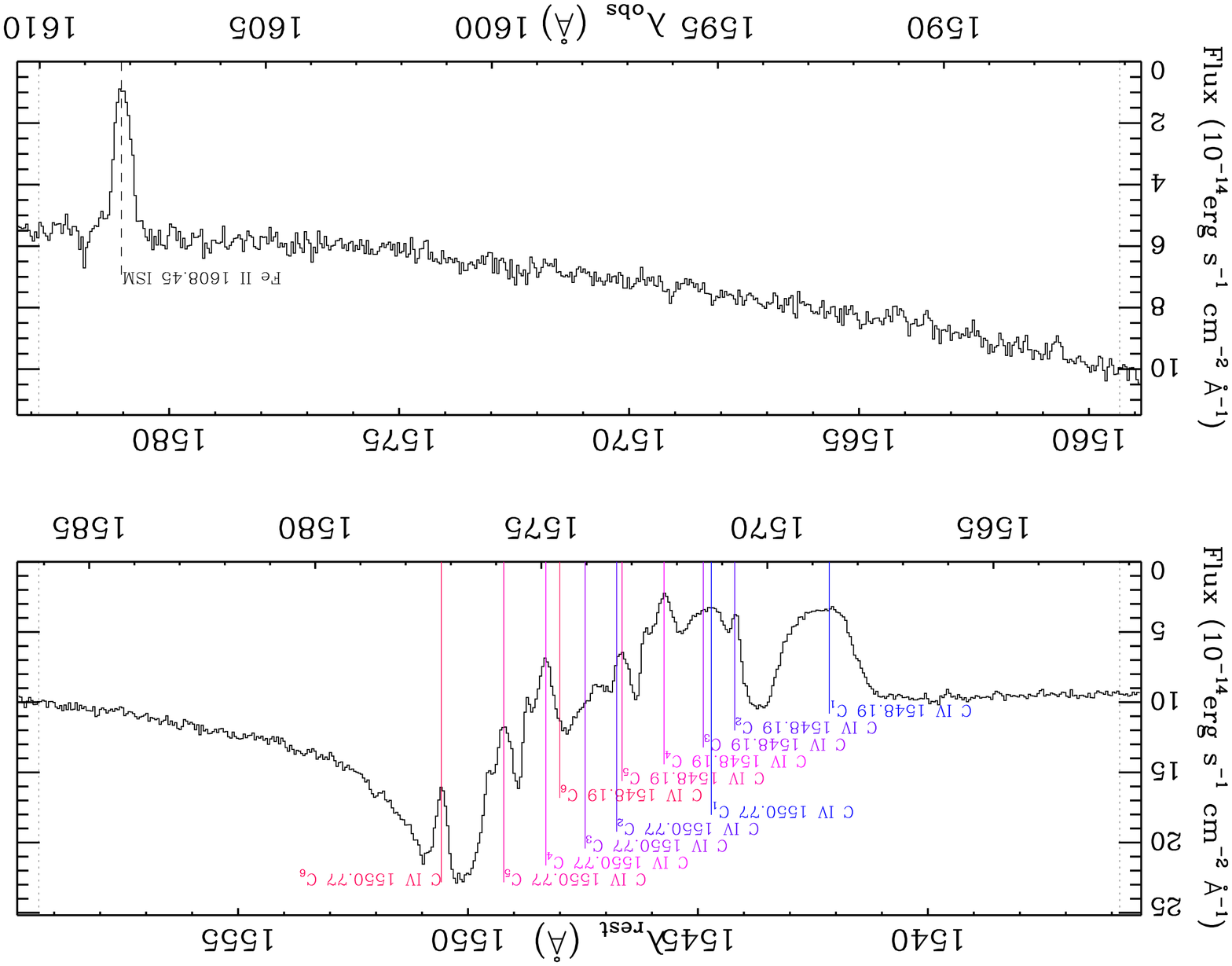}
\caption{continued.}\end{figure}\end{landscape}

\pagebreak
\begin{landscape}\begin{figure}
\includegraphics[width=0.95\textwidth,angle=90]{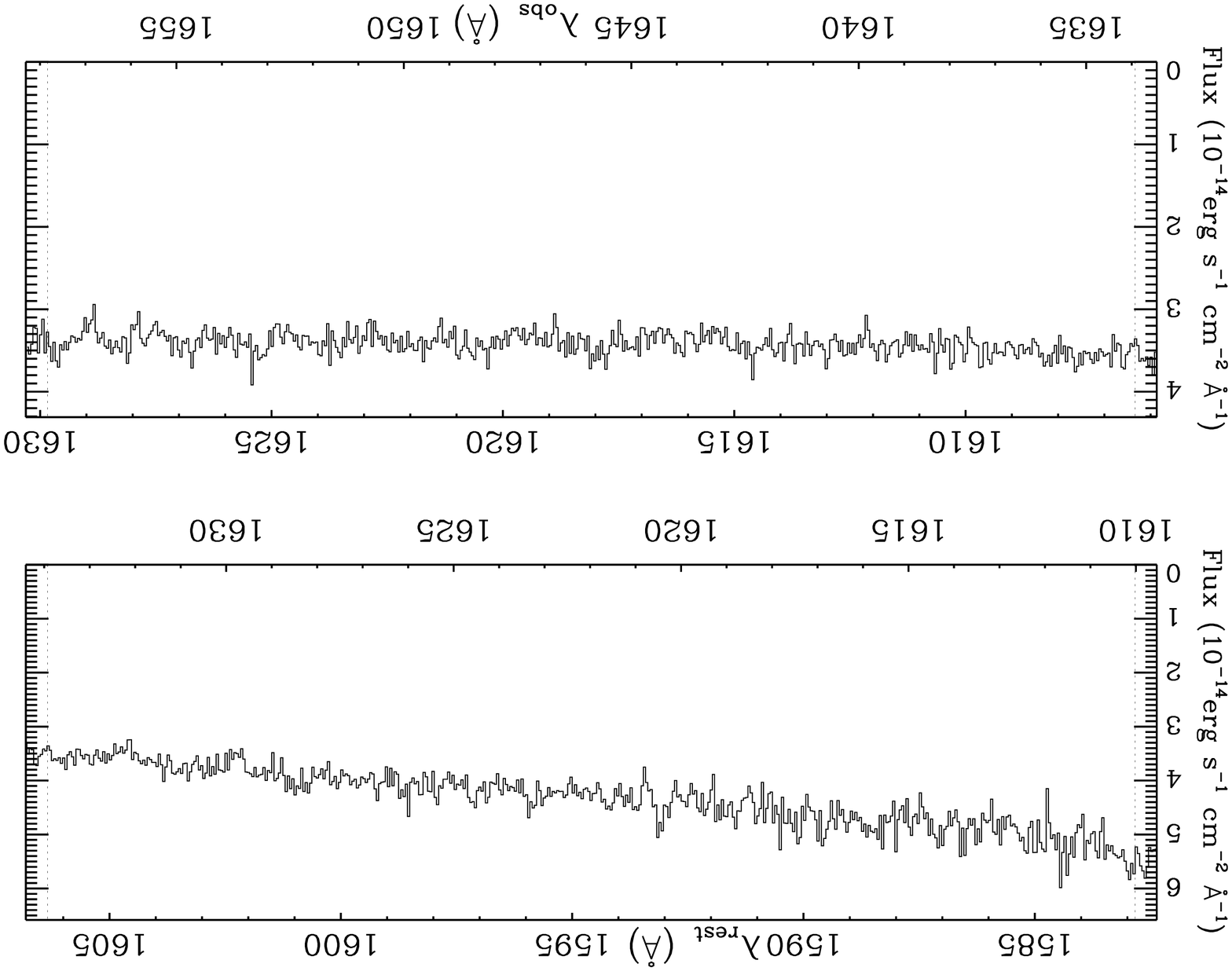}
\caption{continued.}\end{figure}\end{landscape}

\pagebreak
\begin{landscape}\begin{figure}
\includegraphics[width=0.95\textwidth,angle=90]{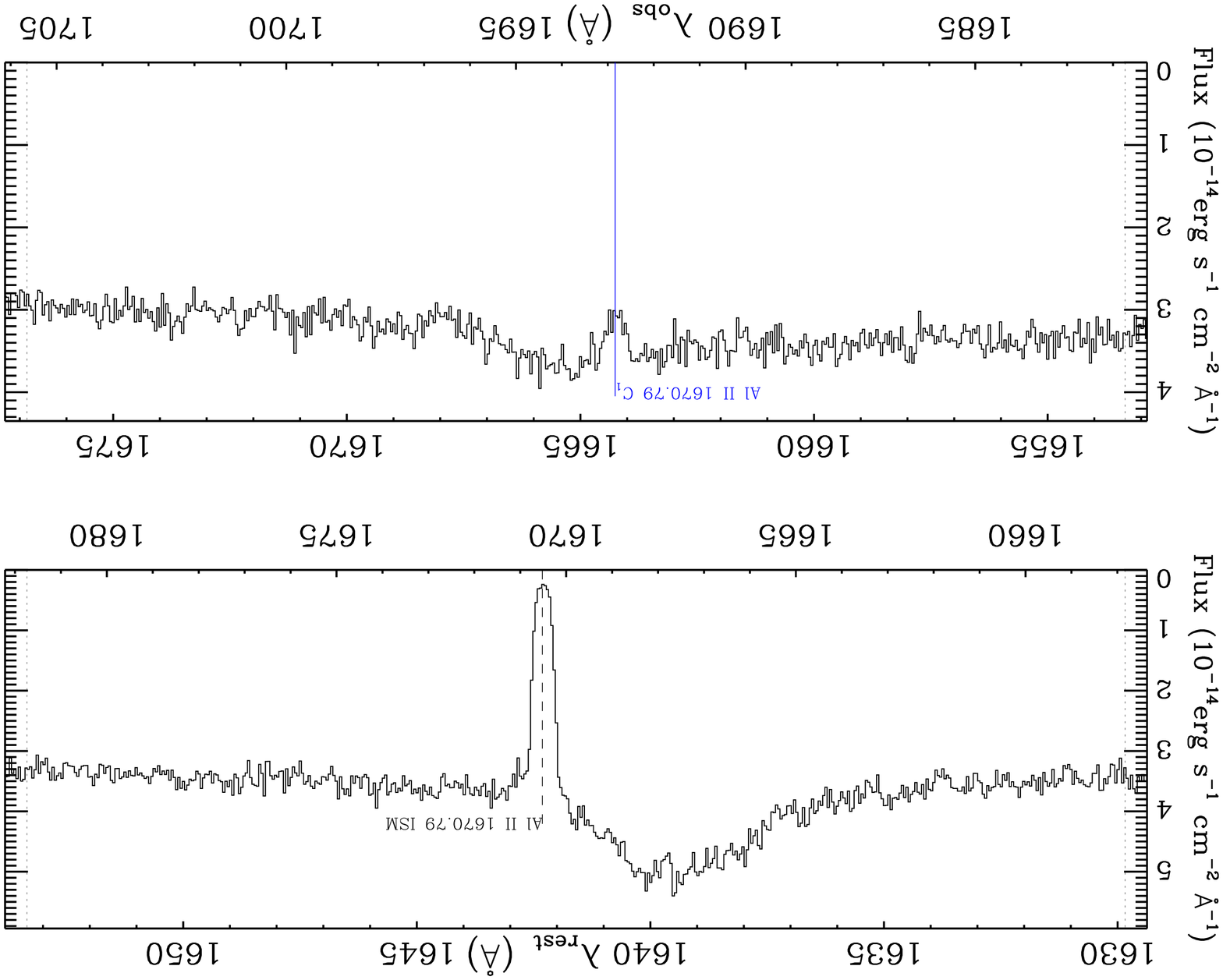}
\caption{continued.}\end{figure}\end{landscape}

\pagebreak
\begin{landscape}\begin{figure}
\includegraphics[width=0.95\textwidth,angle=90]{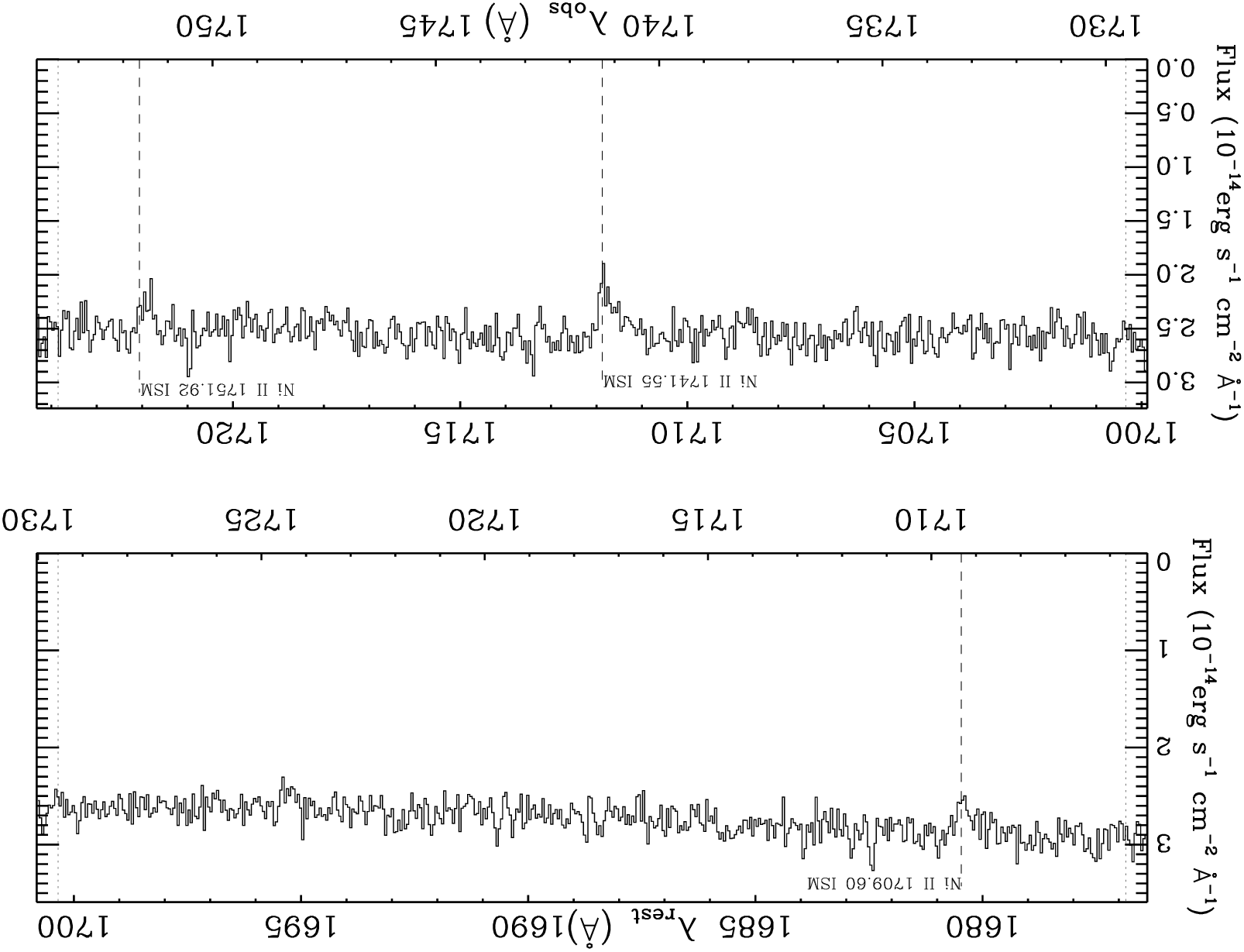}
\caption{continued.}\end{figure}\end{landscape}

\pagebreak
\begin{landscape}\begin{figure}
\includegraphics[width=0.95\textwidth,angle=90]{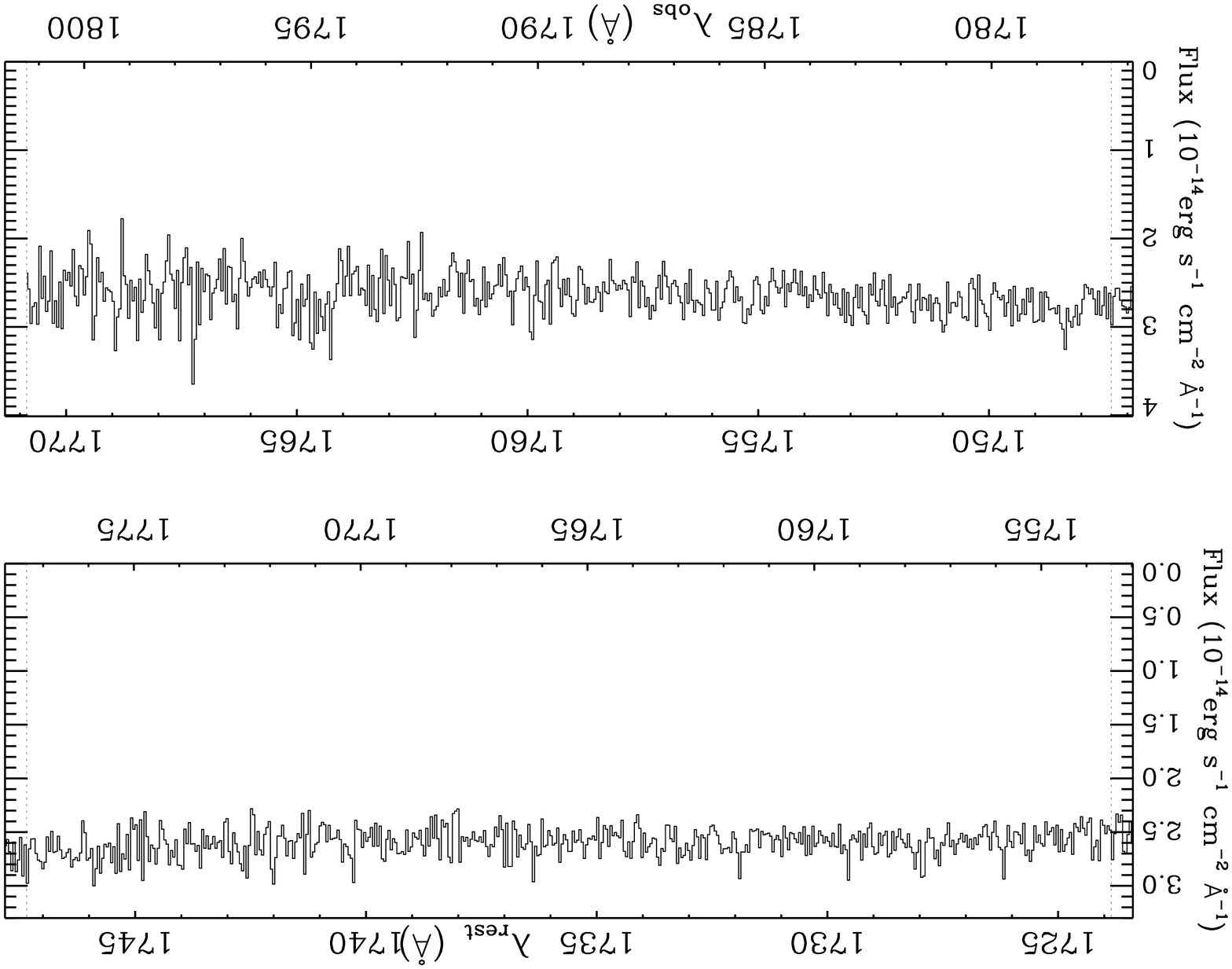}
\caption{continued.}\end{figure}\end{landscape}

\setcounter{figure}{0}
\renewcommand{\thefigure}{A.2.\alph{figure}}

\pagebreak
\begin{landscape}\begin{figure}
\includegraphics[width=0.95\textwidth,angle=90]{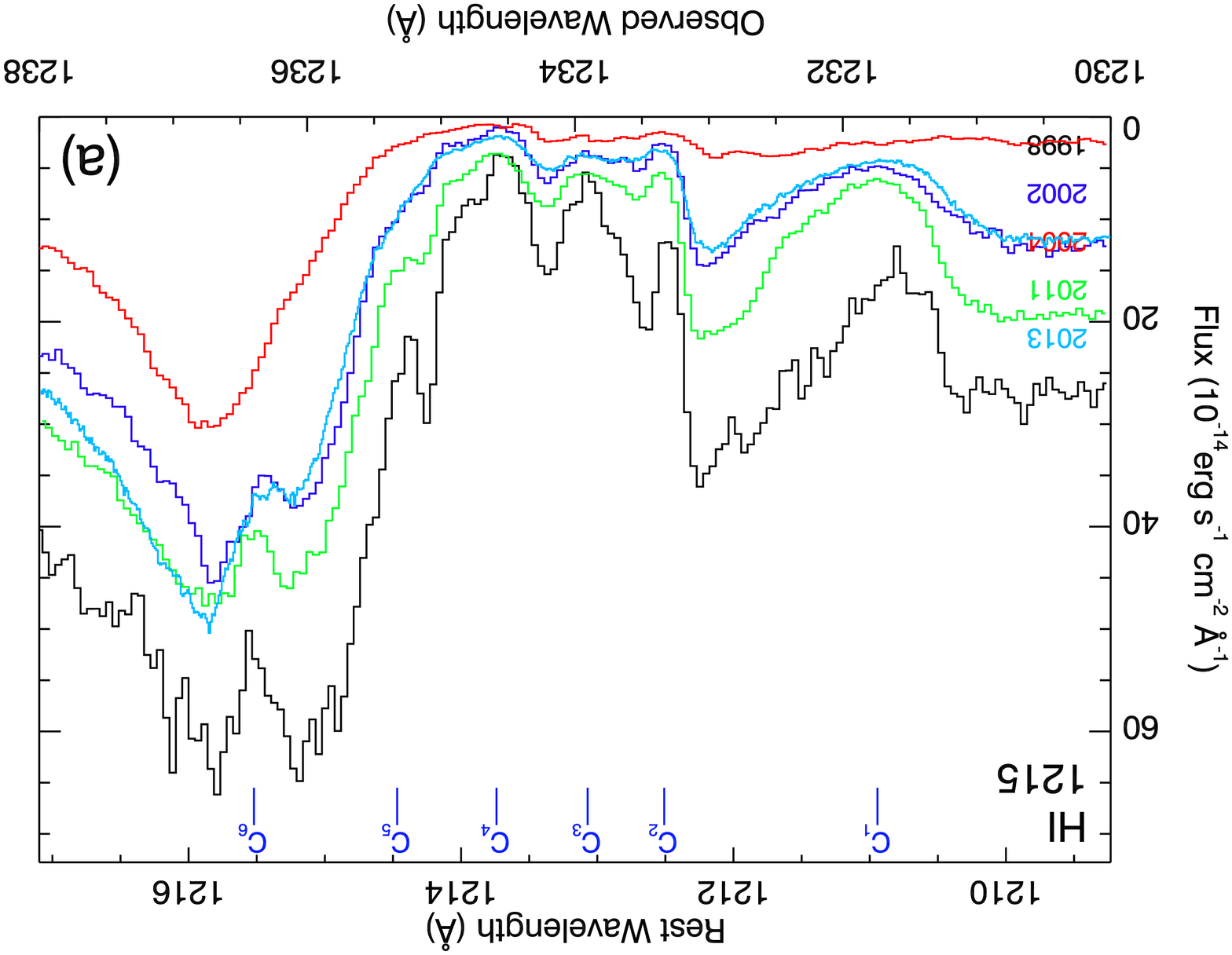}
\parbox{1.25\textheight}{\caption{A plot of the spectrum of NGC 5548 during the five epochs of observation.  The 2013 spectrum is obtained by co-adding visits 1 through 5.  Spectral regions where absorption troughs from five ions are shown in sub-plots (a) through (e) and the six kinematic components associated with such absorption are labelled C$_1$ through C$_6$.  }}\end{figure}\end{landscape}

\pagebreak
\begin{landscape}\begin{figure}
\includegraphics[width=0.95\textwidth,angle=90]{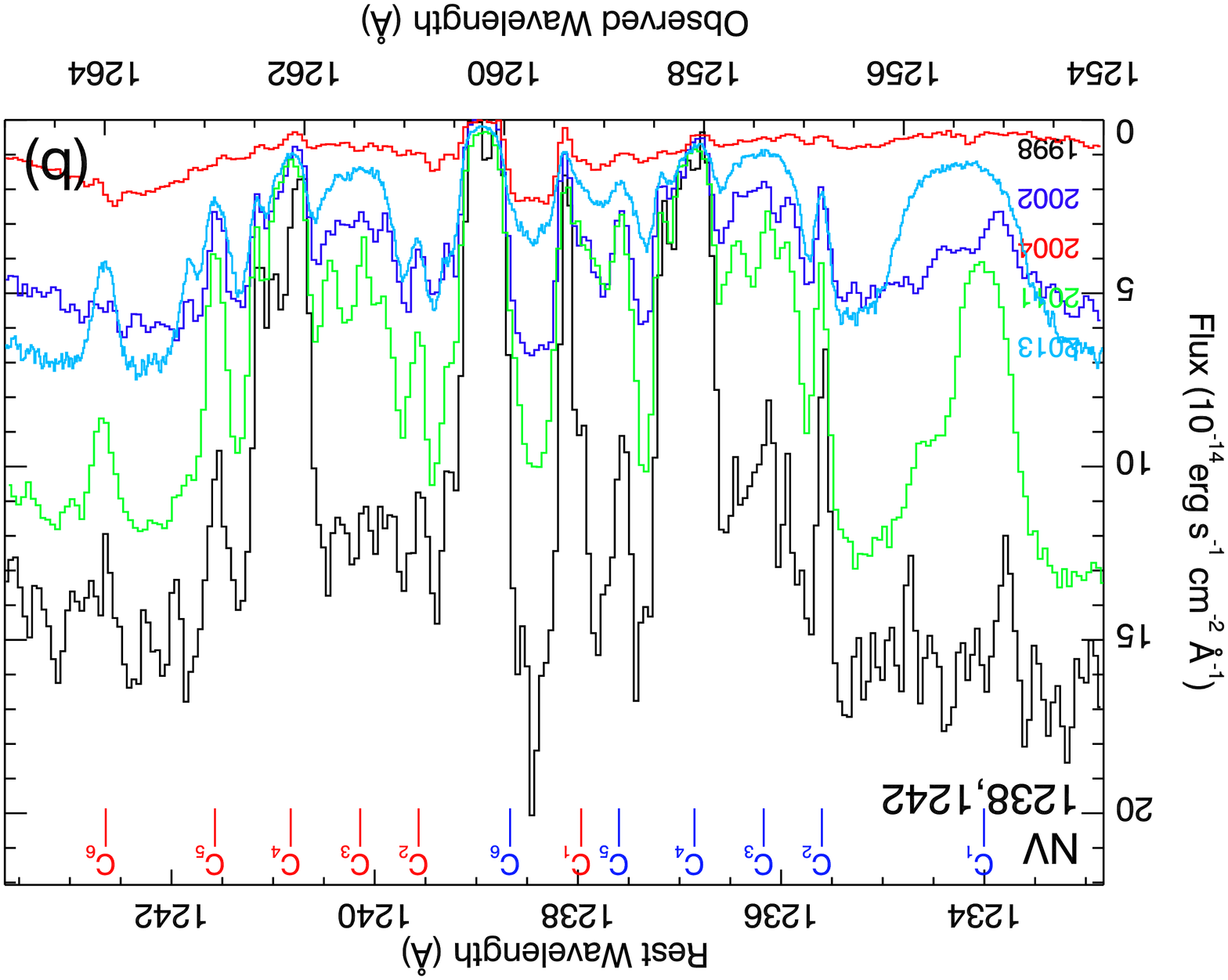}
\caption{continued.}\end{figure}\end{landscape}

\pagebreak
\begin{landscape}\begin{figure}
\includegraphics[width=0.95\textwidth,angle=90]{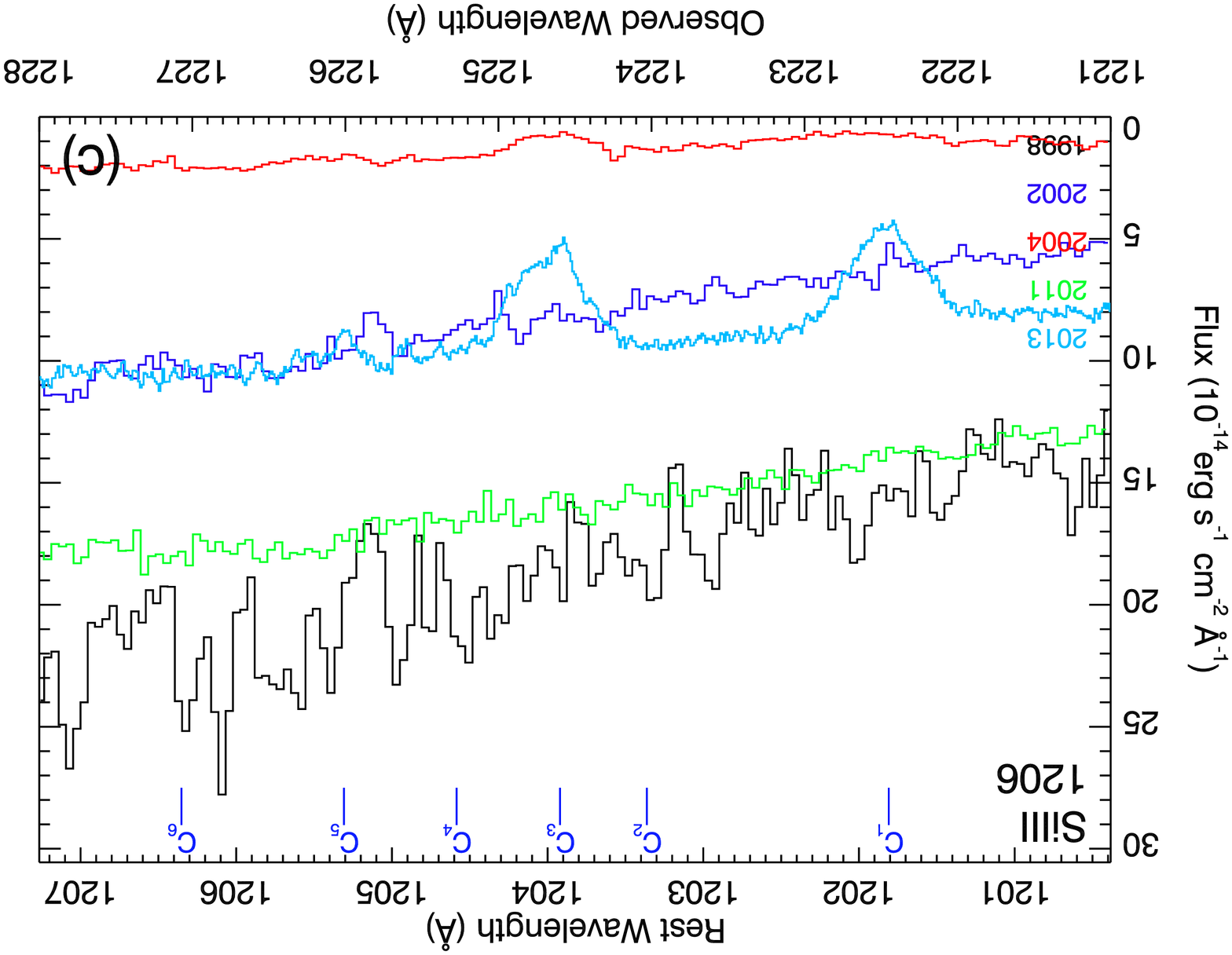}
\caption{continued.}\end{figure}\end{landscape}

\pagebreak
\begin{landscape}\begin{figure}
\includegraphics[width=0.95\textwidth,angle=90]{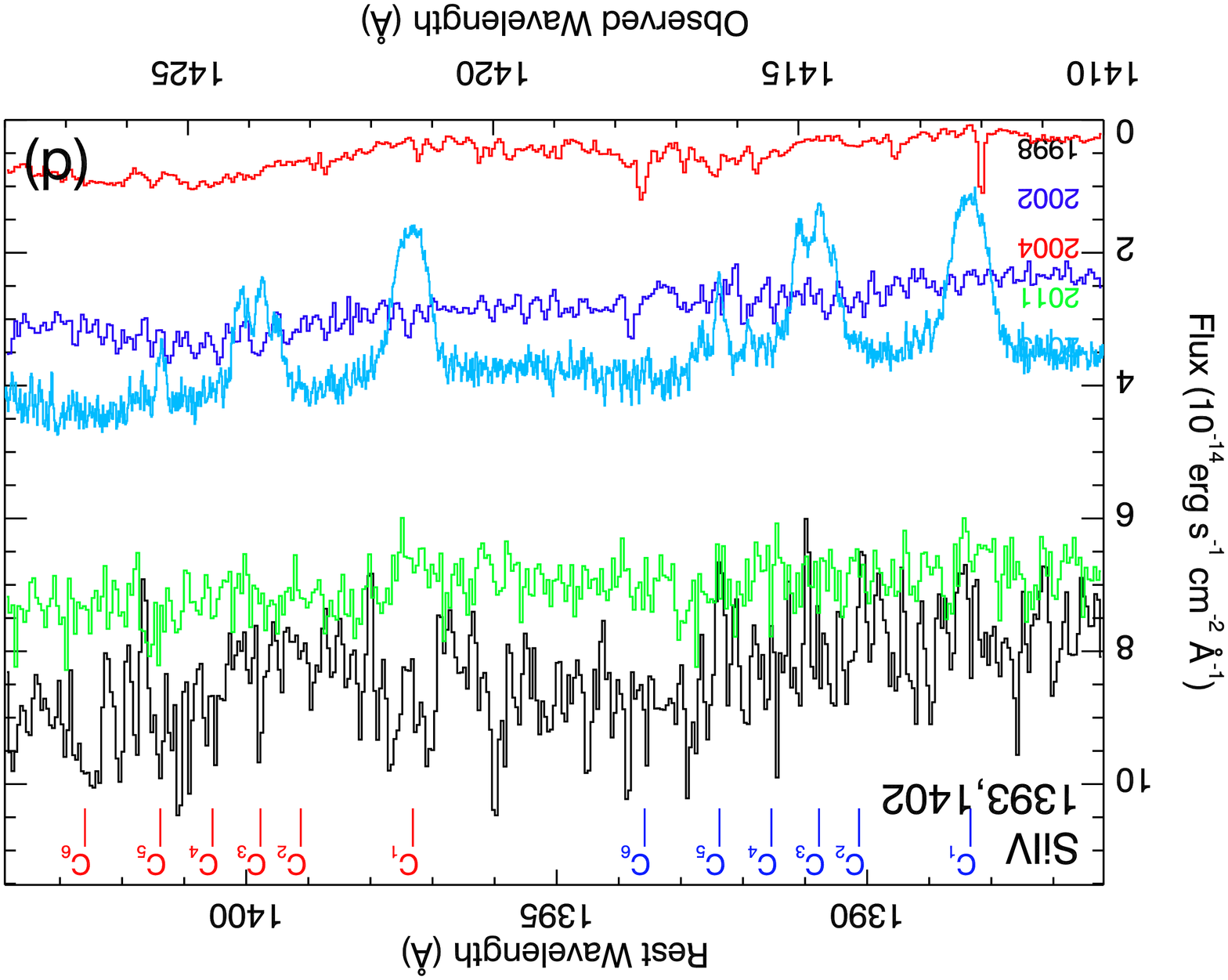}
\caption{continued.}\end{figure}\end{landscape}

\pagebreak
\begin{landscape}\begin{figure}
\includegraphics[width=0.95\textwidth,angle=90]{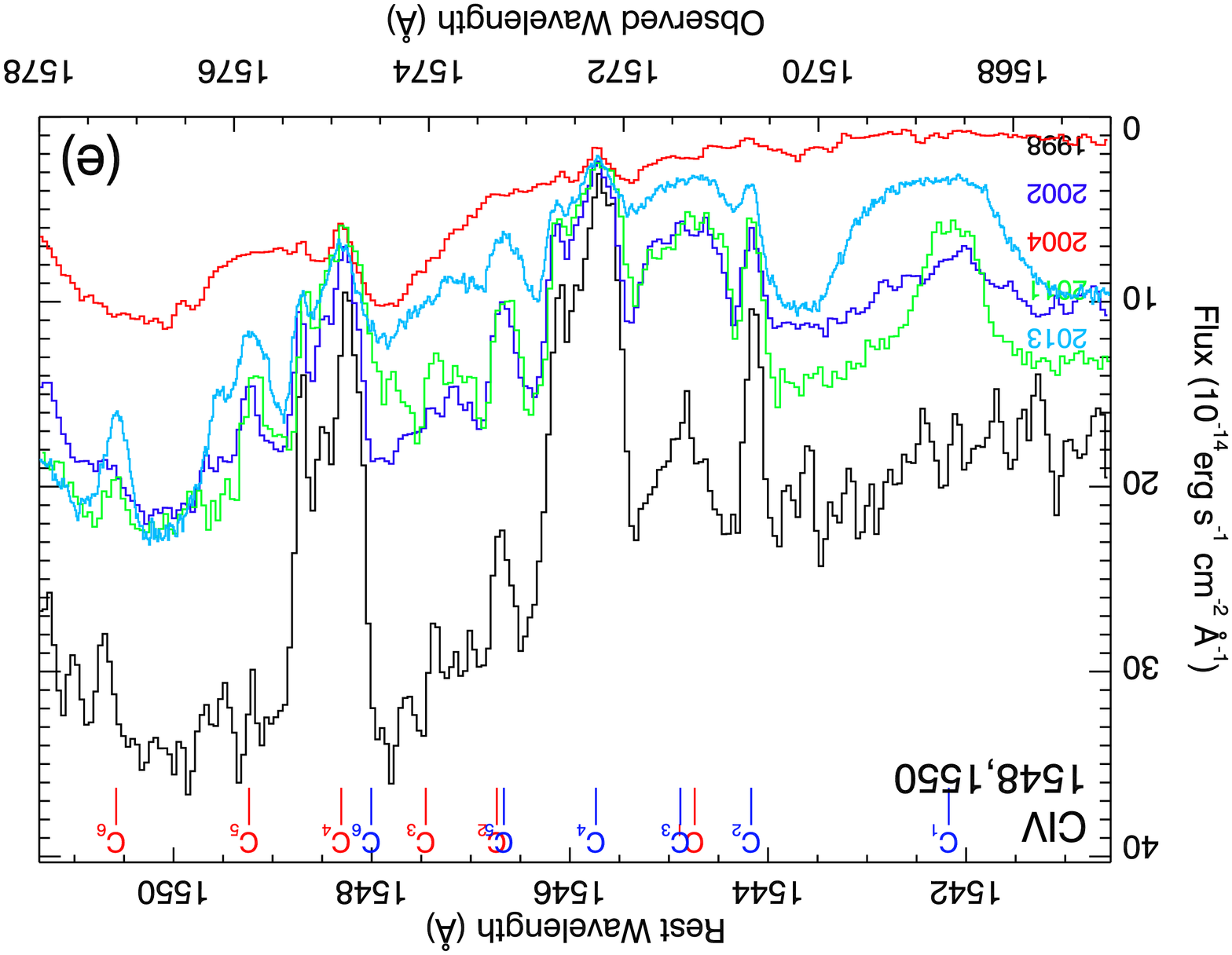}
\caption{continued.}\end{figure}\end{landscape}

\setcounter{figure}{0}
\renewcommand{\thefigure}{A.3.\alph{figure}}

\pagebreak
\begin{landscape}\begin{figure}
\includegraphics[width=0.95\textwidth,angle=90]{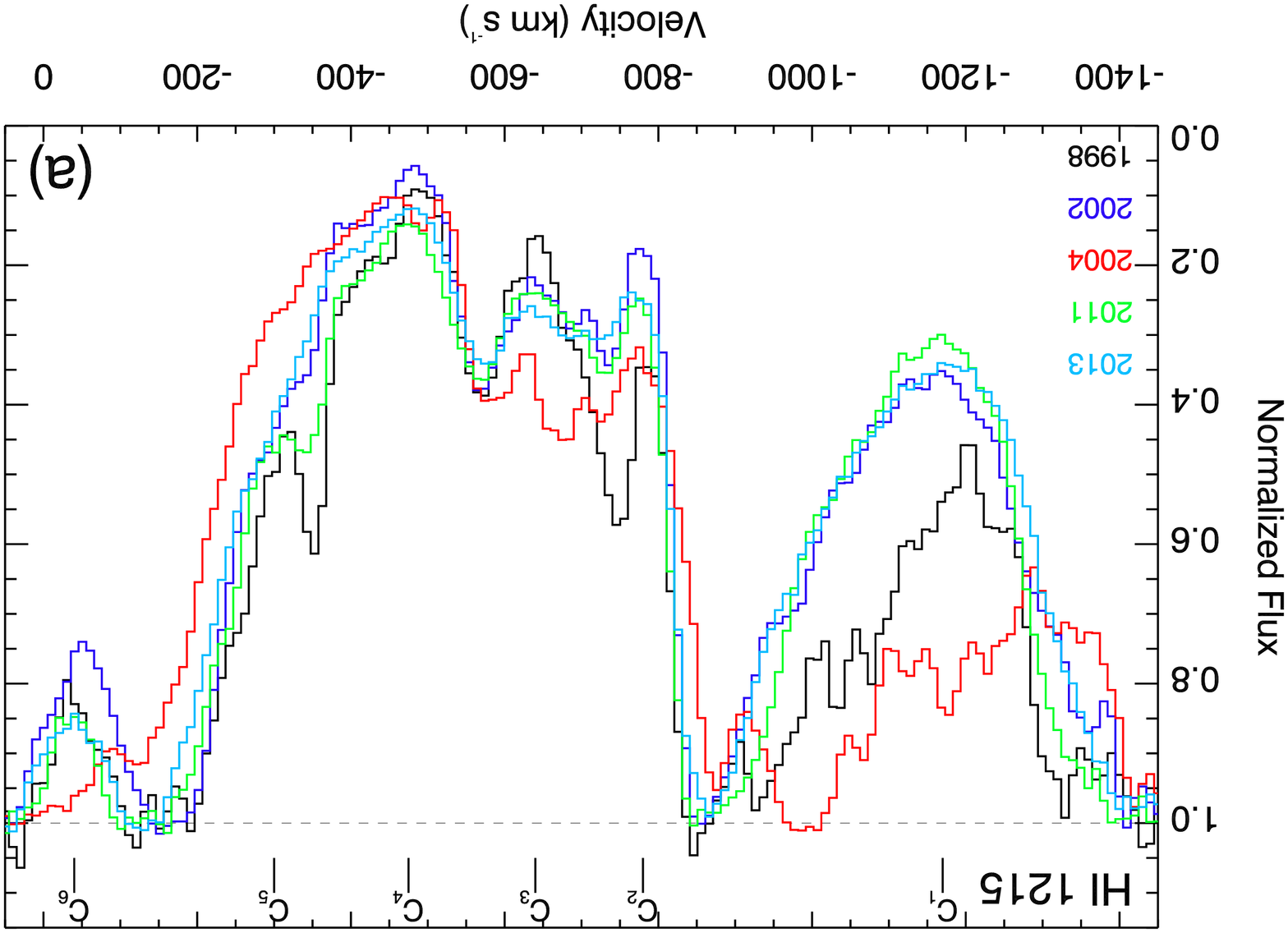}
\parbox{1.25\textheight}{\caption{A plot of the normalized spectrum of NGC 5548 during the five epochs of observation, plotted in the velocity rest-frame of the quasar (same annotation as Fig A.2).}}\end{figure}\end{landscape}

\pagebreak
\begin{landscape}\begin{figure}
\includegraphics[width=0.95\textwidth,angle=90]{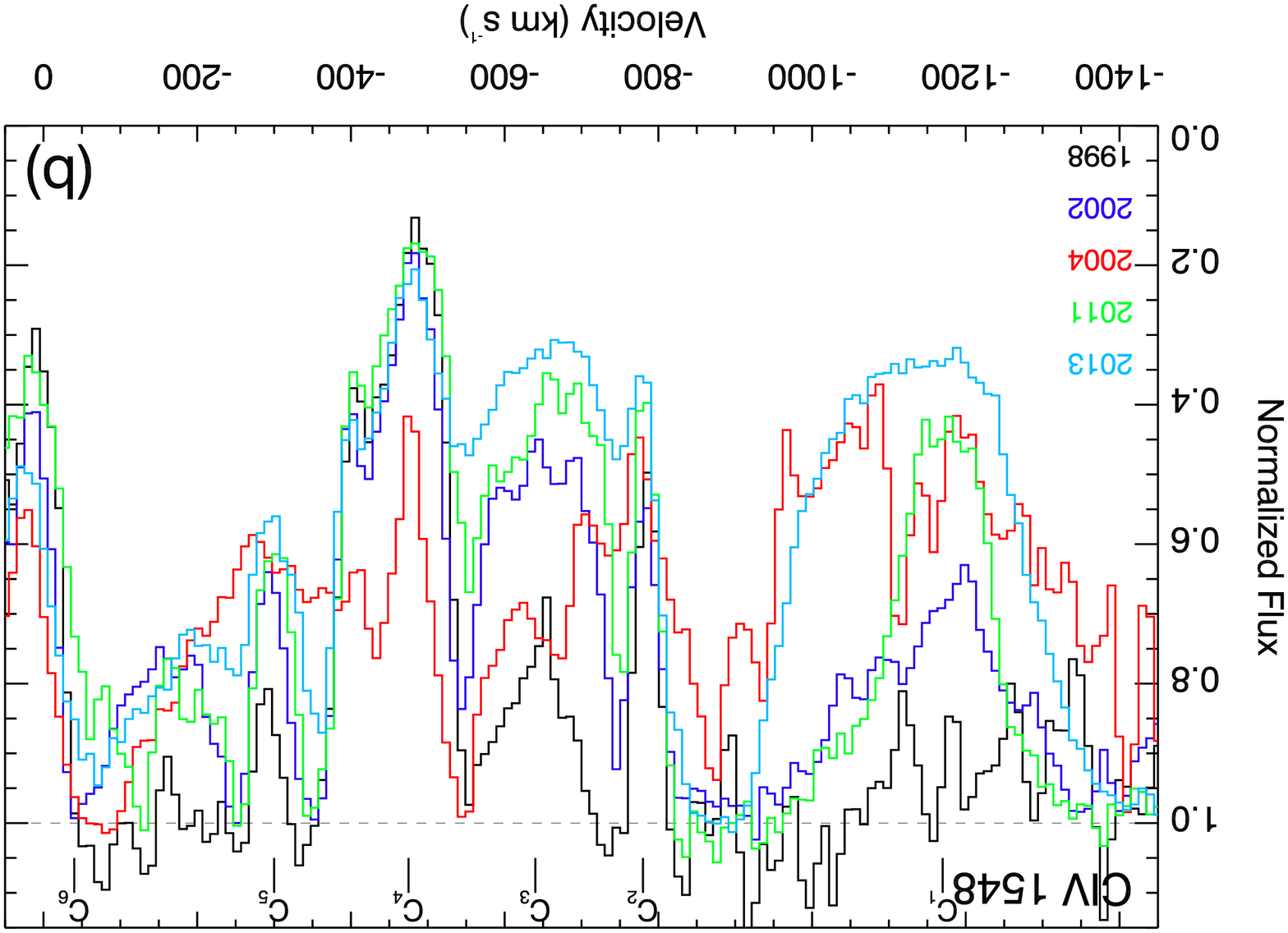}
\caption{continued.}\end{figure}\end{landscape}

\pagebreak
\begin{landscape}\begin{figure}
\includegraphics[width=0.95\textwidth,angle=90]{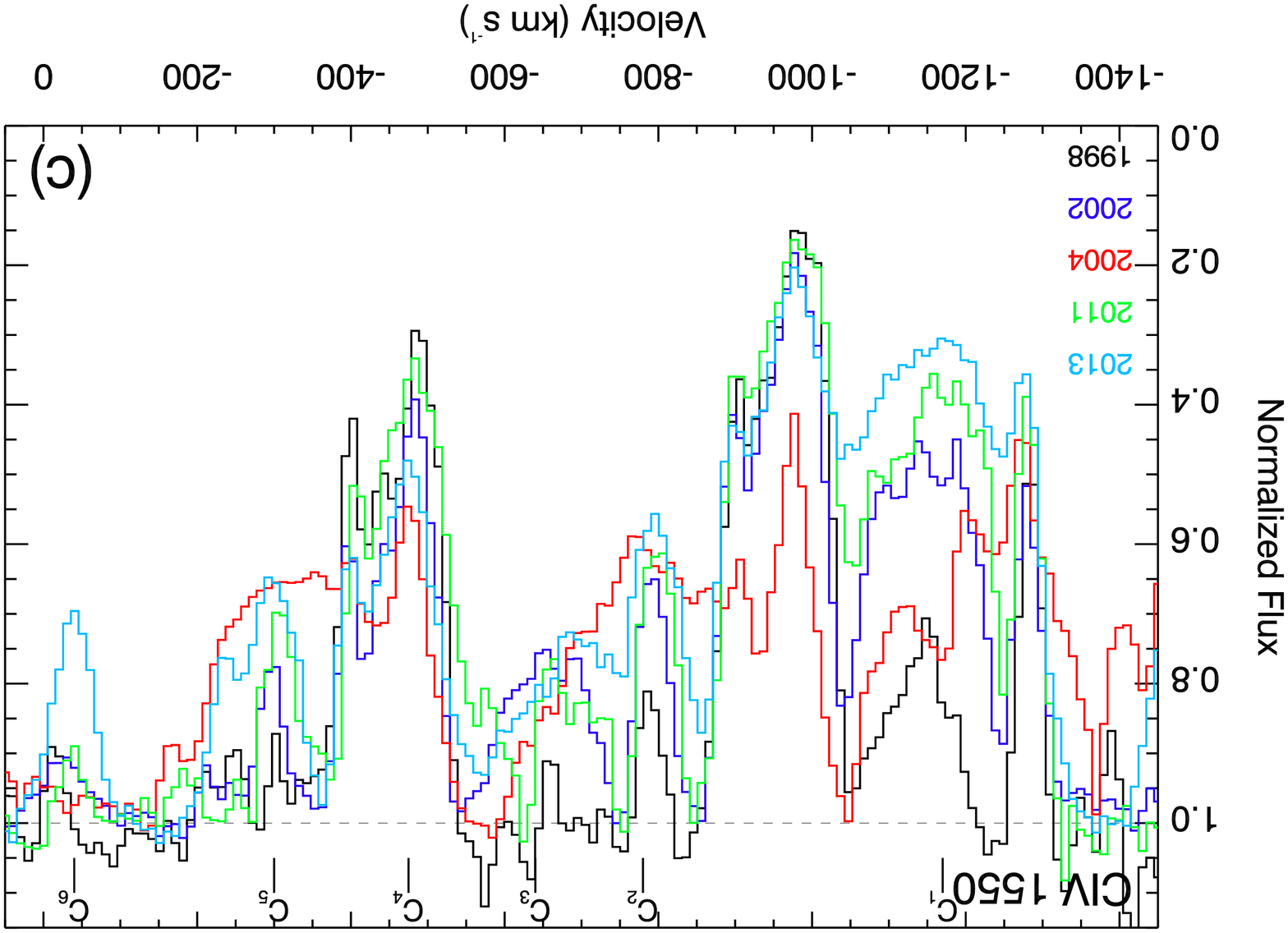}
\caption{continued.}\end{figure}\end{landscape}

\pagebreak
\begin{landscape}\begin{figure}
\includegraphics[width=0.95\textwidth,angle=90]{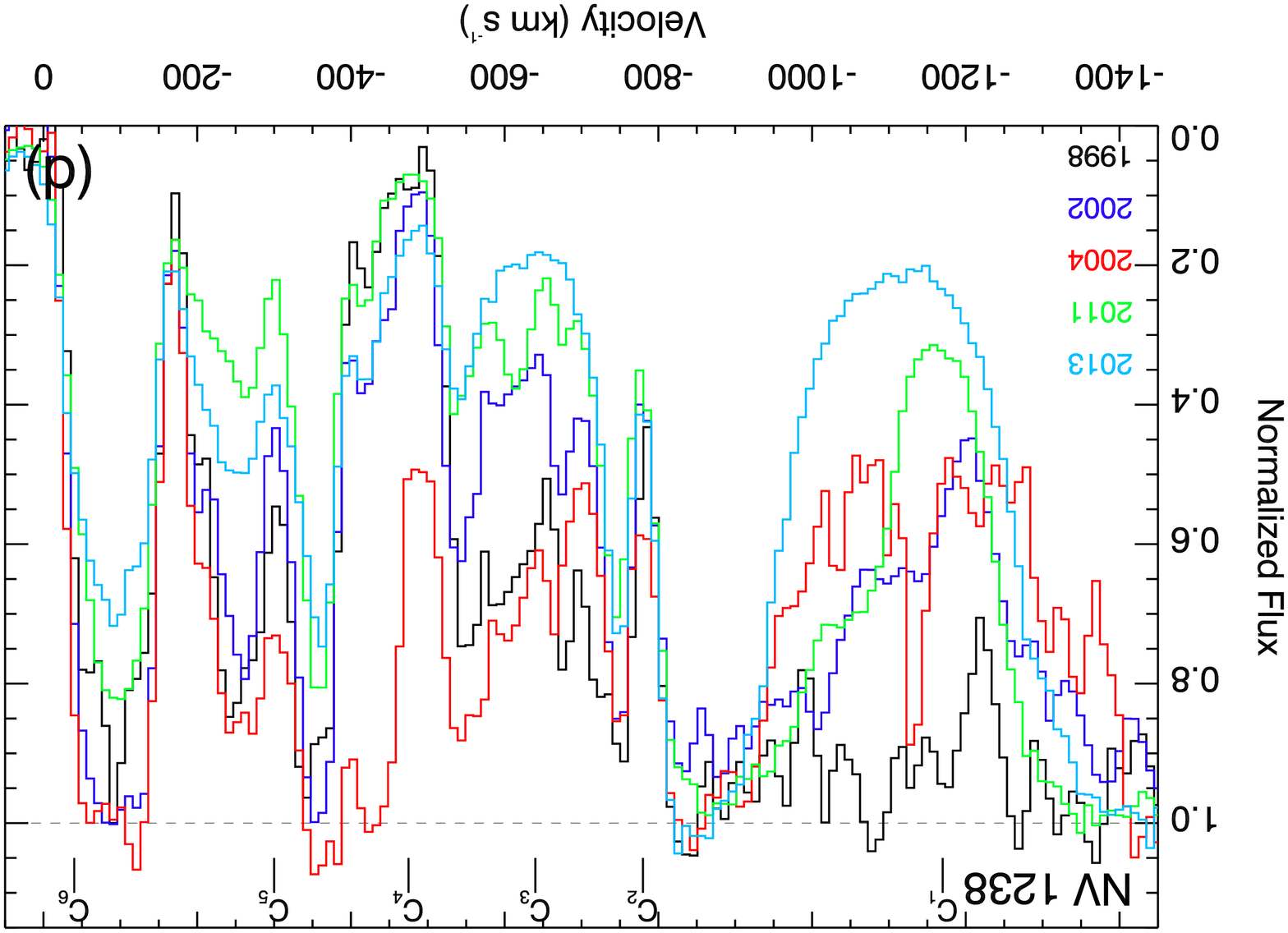}
\caption{continued.}\end{figure}\end{landscape}

\pagebreak
\begin{landscape}\begin{figure}
\includegraphics[width=0.95\textwidth,angle=90]{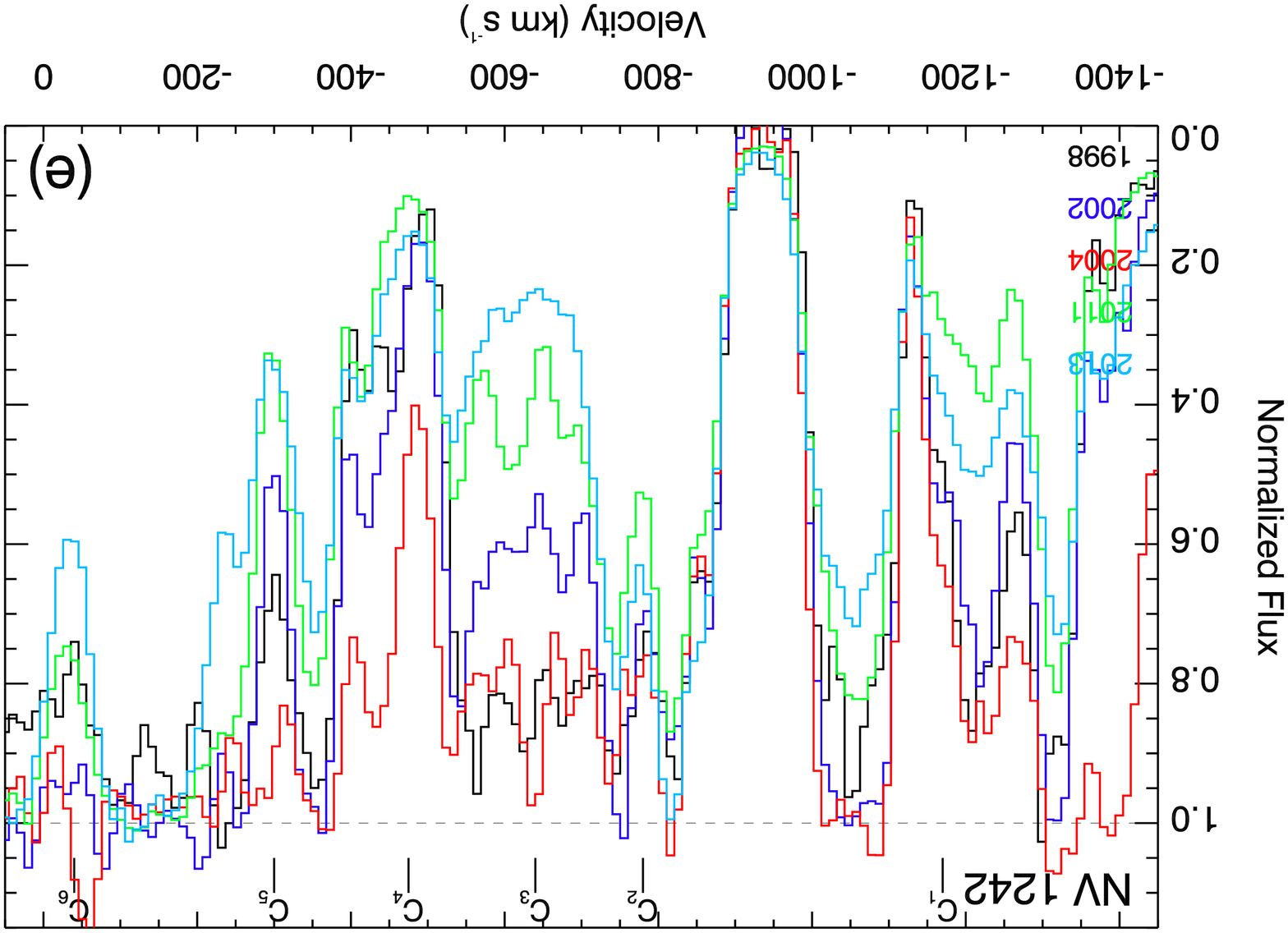}
\caption{continued.}\end{figure}\end{landscape}

\pagebreak
\begin{landscape}\begin{figure}
\includegraphics[width=0.95\textwidth,angle=90]{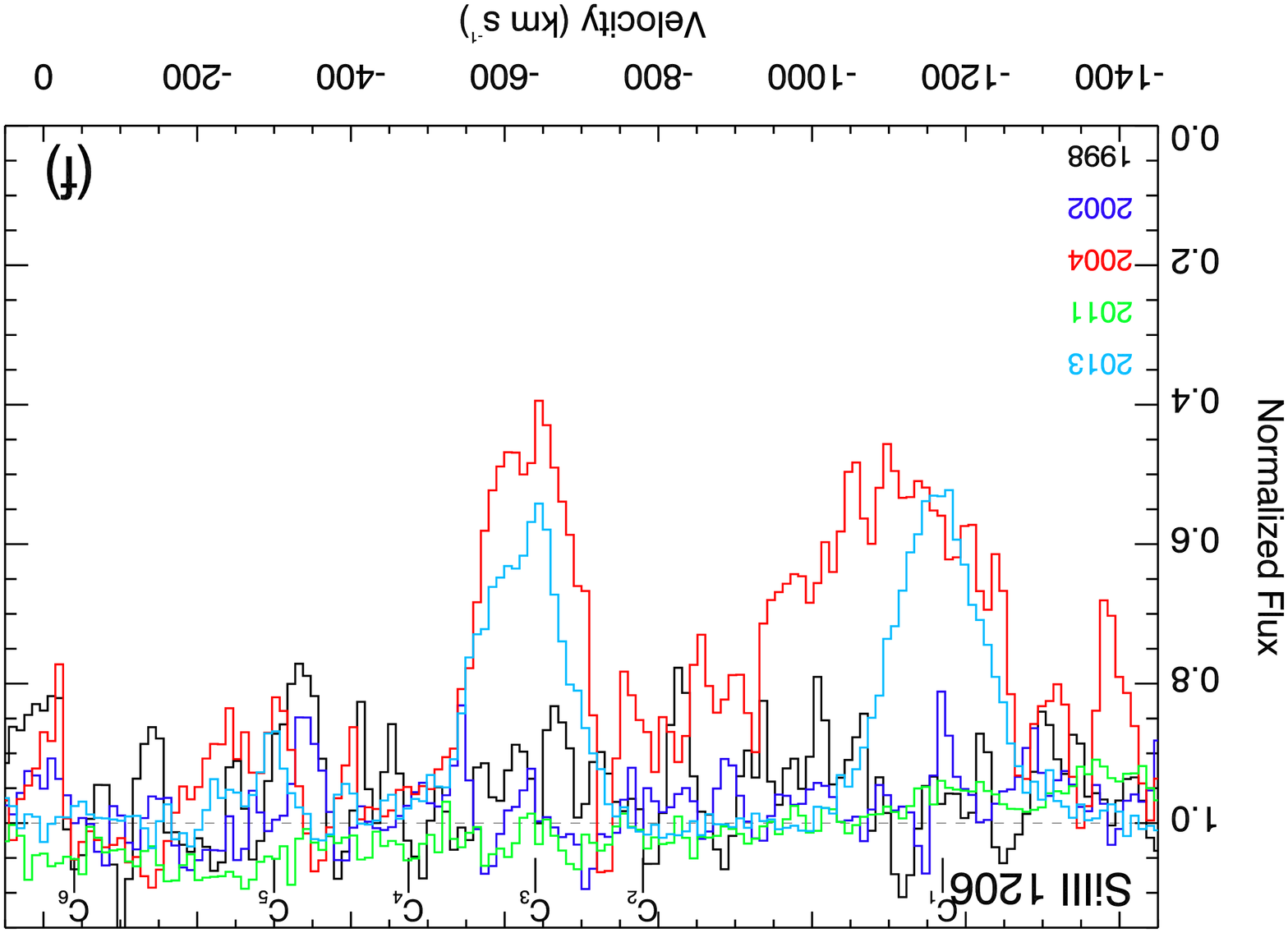}
\caption{continued.}\end{figure}\end{landscape}

\pagebreak
\begin{landscape}\begin{figure}
\includegraphics[width=0.95\textwidth,angle=90]{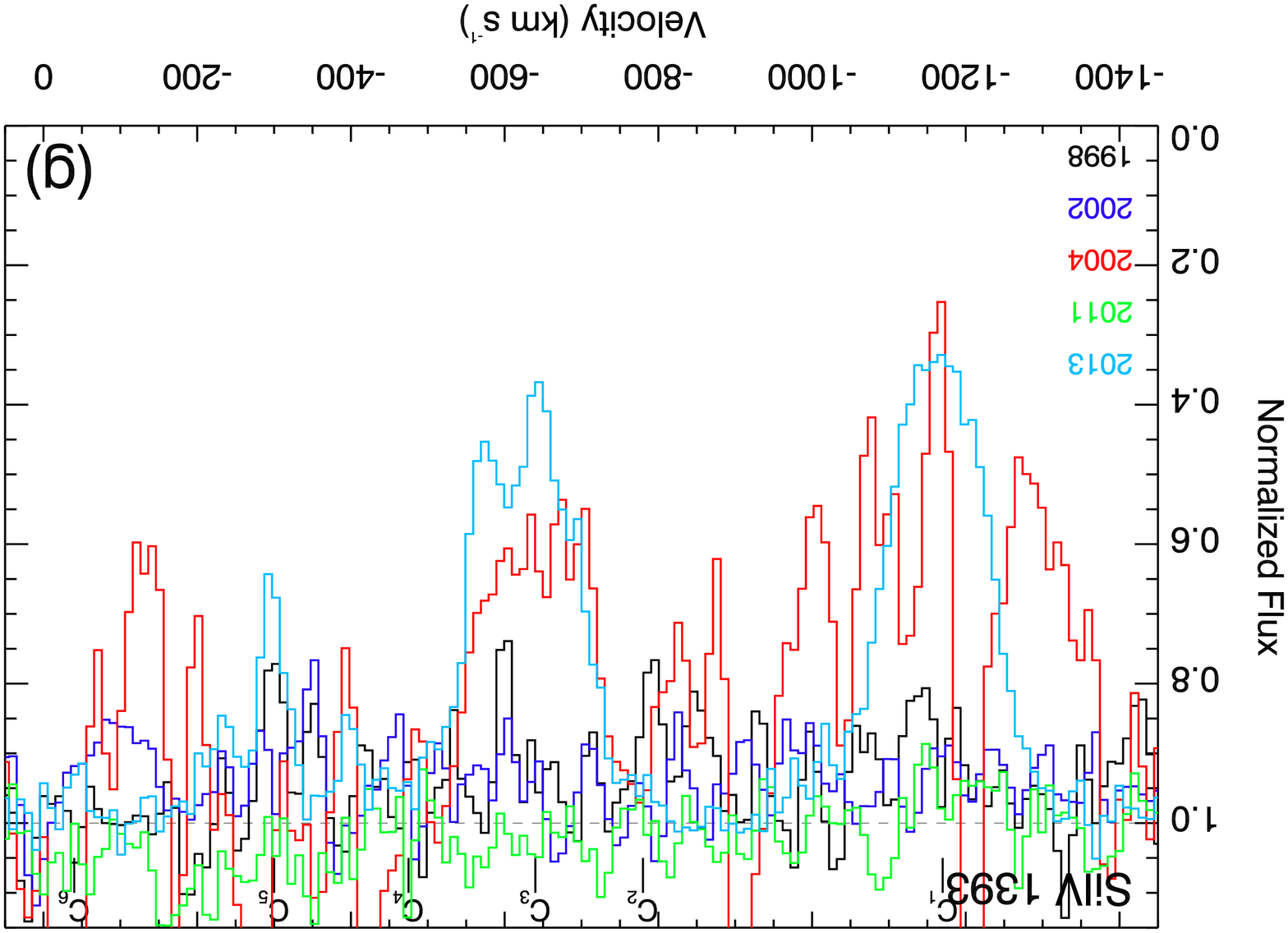}
\caption{continued.}\end{figure}\end{landscape}

\pagebreak
\begin{landscape}\begin{figure}
\includegraphics[width=0.95\textwidth,angle=90]{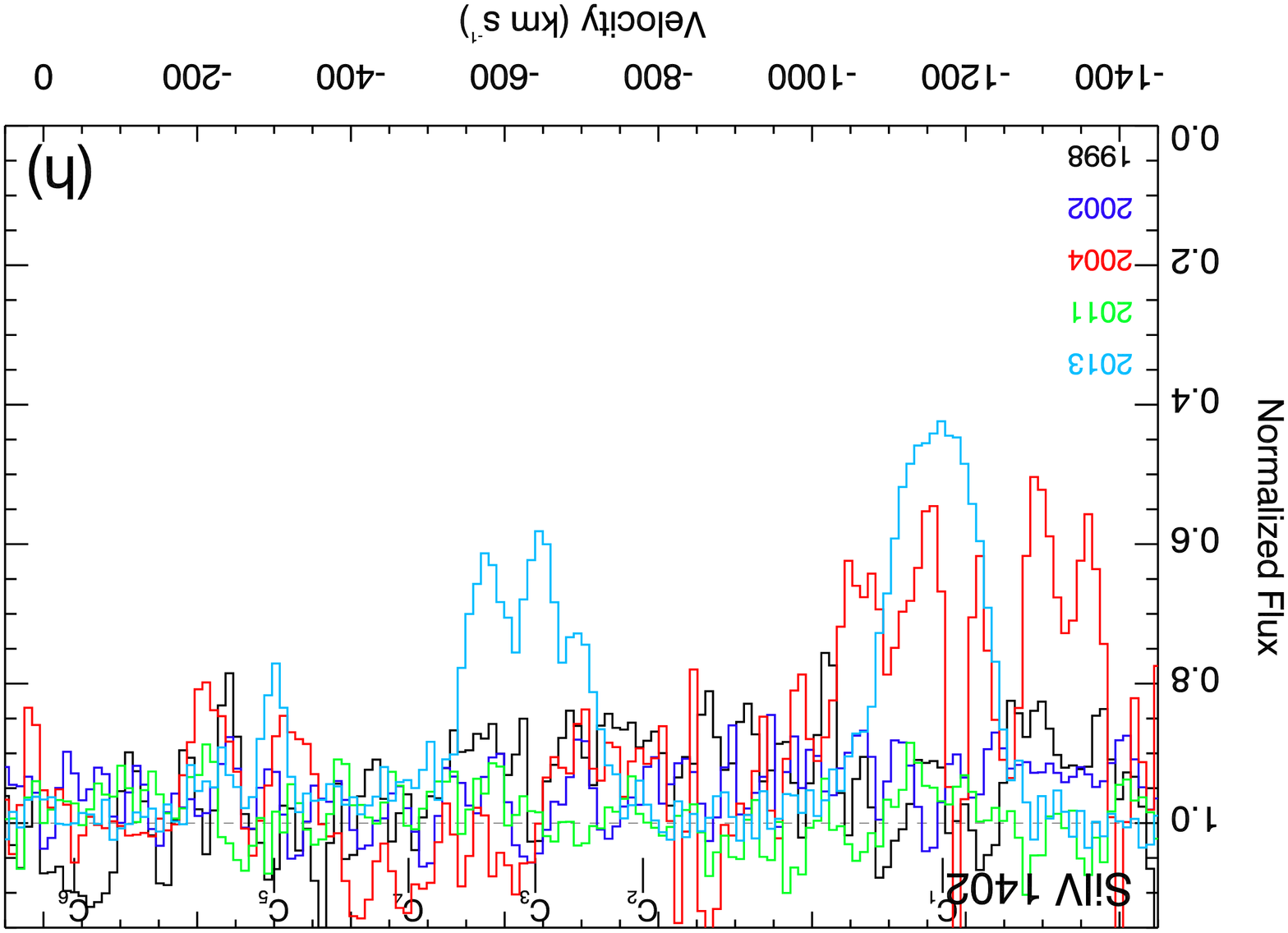}
\caption{continued.}\end{figure}\end{landscape}

\end{document}